\documentclass[aps,prd,amssymb,eqsecnum,nofootinbib,floatfix, a4paper,twocolumn]{revtex4}%{revtex4-1} %showkeys, showpacs showpacs%twocolumn,%

\usepackage{mathrsfs}
\usepackage{latexsym,amsmath,amsfonts,amssymb}%,showkeys}
\usepackage{bm}

\DeclareMathOperator{\arcsinh}{arcsinh}

\usepackage{graphicx}

\allowdisplaybreaks

\newcommand{\be}{\begin{equation}}
\newcommand{\ee}{\end{equation}}
\newcommand{\bea}{\begin{eqnarray}}
\newcommand{\eea}{\end{eqnarray}}

\newcommand{\D}{\partial}
\newcommand{\E}{E_{\rm real}}
\newcommand{\Ef}{{\mathcal E}_{\rm eff}}
\newcommand{\e}{\widehat{\mathcal E}_{\rm eff}}

\newcommand{\hHfS}{{{\widehat H}_{\rm Schw}}}
\newcommand{\p}{{\mathbf p}}
\newcommand{\bP}{{\mathbf P}}
\newcommand{\hQ}{\widehat Q}
\newcommand{\bu}{\bar u}
\newcommand{\br}{\bar r}
\newcommand{\bR}{\bar R}
\newcommand{\bq}{\bar q}

\newcommand{\hh}{ \widehat \hbar}
\newcommand{\x}{{\mathbf x}}
\newcommand{\X}{{\mathbf X}}
\newcommand{\vk}{{\mathbf k}}

\newcommand{\g}{{\gamma}}
\newcommand{\peob}{p_{\rm eob}}
\newcommand{\pinf}{p_{\infty}}

\newcommand{\sQ}{{\sf Q}}

\newcommand{\cM}{{\cal M}}
\newcommand{\s}{{\sigma}}

\begin{document}

\title{Classical and quantum scattering  in post-Minkowskian gravity}

\author{Thibault Damour}
\email{damour@ihes.fr}
\affiliation{Institut des Hautes Etudes Scientifiques, 35 route de Chartres, 91440 Bures-sur-Yvette, France}

\date{\today}

\begin{abstract} 
New structural properties of post-Minkowskian (PM) gravity are derived, notably within its
effective one body (EOB) formulation. Our results concern both the mass dependence,
and the high-energy behavior, of the classical scattering angle. We generalize our previous work  
by deriving, up to the fourth post-Minkowskian (4PM) level included,
 the explicit links between the scattering angle and the two types of potentials entering the Hamiltonian
 description of PM dynamics within EOB theory. We compute the scattering amplitude derived from quantizing 
 the third post-Minkowskian (3PM) EOB radial potential (including the contributions coming from the Born iterations),
 and point out various subtleties in the relation between perturbative amplitudes and classical dynamics.
We highlight an apparent tension between the  classical 3PM dynamics derived 
 by Bern {\it et al.} [Phys.\ Rev.\ Lett.\  {\bf 122}, 201603 (2019)], and previous high-energy self-force results
 [Phys.\ Rev.\ D {\bf 86}, 104041 (2012)],  and propose several possible resolutions of this tension.
 We point out that linear-in-mass-ratio self-force computations can give access to the exact 3PM and 4PM dynamics.
 \end{abstract}

\maketitle

\section{Introduction}

The recent, dramatically successful, beginning of gravitational-wave astronomy
\cite{Abbott:2016blz,Abbott:2016nmj,Abbott:2017vtc,Abbott:2017oio}, and the expected future improvements
in the sensitivity of gravitational-wave detectors, give a renewed motivation for improving our theoretical knowledge of the
gravitational dynamics of two-body systems in General Relativity. Our current knowledge of the dynamics and
gravitational-wave emission of binary systems has been acquired by combining several types of (interrelated) analytical
approximations schemes, and furthermore, by completing analytical results
with the results of a certain number of numerical simulations of coalescing binary black holes. 
The main types of analytical schemes that
have been used are: post-Minkowskian (PM), post-Newtonian (PN), multipolar-post-Minkowskian, effective-one-body (EOB),
black-hole-perturbation, gravitational self-force (SF), and effective-field-theory (EFT).

Recently, a new avenue for improving our theoretical knowledge of gravitational dynamics\footnote{We shall not discuss here
the related issue of improving our knowledge of gravitational-wave emission by amplitude methods; see Ref. \cite{Laddha:2018rle}
and references therein.} has been actively pursued.
It consists of translating the (classical or quantum) scattering observables of gravitationally interacting two-body
systems into some Hamiltonian counterpart. The idea of mapping quantum gravitational scattering amplitudes
onto some type of gravitational potential had been first explored long ago 
\cite{Corinaldesi:1956,Barker:1966zz,Barker:1970zr,Corinaldesi:1971sz,Iwasaki:1971vb,Okamura:1973my}.
The idea of these works was to construct a two-body Hamiltonian of the type
\begin{eqnarray} \label{Husual}
H(\x_1,\x_2,\p_1,\p_2) &=& c^2\sqrt{ m_1^2 + \frac{\p_1^2}{c^2}} +
 c^2\sqrt{ m_2^2 + \frac{\p_2^2}{c^2}} \nonumber \\ &+& V(\x_1-\x_2,\p_1,\p_2)\,,
\end{eqnarray}
such that the scattering amplitude in the momentum-dependent potential $V(\x_1-\x_2,\p_1,\p_2)$ 
(given by a usual Born-type expansion) is equal to the scattering amplitude computed by means of the
Feynman-diagrams defined by a (perturbative) quantum field theory comprising two scalar fields
$\phi_1$, $\phi_2$ (of masses $m_1$ and $m_2$) interacting via perturbatively quantized Einstein gravity.
This was done within the framework of the PN approximation scheme, {\it i.e.}, using a small-velocity expansion, 
and working actually with the PN-expanded form of the Hamiltonian, up to some finite (and rather low) accuracy:
\begin{eqnarray} \label{HPN}
&&H(\x_1,\x_2,\p_1,\p_2) = (m_1+m_2)c^2 + \frac{\p_1^2}{2 m_1}+ \frac{\p_2^2}{2 m_2} 
\nonumber \\ & & -\frac{\p_1^4}{8 m_1^3c^2}   - \frac{\p_2^4}{8 m_2^3c^2} +\cdots + V^{\rm PN}(\x_1-\x_2,\p_1,\p_2)\,,\nonumber\\
\end{eqnarray}
with
\be
V^{\rm PN}(\x_1-\x_2,\p_1,\p_2)= - \frac{G m_1 m_2}{|\x_1-\x_2|} + {\rm PN \,corrections}\,.
\ee
This did not yield at the time
results that could not be (often more efficiently) obtained by conventional PN classical computations\footnote{Let us note that
Corinaldesi \cite{Corinaldesi:1956}  incorrectly concluded that the full 1PN Einstein-Infeld-Hoffmann equations
of motion could be derived from the one-graviton-exchange amplitude. The first formally correct and complete derivation of the
1PN Hamiltonian from the {\it one-loop} scattering amplitude of two scalar particles is due to Iwasaki \cite{Iwasaki:1971vb}}.
A similar approach was also
used in Quantum Electrodynamics to derive the $(v^2/c^2)$-accurate (first post-Coulombian)  Breit Hamiltonian. See, notably,
the fourth volume of the Landau-Lifshitz treatise of theoretical physics \cite{Landau4} which derives the 
 Breit Hamiltonian by starting from the scattering amplitude ${\cal A}$ of two massive, charged particles.

The idea of extracting classical gravitational dynamics from the scattering amplitude $\cM$ of two gravitationally interacting
massive particles has been further explored and extended in more recent papers 
\cite{Donoghue:1993eb,Donoghue:1994dn,BjerrumBohr:2002kt,Bjerrum-Bohr:2013bxa,Neill:2013wsa,Cachazo:2017jef}.
However, these works limited their ambition to extracting leading terms in the PN expansion of the dynamics.

It is only  recently that the issue of linking the gravitational scattering amplitude $\cM$ to PM gravity, {\it i.e.}, without using a small-velocity expansion, has been explored. This was done at the second post-Minkowskian (2PM) level ({\it i.e.}, $O(G^2)$ or one-loop) in Refs. \cite{Guevara:2017csg,Damour2018,Bjerrum-Bohr:2018xdl,Cheung:2018wkq,KoemansCollado:2019ggb}, and at the third 
post-Minkowskian (3PM) level ({\it i.e.}, $O(G^3)$ or two-loop) in the  breakthrough work of Bern {\it et al.} \cite{Bern:2019nnu,Bern:2019crd}. Before the latter work, the only extant two-loop result was the trans-Planckian,
eikonal-approximation two-loop result of Amati, Ciafaloni and Veneziano (ACV) \cite{Amati:1990xe}
(which was recently generalized \cite{DiVecchia:2019kta,Bern2020}, and confirmed \cite{Bern2020}). 
[Ref. \cite{Damour2018} has extracted both 3PM and 4PM classical information from the result of ACV.]
Let us also mention some further (partly conjectural) work concerning the link between the gravitational scattering amplitude
of spinning particles and the  classical  gravitational interaction of Kerr black holes \cite{Guevara:2018wpp,Chung:2018kqs,Guevara:2019fsj,Arkani-Hamed:2019ymq,Siemonsen:2019dsu}, as well as work on the computation of classically measurable quantities
from on-shell amplitudes \cite{Kosower:2018adc,Maybee:2019jus}.

Those recent works dealing with PM gravity in connection with the quantum amplitude $\cM$ 
have been preceded by older investigations,
using purely classical methods, of the PM expansion of the gravitational dynamics of two-body systems.  The  
first post-Minkowskian (1PM; $O(G^1)$) dynamics was  studied in Refs. \cite{Bertotti1956,Portilla:1980uz,Ledvinka:2008tk},
while the second post-Minkowskian (2PM; $O(G^2)$) one was tackled in Refs. \cite{Westpfahl:1979gu,Bel:1981be,Westpfahl:1985,Westpfahl:1987}). More recently, the investigation of classical PM gravity has been revived by showing how the
 EOB formalism \cite{Buonanno:1998gg,Buonanno:2000ef,Damour:2000we} was able to provide a much simplified description
 of PM gravity, based on the gauge-invariant information contained in the scattering function $\frac12 \chi(E,J)$.
  In particular: (i) Ref. \cite{Damour:2016gwp} has shown how the 1PM-accurate classical scattering of two non-spinning bodies could be transcribed, within the EOB formalism  into the geodesic dynamics of 
a particle of mass $\mu=m_1 m_2/(m_1+m_2)$ in a (linearized) Schwarzschild background of mass $M=m_1+m_2$.
[This EOB formulation of the 1PM dynamics is much simpler than the previously obtained Arnowitt-Deser-Misner one \cite{Ledvinka:2008tk}.];
(ii) Ref. \cite{Vines:2017hyw} has shown how to transcribe within the EOB formalism the 1PM gravitational interaction of  
spinning bodies at all orders in the spins (see also \cite{Bini:2017xzy});
(iii) Ref. \cite{Damour2018} derived, for the first time, a
next-to-leading-order, $O(G^2)$ (second-post-Minkowskian, 2PM) Hamiltonian EOB 
description of the (non-spinning) two-body dynamics from
the classical 2PM scattering angle \cite{Westpfahl:1985} (This EOB description of the 2PM dynamics is equivalent, but simpler,
than the one later derived in \cite{Cheung:2018wkq}, using a potential of the form of Eq. \eqref{Husual}); 
(iv) Ref. \cite{Bini:2018ywr} derived (by using the 2PM-accurate metric of Ref. \cite{Bel:1981be}) a 2PM-accurate Hamiltonian EOB
 description of the gravitational interaction of two spinning bodies at linear order in spins; and (v) a conjectural 2PM-level generalization 
 of the 1PM result of Ref. \cite{Vines:2017hyw} concerning the non-linear-in-spin dynamics of aligned-spin bodies was proposed in 
 Ref. \cite{Vines:2018gqi}.
  In addition, the 5PN-level truncation of the 
 classical 3PM dynamics extracted from the two-loop result of Bern {\it et \,al.} 
 \cite{Bern:2019nnu,Bern:2019crd} (see also Ref. \cite{Antonelli:2019ytb}) has been confirmed
 by an independent, purely classical computation \cite{Bini:2019nra}. See below for the discussion
 of  more recent, classical and quantum, 6PN-level confirmations.

The main aim of the present work is to derive some structural properties of the classical scattering angle, $\chi$, considered
as a function of the various arguments in which it can be expressed: energy, angular momentum, impact parameter, and masses.
This will allow us to derive several new results of direct importance for improving our current knowledge of the
dynamics of two-body systems. In particular, we shall derive a property of the dependence of  $\chi$ on the masses
which was crucially used in Ref. \cite{Bini:2019nra} for determining most of the mass dependence of the 5PN-level dynamics.
We shall also discuss a constraint  on the high-energy behavior of $\chi$ that follows from the SF result of Ref. \cite{Akcay:2012ea}.
The latter high-energy constraint seems to be discrepant with  the high-energy (or massless) limit of the
3PM-level results of Bern {\it et al.} \cite{Bern:2019nnu,Bern:2019crd}. We will suggest two types of possible resolutions
of this apparent discrepancy. One resolution consists in conjecturing that the 3PM dynamics is described by another
classical Hamiltonian, yielding the same 5PN-level $O(G^3)$ scattering angle (which was recently independently
obtained \cite{Bini:2019nra}), but a softer high-energy behavior than that of Refs. \cite{Bern:2019nnu,Bern:2019crd}.
Another resolution consists in conjecturing a special structure of the 4PM ($O(G^4)$) dynamics,
such that its high-energy behavior modifies the consequences drawn from considering the high-energy behavior
of the  3PM-level-only result of Bern {\it et al.}. Both types of resolutions will be shown to lead to a classical massless
scattering angle that disagrees with the one derived from the eikonal-approximated quantum two-loop massless 
amplitude \cite{Amati:1990xe,Bern2020}.

A secondary aim of the present work is to 
clarify the various links between the  physical quantities 
involved in the maps that have been recently used to relate classical and quantum dynamics. These three quantities are: the classical scattering angle $\chi$, the quantum scattering amplitude $\cM$ (considered in a limit formally
corresponding to classical scattering), and the two different potentials (EOB-type \cite{Damour2018} or 
EFT-type \cite{Cheung:2018wkq}) used to transcribe
(classical or quantum) scattering observables into an Hamiltonian description. In this connection, we will explicitly derive below
the map going from the 3PM-level classical Hamiltonian to the corresponding piece of the  two-loop amplitude.
 [Some of the results derived below (which have been presented in various talks \cite{Damourtalks}), have been recently
discussed from quite different (non-EOB-based) perspectives in two papers \cite{Kalin:2019rwq,Bjerrum-Bohr:2019kec}.]
Our 3PM-level map will be found to be fully compatible with the corresponding results in section 10 of Ref. \cite{Bern:2019crd},
but are more complete in that they detail the IR-divergent contributions coming from iterating the 1PM and 2PM levels, which
also contribute IR-finite terms.

In addition, we will point out various subtleties in the relation between perturbative amplitudes and classical dynamics.
 Several tools, concerning the link
between the classical PM dynamics and the quantum amplitude $\cM$,  have been presented in the recent literature 
 \cite{Amati:1990xe,
 Donoghue:1993eb,
 Donoghue:1994dn,
 BjerrumBohr:2002kt,
 Bjerrum-Bohr:2013bxa,
 Neill:2013wsa,
 Cachazo:2017jef,
 Guevara:2017csg,
 Damour2018,
 Bjerrum-Bohr:2018xdl,
 Cheung:2018wkq,
 Bern:2019nnu,
 Bern:2019crd,
 KoemansCollado:2019ggb,
 Kosower:2018adc,
 Maybee:2019jus}. These tools have been checked to give
 a correct result at the 2PM (one-loop) level \cite{Guevara:2017csg,
 Damour2018,
 Bjerrum-Bohr:2018xdl}. As the 3PM-level classical dynamics of  Refs. \cite{Bern:2019nnu,Bern:2019crd}
 has not yet been confirmed by an independent {\it classical} derivation, it might be useful to point out the
 existence of conceptual subtleties in the  map going from   the quantum $\cM$ towards the classical dynamics
 (which is the inverse of the classical-to-quantum map that we shall discuss below). We shall recall in this respect a classic result
 of Niels Bohr \cite{Bohr1948} highlighting the lack of overlap between the domains of validity of classical and quantum 
 (perturbative) scattering theory.

Technically speaking, we will be dealing below with the 3PM-accurate expansions ({\it i.e.}, the
expansions in powers of the gravitational constant $G$ up to $G^3$ included) of various
physical quantities: the classical (half) scattering angle
expressed as a function of (center-of-mass) energy ($E= \sqrt{s}$) and angular momentum ($J$),
\be \label{chiPM}
\frac12 \chi^{\rm }(E, J) = \frac{\chi_{1}(\e, \nu)}{j} + \frac{ \chi_{2}(\e, \nu)}{j^2} + 
 \frac{\chi_{3}(\e, \nu)}{j^3}  + O(G^4),
\ee
(see below the definitions of the dimensionless variables $\e$, $j$ and $\nu$);
the (relativistic) quantum scattering amplitude expressed as a function of Mandelstam
invariants $s= -(p_1+p_2)^2$ and $t= -(p'_1-p_1)^2$ (in the mostly-plus signature we use),
\be \label{cMPM}
\cM(s,t) = G \cM_{1}(s,t) + G^2 \cM_{2}(s,t)+ G^3 \cM_{3}(s,t) + O(G^4)\,;
\ee
and the PM expansions of the two (closely connected) types of EOB potentials describing 
the gravitational interaction of two classical masses. Namely, with $u\equiv GM/R_{\rm EOB}$,
and now including the 4PM, $O(G^4)$, contribution,
\be 
\hQ(p,u)= u^2 q_2(p) + u^3 q_3(p) + u^4 q_4(p) + \cdots \,,
\ee
and (with $\bu \equiv GM/{\bar R}_{\rm EOB}$; in isotropic coordinates)
\be
w(\g,\bu) = w_1(\g) \bu +  w_2(\g) \bu^2 +  w_3(\g) \bu^3 +  w_4(\g) \bu^4 + \cdots
\ee
As we will explicitly discuss, these EOB potentials are equivalent (and simpler) than the 
more traditional type of potential  $V(\x_1-\x_2,\p_1,\p_2)$ entering Eq. \eqref{Husual},
and used in the EFT-type formalism of Refs. \cite{Cheung:2018wkq,Bern:2019nnu,Bern:2019crd}.
We briefly discuss  in Appendix \ref{A}  the link between the EOB potentials and the  PM expansion
of the isotropic-gauge EFT-type potential \cite{Cheung:2018wkq} in the center of mass (c.m.) frame,
\be \label{VPM}
V(\bP, \X)= G \frac{c_1(\bP^2)}{|\X|}+ G^2 \frac{c_2(\bP^2)}{|\X|^2}+ G^3 \frac{c_3(\bP^2)}{|\X|^3}+ \cdots 
\ee
The precise technical meaning of the EOB potentials, $\hQ(p,u)$ and $w(\g,\bu)$, will be presented below.
On the right-hand side of Eq. \eqref{chiPM} we have replaced the total c.m. energy of the two-body system ,
$E = \E = \sqrt{s}$, by the corresponding
 dimensionless EOB ``effective energy" \cite{Buonanno:1998gg,Buonanno:2000ef,Damour:2000we,Damour:2016gwp},
\be \label{Ef}
\e \equiv \frac{{\mathcal E}_{\rm eff}}{\mu} \equiv  \frac{({E}_{\rm real})^2 - m_1^2  -m_2^2 }{2 \, m_1 m_2} =  \frac{s - m_1^2  -m_2^2 }{2 \, m_1 m_2}.
\ee
Let us note in advance that, in scattering situations, $\e$ is equal to the relative Lorentz gamma factor of the incoming worldlines,
denoted $\g$ below (and $\sigma$ in Refs. \cite{Bern:2019nnu,Bern:2019crd}).
In addition, we have replaced the total (c.m.) angular momentum $J$ by the dimensionless variable
\be \label{j}
  j \equiv \frac{J}{G m_1 m_2} = \frac{ J}{G \mu M}\,,
\ee
with
\be
 M \equiv m_1 +m_2;\:
 \mu \equiv \frac{m_1 m_2}{m_1+m_2};\:
 \nu  \equiv \frac{\mu}{M} = \frac{m_1 m_2}{(m_1+m_2)^2}.
\ee
As $1/j= Gm_1m_2/J$, the perturbative expansion of the (classical) scattering function in powers of the gravitational constant $G$ (i.e. its PM expansion) is  seen to be equivalent to an expansion in inverse powers of the angular momentum.

%%%%%%
\section{On the mass dependence of the classical two-body scattering function.} \label{sec2}
  
 The aim of the present section is to extract from PM perturbation theory simple rules constraining
 the mass dependence of the scattering function at each PM order. Though their technical origin is
rather simple, these rules turn out to give very useful constraints on the functional structure of the scattering 
function. The PM perturbation theory of interacting point masses has been worked out at the 2PM (one-loop)
level long ago \cite{Westpfahl:1979gu,Bel:1981be,Westpfahl:1985}. Recently,
 Refs. \cite{Damour:2016gwp,Damour2018,Bini:2018ywr} have outlined a formal iteration scheme for computing the
 PM expansion of the scattering function to all PM orders, and showed how it could be naturally expressed as a sum
 of Feynman-like diagrams (see Fig. 1 in \cite{Damour:2016gwp}, and Figs. 1 and 2 in \cite{Damour2018}).
 Let us recall this construction. The PM expansion of  the classical momentum transfer 
 (dubbed the ``impulse'' in Ref. \cite{Kosower:2018adc}), 
 {\it i.e.}, the total change $\Delta p_{a \mu}$, between the infinite past
and the infinite future, of the 4-momentum $p_{a \mu} = m_a u_{a \mu}$ of the particle labelled by $a=1,2$
, is obtained by  inserting on the right-hand
side of the  integral expression
\be \label{deltapmu}
\Delta p_{a \mu}
= -\frac{m_a}{2}\int_{- \infty}^{+\infty} d s_a \,\D_\mu g^{\alpha \beta}(x_a) \, u_{a \alpha} u_{a \beta}\,,
\ee
 the iterative solutions (in successive powers of $G$) of the combined system of equations describing the coupled evolution
 of the two worldlines
 \begin{align} \label{eoma}
&\frac{d x_a^{ \mu}}{d s_a}=   g^{\mu \nu}(x_a) u_{a \nu}\,, \nonumber \\
&\frac{d u_{a \mu}}{d s_a} = -\frac12 \,\D_\mu g^{\alpha \beta}(x_a) \, u_{a \alpha} u_{a \beta}\,,
\end{align}
and of the metric $g_{\mu \nu}$. The latter mediates the interaction between the two worldlines,
and is  generated by them via Einstein's equations,
\be \label{Eeqs}
R^{\mu \nu}-\frac12 R g^{\mu \nu}= 8\pi G T^{\mu \nu}\,,
\ee
with
\be
T^{\mu \nu}(x)= \sum_{a=1,2} m_a \int ds_a   u_a^{\mu} u_a^{\nu} \, \frac{\delta^4(x-x_a(s_a))}{\sqrt{g}}\,,
\ee
where $ u_a^{\mu} \equiv g^{\mu \nu} u_{a \nu}$ and $g=- \det g_{\mu \nu}$.

Here we need to work in some gauge (say in harmonic gauge), and, as we are discussing the conservative dynamics
of two particles, we iteratively solve Einstein equations \eqref{Eeqs} by means of the time-symmetric classical graviton
propagator (in Minkowski spacetime)
\be \label{gravitonpropagator}
{\mathcal P}^{\alpha \beta ;\alpha' \beta'}(x-y) = \left(\eta^{\alpha \alpha'} \eta^{\beta \beta'} -\frac12 \eta^{\alpha \beta}  \eta^{\alpha' \beta'}\right) {\mathcal G}_{\rm sym}(x-y),
\ee
with ${\mathcal G}_{\rm sym}(x-y)= \delta\left[ \eta_{\mu \nu} (x^\mu-y^\mu) (x^\nu-y^\nu)\right]$.

The crucial point for our present purpose is that this iterative procedure, which involves expanding in powers of $G$
both the worldlines, say
\begin{align}
x_a^{\mu}(s_a) &={}_0x_a^{\mu}(s_a) + G\, {}_1x_a^{\mu}(s_a) + G^2\, {}_2x_a^{\mu}(s_a)+\cdots, \nonumber \\
 u_{a \mu}(s_a) &={}_0u_{a \mu}(s_a) + G \,{}_1u_{a \mu}(s_a) + G^2 \,{}_2u_{a \mu}(s_a)+\cdots 
\end{align}
and the metric
\be
g^{\mu \nu}(x)=\eta^{\mu \nu}- G h_1^{\mu \nu}(x)  - G^2 h_2^{\mu \nu}(x)- \cdots\,,
\ee
yields, at each order  $G^n$, expressions that are {\it homogeneous polynomials} of degree $n$ in
the  masses $m_a$. E.g. 
\begin{align}
 h_1^{\mu \nu}(x) &= m_1 h_{m_1}^{\mu \nu}(x)+ m_2 h_{m_2}^{\mu \nu}(x), \nonumber \\
 h_2^{\mu \nu}(x) &= m_1^2 h_{m_1^2}^{\mu \nu}(x)+  m_2^2 h_{m_2^2}^{\mu \nu}(x)
+ m_1 m_2 h_{m_1 m_2}^{\mu \nu}(x).
\end{align}
Here, we assume that the iterative solutions are systematically expressed in terms of the mass-independent data
describing the two asymptotic incoming worldlines, say ${}_0x_a^{\mu}(s_a)= x_{a 0}^{\mu}+ u_{a 0}^{\mu} s_a$.
See, e.g., section IV of Ref. \cite{Bini:2018ywr} for an explicit example of the structure of the PM-expanded metric, and
worldlines, expressed as explicit functionals of the  incoming worldline data (and for a discussion of the logarithmic
asymptotic corrections to the asymptotic free motions). From a geometric perspective, the latter 
incoming worldline data can be described by the two incoming 4-velocity vectors  $u_{1 0}^{\mu}$ and 
$u_{2 0}^{\mu}$, and by the vectorial impact parameter $b^\mu= x_{1 0}^{\mu}- x_{2 0}^{\mu}$ (chosen so as to be orthogonal to $u_{1 0}^{\mu}$ and $u_{2 0}^{\mu}$). 

At the end of the day, one gets a PM expansion for $\Delta p_{1 \mu}= -\Delta p_{2 \mu}$ (expressed
in terms of $b^\mu/b$, $u_{1 0}^{\mu}$ and $u_{2 0}^{\mu}$) that is,
at each order in $G$, a polynomial in the masses. It can be written as 
\be \label{deltapmu2}
\Delta p_{1 \mu}= - 2G m_1 m_2 \frac{2 (u_{1 0} \cdot u_{2 0})^2-1}{\sqrt{(u_{1 0} \cdot u_{2 0})^2-1}} \frac{b_\mu}{b^2} + 
\frac{G m_1 m_2}{b} \Delta_\mu.
\ee
Here we displayed the leading-order term \cite{Portilla:1980uz,Westpfahl:1985,Damour:2016gwp} and indicated that
the higher PM contributions (described by the term $\frac{G m_1 m_2}{b} \Delta_\mu$ with 
$\Delta_\mu = G \Delta^{(1)}_\mu+ G^2 \Delta^{(2)}_\mu+ \cdots$)
all contain $m_1 m_2$ as a common factor. Each PM contribution $\Delta^{(n)}_\mu$
is a combination of the three vectors $b^\mu/b$, $u_{1 0}^{\mu}$ and 
$u_{2 0}^{\mu}$, with coefficients that are, at each order in $G$,  homogeneous polynomials in $Gm_1$ and $Gm_2$.
By dimensional analysis, as the only length scale entering each order in the PM expansion\footnote{This contrasts with the PN expansion where one has two different length scales: $b$ and the characteristic wavelength of the gravitational radiation
$\lambda \sim c b/v$.} is the impact parameter $b$, we can write the three vectorial coefficients of the dimensionless 
$\Delta^{(n)}_\mu$ as  polynomials in $Gm_1/b$ and $Gm_2/b$, with coefficients depending only on the dimensionless
quantity
\be
\gamma \equiv - u_{1 0} \cdot u_{2 0} \; .
\ee
The latter quantity (denoted $\sigma$ in Refs. \cite{Bern:2019nnu,Bern:2019crd}), which is the relative Lorentz factor between the two incoming particles, will play a central role in the
following. Let us immediately note that it is equal to the dimensionless effective EOB energy of the binary system:
\be \label{g=e}
\gamma = \e .
\ee
Indeed,
\begin{align}
\gamma &= - \frac{p_{1 0} \cdot p_{2 0}}{m_1 m_2}= - \frac{(p_{10}+p_{20})^2 - p_{10}^2 -p_{20}^2}{2 m_1 m_2} \nonumber\\
&= \frac{s - m_1^2 -m_2^2}{2 m_1 m_2},
\end{align}
to be compared with the EOB definition \eqref{Ef}.

Let us now consider the magnitude of the (classical) momentum transfer, namely
\be
{\sf Q} \equiv \sqrt{-t} \equiv \sqrt{\eta^{\mu \nu}\Delta p_{1 \mu} \Delta p_{1 \nu}}\,,
\ee
which is related to the center-of-mass (c.m.) scattering angle $\chi$, and the c.m. three-momentum $P_{\rm c.m.}$, via
\be
 {\sf Q}= 2 P_{\rm c.m.} \sin \frac{\chi}{2}\,.
\ee
The structure of the PM expansion of the vectorial momentum transfer \eqref{deltapmu2} is easily seen to imply that
\begin{eqnarray} \label{QPM}
{\sf Q}&=& \frac{2 G m_1 m_2}{b} \left[  {\sf Q}^{1 \rm PM}(\g) \right.\nonumber \\ 
&&+\left({\sf Q}^{2 \rm PM}_{1}(\g) \frac{G m_1}{b}+  {\sf Q}^{2 \rm PM}_{2}(\g) \frac{G m_2}{b}\right) \nonumber \\ 
&& +\left({\sf Q}^{3 \rm PM}_{11}(\g) \left(\frac{G m_1}{b}\right)^2+  {\sf Q}^{3 \rm PM}_{22}(\g) \left(\frac{G m_2}{b}\right)^2  \right.\nonumber \\ 
&&+ \left. \left. {\sf Q}_{12}^{3 \rm PM}(\g)\frac{G m_1}{b}\frac{G m_2}{b}\right)  \right.\nonumber \\ 
&&+ \left({\sf Q}^{4 \rm PM}_{111}(\g) \left(\frac{G m_1}{b}\right)^3+  {\sf Q}^{4 \rm PM}_{222}(\g) \left(\frac{G m_2}{b}\right)^3  \right.\nonumber \\ 
&&+ \left. \left. {\sf Q}_{112}^{4 \rm PM}(\g)\left(\frac{G m_1}{b}\right)^2\frac{G m_2}{b}   \right. \right.\nonumber \\ 
&&+ \left. \left. {\sf Q}_{122}^{4 \rm PM}(\g) \frac{G m_1}{b}\left(\frac{G m_2}{b}\right)^2 \right) + \cdots \right]
\end{eqnarray}
where
\be
 {\sf Q}^{1 \rm PM}(\g)= \frac{2 \g^2-1}{\sqrt{\g^2-1}}\,.
\ee
Three apparently trivial, but quite useful, pieces of information controlling the structure of this PM expansion are: (i) the homogeneous polynomial
dependence in $m_1$ and $m_2$ (and therefore, by dimensional analysis, in $Gm_1/b$ and $Gm_2/b$) at each PM order;
 (ii)  the exchange symmetry between the two masses; and (iii) the consideration of the test-particle limit where, say, $m_1 \ll m_2$.
The exchange symmetry tells us that, for instance, 
$\sQ_{1}^{2 \rm PM}(\g) =\sQ_{2}^{2 \rm PM}(\g) $, $\sQ_{11}^{3 \rm PM}(\g) =\sQ_{22}^{3 \rm PM}(\g) $, 
$\sQ_{111}^{4 \rm PM}(\g) =\sQ_{222}^{4 \rm PM}(\g) $,  $\sQ_{112}^{4 \rm PM}(\g) =\sQ_{122}^{4 \rm PM}(\g) $,  etc.
In other words, at each PM order, we will have a symmetric polynomial in $m_1$ and $m_2$, with $\g$-dependent coefficients.
In addition, the test-mass limit tells us that all the functions involving only one mass are equal to the corresponding function of $\g$
appearing in the scattering of a test mass around a Schwarzschild black hole. Therefore, we have
 \begin{eqnarray} \label{QPMvsQS}
  \sQ^{1 \rm PM}(\g) &=& \sQ_{S}^{1 \rm PM}(\g)\,, \nonumber \\
 \sQ_{1}^{2 \rm PM}(\g) &=&\sQ_{2}^{2 \rm PM}(\g)= \sQ_{S}^{2 \rm PM}(\g) \,,\nonumber \\
  \sQ_{11}^{3 \rm PM}(\g) &=&\sQ_{22}^{3 \rm PM}(\g)= \sQ_{S}^{3 \rm PM}(\g) \,,\nonumber \\
   \sQ_{111}^{4 \rm PM}(\g) &=&\sQ_{222}^{4 \rm PM}(\g)= \sQ_{S}^{4 \rm PM}(\g) \,,
\end{eqnarray}
where the subscript $S$ refers to the Schwarzschild limit.

The 1PM-level result (first line of Eq. \eqref{QPMvsQS}) was already used in \cite{Damour:2016gwp} to show that the 1PM dynamics is
equivalent (after using the EOB energy map) to geodesic motion in a linearized Schwarzschild metric of mass $M=m_1+m_2$.
Let us emphasize that the 2PM-level result (second line of \eqref{QPMvsQS}) gives a one-line proof that
the 2PM fractional contribution to the momentum transfer (considered as a function
of the impact parameter) of a two-body system is simply given by the formula,
\be
{\sf Q}_S^{2 \rm PM}(\g) \frac{G (m_1+ m_2)}{b}\,,
\ee
where ${\sf Q}^{2 \rm PM}_S(\g)$ denotes the function of $\g$ obtained by computing the 2PM-accurate
scattering of a test particle around a Schwarzschild black hole, namely (see, {\it e.g.}, \cite{Damour2018})
\be \label{Q2pmS}
{\sf Q}^{2 \rm PM}_S(\g)= \frac{3 \pi}{8} \frac{5 \g^2-1}{\sqrt{\g^2-1}}\,.
\ee
The test-mass computation yielding \eqref{Q2pmS} (equivalent to Eq. (3.19) in \cite{Damour2018})
 is much simpler than the full, two-body 2PM scattering
computation (involving complicated  nonlinear terms and recoil effects) first done by Westpfahl \cite{Westpfahl:1985} 
(and recently redone in \cite{Bini:2018ywr}).
The simple link between the 2PM test-mass result and the two-body one was also recently discussed in Ref. \cite{Vines:2018gqi}, 
but in a  different context, and arguing from the
structure of the so-called classical part of the one-loop amplitude \cite{Guevara:2017csg,Bjerrum-Bohr:2018xdl}, instead of our
purely classical analysis above.
Note that the mass-dependence we are talking about here
has taken an especially simple form because we focussed on the variable $ {\sf Q}$ as a function of $\g$ and $b$.
As we shall see next, the mass-dependence of the scattering angle $\chi$ as a function of $\g$ and either $b$ 
or $ j \equiv \frac{J}{G m_1 m_2}$ is more involved. 

Summarizing so far, we conclude that both the 1PM and 2PM two-body scattering can be deduced (without any extra calculation) from
the 1PM and 2PM test-mass scattering. 

Let us now consider what happens at higher PM orders. 
At the 3PM order, $O(G^3)$, we  conclude from the above results that the  scattering depends not only on the test-mass-derivable
function $\sQ_{11}^{3 \rm PM}(\g) =\sQ_{22}^{3 \rm PM}(\g)= \sQ_{S}^{3 \rm PM}(\g)$,
but also on a {\it single} further function of $\g$, namely  $\sQ_{12}^{3 \rm PM}(\g)$.
Similarly, at the 4PM order, the full two-body scattering depends, besides the test-mass-derivable
function  $\sQ_{111}^{4 \rm PM}(\g) =\sQ_{222}^{4 \rm PM}(\g)= \sQ_{S}^{4 \rm PM}(\g) $, on a
 {\it single} further function of $\g$, namely  $\sQ_{112}^{4 \rm PM}(\g)= \sQ_{122}^{4 \rm PM}(\g)$.

It is easy to generalize this result to higher PM orders. {\it E.g.}, at 5PM, modulo the $1 \leftrightarrow 2$ symmetrization,
there will be terms $\propto m_1^4$,  $m_1^3m_2$ and $m_1^2m_2^2$. The first one of these is deducible from the test-mass limit,
so that the full two-body 5PM scattering depends on only {\it two} non-trivial extra functions of $\g$. The same counting applies
at the 6PM level where there will be (modulo $1 \leftrightarrow 2$ symmetrization) terms 
$\propto m_1^5$ (test-mass-deducible),  $m_1^4m_2$ and $m_1^3m_2^2$.
The general rule is that, at the $n$PM order, there will appear only (using $[\cdots]$ to denote the integer part)
\be
d(n) \equiv \left[\frac{n-1}{2}\right]\,,
\ee 
non-test-mass-deducible functions of $\g$.

The latter result can be translated into a dependence on the symmetric mass ratio $\nu \equiv m_1 m_2/(m_1+m_2)^2$
if one expresses $m_1$ and $m_2$ (with, say, $m_1\leq m_2$) in terms of the total mass $M= m_1+m_2$,
and of the two dimensionless mass ratios
\begin{eqnarray}
X_1 &\equiv& \frac{m_1}{m_1+m_2}= \frac{1- \sqrt{1-4 \nu}}{2} \,,\nonumber \\
X_2 &\equiv& \frac{m_2}{m_1+m_2}= 1-X_1=\frac{1+ \sqrt{1-4 \nu}}{2} \,,
\end{eqnarray}
such  that $\nu \equiv X_1 X_2$. Indeed, an homogeneous, symmetric polynomial of degree $n$ in the masses
yields (after division by $M^n$)  a sum $\sum_k c_k X_1^k X_2^{n-k}$. Using $X_2 \equiv 1-X_1$ and symmetrizing over
$1 \leftrightarrow 2$ yields a sum $\sum_k c'_k (X_1^k + X_2^k)$ over $0\leq k\leq n$.
What will be important here is the maximum power of $\nu$ entering such symmetric polynomials in the mass ratios.
We note the following results
\begin{eqnarray} \label{Xn}
X_1^2+ X_2^2 &=& 1- 2 \nu \,, \nonumber \\
X_1^3+ X_2^3&=& 1- 3 \nu \,,  \nonumber \\
X_1^4+ X_2^4&=& 1-4 \nu+ 2 \nu^2 \,,  \nonumber \\
X_1^5+ X_2^5&=& 1-5 \nu+ 5 \nu^2\,.
\end{eqnarray}
More generally, $X_1^k + X_2^k$ is a polynomial in $\nu$ of degree  $\left[\frac{k}{2}\right]$.
At the $n$PM order, after having factored the prefactor,
\be
\frac{2 G m_1 m_2}{b} \left(\frac{G M}{b}\right)^{n-1}\,,
\ee
there appears such an homogeneous, symmetric polynomial of degree $n-1$ in $X_1$ and $X_2$.

Finally, the PM expansion of the momentum transfer can be written as:
\be \label{Qvsnu}
\sQ=\frac{2 G m_1 m_2}{b} \sum_{n\geq 1} \left(\frac{G M}{b}\right)^{n-1} \sQ^{n \rm PM}(\g,\nu) \,,
\ee
where $\sQ^{n \rm PM}(\g,\nu)$ is a polynomial in $\nu$ of degree $d(n) \equiv \left[\frac{n-1}{2}\right]$:
\be
\sQ^{n \rm PM}(\g,\nu)= Q_0^{n \rm PM}(\g) + \nu Q_1^{n \rm PM}(\g) +\ldots+ \nu^{d(n)} Q_{d(n)}^{n \rm PM}(\g) .
\ee
For instance, at the 3PM level, we have explicitly
\begin{eqnarray}
\sQ^{3 \rm PM}(\g,\nu)&=&{\sf Q}^{3 \rm PM}_{11}(\g) (X_1^2+ X_2^2) + {\sf Q}^{3 \rm PM}_{12}(\g) X_1 X_2 \phantom{xxxxxx}\nonumber \\
&=&  {\sf Q}^{3 \rm PM}_{S}(\g) (1-2 \nu) + {\sf Q}^{3 \rm PM}_{12}(\g) \nu \,.
\end{eqnarray}
It is easily seen that, at all PM orders, the coefficient of $\nu^0$ is simply the result given by the test-mass computation:
\be
Q_0^{n \rm PM}(\g)=\sQ_S^{n \rm PM}(\g)\,.
\ee

Let us now translate the above structural information into an information about the classical scattering function itself, i.e.
the half scattering angle $\chi/2$ considered as a function of the energy and angular momentum of the system.
As indicated in Eq. \eqref{chiPM}, it is convenient to measure the total c.m. energy of the system by means of 
the dimensionless effective energy $\e= \g$ given by Eq. \eqref{Ef}, and to measure the total c.m. angular momentum
by means of the dimensionless variable $j= J/(Gm_1 m_2)$, Eq. \eqref{j}. We also need the relations connecting the
c.m. linear momentum $P_{\rm c.m.}$ both to $b$, to $J$ and to $\g$. These are 
(see Eqs. (7.6) and (10.27) in \cite{Damour2018})
\begin{eqnarray}
 b P_{\rm c.m.}&=& J = G m_1 m_2 j \,,\nonumber \\
\E P_{\rm c.m.}&=& \sqrt{(p_{1 0}\cdot p_{2 0})^2- p_{1 0}^2 p_{2 0}^2} = m_1 m_2 \sqrt{\g^2-1}\,. \nonumber\\
\end{eqnarray}
From these links follows the relation
\be
\frac{GM}{b} = \frac{\sqrt{\g^2-1}}{h(\g,\nu) j} = \frac{\peob}{h(\g,\nu) j} \,.
\ee
Here we introduced some abbreviated notation for two dimensionless quantities
crucially entering many equations, namely
\begin{eqnarray}
h(\g,\nu) &&\equiv \frac{\E}{M} = \frac{\sqrt{s}}{M}= \sqrt{1+ 2\nu (\g-1)} \,,\nonumber \\
p_{\rm eob} &&\equiv \sqrt{\g^2-1} \equiv \pinf\,.
\end{eqnarray}
[We will indifferently use the notation $\peob$ or $\pinf$.]
Inserting these relations in the above expression of the momentum transfer $\sf Q$,
and computing
\be
\sin \frac{\chi}{2}=\frac{{\sf Q}}{2 P_{\rm c.m.}}\,,
\ee
 yields
\be
\sin \frac{\chi}{2}=\frac1j \sum_{n\geq 1} \left( \frac{\peob}{h(\g,\nu) j}  \right)^{n-1} \sQ^{n \rm PM}(\g,\nu)\,. 
\ee
This reads more explicitly
\begin{eqnarray}\label{sinchivsQ}
\sin \frac{\chi}{2}&=&  \frac{ {\sf Q}^{1 \rm PM}(\g)}{j} + \frac{\peob {\sf Q}^{2 \rm PM}(\g)}{h(\g,\nu) j^2}   \nonumber \\ 
&+& \frac{\peob^2 {\sf Q}^{3 \rm PM}(\g,\nu)}{h^2(\g,\nu) j^3} +   \frac{\peob^3 {\sf Q}^{4 \rm PM}(\g,\nu)}{h^3(\g,\nu) j^4} \nonumber \\ 
&+&  \cdots  
\end{eqnarray}
Let us compare this expression to the usual way of writing the scattering function, namely
(using $\g \equiv \e$ as energy variable and $j \equiv J/(G m_1 m_2)$ as angular momentum variable)
\begin{eqnarray} \label{chiPMbis}
\frac12 \chi^{\rm }(\E, J) &=& \frac{\chi_{1}(\g, \nu)}{j} + \frac{ \chi_{2}(\g, \nu)}{j^2} 
+  \frac{\chi_{3}(\g, \nu)}{j^3} \nonumber\\ &&+  \frac{\chi_{4}(\g, \nu)}{j^4}  + \cdots,
\end{eqnarray}
 which implies
 \begin{eqnarray} \label{chiPMter}
\sin \frac12 \chi^{\rm }(\g, j,\nu) &=& \frac{\widetilde \chi_{1}(\g, \nu)}{j} + \frac{ \widetilde \chi_{2}(\g, \nu)}{j^2} 
+  \frac{\widetilde \chi_{3}(\g, \nu)}{j^3} \nonumber\\ &&+  \frac{\widetilde \chi_{4}(\g, \nu)}{j^4}  + \cdots.
\end{eqnarray}
where
\bea\label{tildechivschi}
\widetilde \chi_{1}&=&  \chi_{1} \,,\nonumber\\
\widetilde \chi_{2}&=&  \chi_{2} \,,\nonumber\\
\widetilde \chi_{3}&=&  \chi_{3} - \frac16 \chi_1^3 \,, \nonumber\\
\widetilde \chi_{4}&=&\chi_{4}-\frac12 \chi_1^2 \, \chi_2 \,.
\eea
When comparing the definitions of the expansion coefficients $\chi_n$ and $\widetilde \chi_{n}$ to the structural result
\eqref{sinchivsQ} we find
\be
\widetilde \chi_n(\g,\nu) =  \frac{\peob^{n-1} {\sf Q}^{n \rm PM}(\g,\nu)}{h^{n-1}(\g,\nu)}\,.
\ee
Remember the fact that ${\sf Q}^{n \rm PM}(\g,\nu)$ was proven above to be a polynomial in $\nu$ of degree $d(n)$
(with $\g$-dependent coefficients). 
We then get the rule that
\be \label{ruletildechi}
h^{n-1}(\g,\nu) \widetilde \chi_n(\g,\nu)= \widetilde P_{d(n)}^\g(\nu)\,,
\ee
where $\widetilde P_{d(n)}^\g(\nu)$ denotes a polynomial in $\nu$ of degree $d(n)$
with $\g$-dependent coefficients. When transferring this information into a corresponding information
for the expansion coefficients $\chi_n(\g,\nu)$ of $\frac12 \chi(\g,j)$, using Eqs. \eqref{tildechivschi},
it is easily seen that we have the same structure for them, namely
\be \label{rulechi}
h^{n-1}(\g,\nu)  \chi_n(\g,\nu)=  P_{d(n)}^\g(\nu)\,,
\ee
where $P_{d(n)}^\g(\nu)$ denotes another degree-$d(n)$ polynomial in $\nu$ 
with $\g$-dependent coefficients.

We can combine this structural information with the knowledge of the test-mass limit of the  $\chi_n(\g,\nu)$'s.
In the context of the functions $\chi_n(\g,\nu)$, the test-mass limit is simply the  $\nu \to 0$ limit.
Therefore, the $\nu \to 0$ limit of the various $\chi_{n}(\g, \nu)$'s must coincide with the values  $\chi^{\rm Schw}_{n}(\g)$ of the
scattering coefficients for a test particle in a Schwarzschild background. The latter values were computed in \cite{Damour2018}
with the results
\begin{align} 
\label{chischw1}
& \chi_1^{\rm Schw}(\peob) = \frac{2 \, \peob^2+1}{\peob}= \frac{2 \, \g^2-1 }{\sqrt{\g^2-1}},\\
\label{chischw2}
&\chi_2^{\rm Schw}(\peob)= \frac{3 \pi}{8} (5 \, \peob^2+4)= \frac{3 \pi}{8} (5 \, \g^2-1), \\
&\chi_3^{\rm Schw}(\peob)=\frac{64\, \peob^6 + 72 \,\peob^4 + 12\, \peob^2 -1}{3 \, \peob^3}, \\
&\chi_4^{\rm Schw}(\peob) = \frac{105 \pi}{128} (33 \, \peob^4 + 48 \, \peob^2+16).
\end{align}
We then get the information that
\be
 P_{d(n)}^\g(0)= \chi_n^{\rm Schw}(\peob)\,.
\ee
As already implied by the discussion above, this fully determines the 1PM \cite{Portilla:1980uz,Damour:2016gwp}
and 2PM \cite{Westpfahl:1985,Bini:2018ywr}  scattering coefficients, namely
\be \label{chi1pm}
\chi_{1}(\g,\nu)=  \chi_1^{\rm Schw}(\g) = \frac{2 \g^2-1}{\sqrt{\g^2-1}} \,,
\ee
and
\be \label{chi2pm}
\chi_{2}(\g, \nu)= \frac{\chi_{2}^{\rm Schw}(\g)}{h(\g,\nu)} =   \frac{3 \pi}{8} \frac{(5 \, \g^2-1)}{h(\g,\nu)}\,.
\ee
Note in passing that it is crucial,
in order to find the $\nu$-independence of $\chi_{1}(\g,\nu)$, to measure the energy by means of $\g$ (i.e. the EOB effective energy),
and not by means of the total c.m. energy $\E = \sqrt{s}= M h(\g,\nu)$.

Concerning the higher-order expansion coefficients, using the fact that  $h^2(\g,\nu)= 1 + 2 \nu (\g-1)$
is a linear function of $\nu$ (so that a polynomial in $\nu$ can be reexpressed as
a polynomial in $h^2(\g,\nu)$) they can be written in the following form
\begin{eqnarray} \label{chi3456}
\chi_{3}(\g, \nu) &=& \widehat \chi_{3}^{(0)}(\g) +  \frac{\widehat \chi_{3}^{(2)}(\g)}{h^2(\g,\nu)}, \nonumber\\
\chi_{4}(\g, \nu) &=& \frac{\widehat \chi_{4}^{(1)}(\g)}{h(\g,\nu)} +  \frac{\widehat \chi_{4}^{(3)}(\g)}{h^3(\g,\nu)},  \nonumber\\
\chi_{5}(\g, \nu) &=& \widehat \chi_{5}^{(0)}(\g) +  \frac{\widehat \chi_{5}^{(2)}(\g)}{h^2(\g,\nu)} +  \frac{\widehat \chi_{5}^{(4)}(\g)}{h^4(\g,\nu)},\nonumber\\
\chi_{6}(\g, \nu) &=& \frac{\widehat \chi_{6}^{(1)}(\g)}{h(\g,\nu)} +  \frac{\widehat \chi_{6}^{(3)}(\g)}{h^3(\g,\nu)} +  \frac{\widehat \chi_{6}^{(5)}(\g)}{h^5(\g,\nu)}\,,
\end{eqnarray}
with the information that, at each PM order, the sum over $k$ of the various numerators $\widehat \chi_{n}^{(k)}(\g)$ is
equal to the Schwarzschild limit $\chi_{n}^{\rm Schw}(\g)$. This implies, for instance, that at the 3PM level we can
also write
\be \label{chi3pm}
\chi_{3}(\g, \nu)= \chi_{3}^{\rm Schw}(\g) + \widehat \chi_{3}^{(2)}(\g) \left(  \frac{1}{h^2(\g,\nu)} -1\right),
\ee
where the last term vanishes when $\nu \to 0$. A similar structure describes the 4PM-level scattering, namely
\be \label{chi4pm}
\chi_{4}(\g, \nu)= \frac{\chi_{4}^{\rm Schw}(\g)}{h(\g,\nu)} + \frac{\widehat \chi_{4}^{(3)}(\g)}{h(\g,\nu)} \left(  \frac{1}{h^2(\g,\nu)} -1\right)\,.
\ee
In both cases, we see that the full 3PM and 4PM dynamical information is encapsulated in a single function of $\g$,
namely $\widehat \chi_{3}^{(2)}(\g)$ and $\widehat \chi_{4}^{(3)}(\g)$, respectively.

Let us  note that in the high-energy (HE) limit ($\g \to \infty$, i.e. $\peob \to \infty$) we have
the following asymptotic behavior of the test-mass-limit scattering coefficients
\be
\chi_n^{\rm Schw}(\peob)  \overset{\rm HE}{=} c_n^{\chi \rm  Schw} \, \peob^n    \overset{\rm HE}{=} c_n^{\chi \rm Schw} \, \g^n \,,
\ee
where $c_n^{\chi \rm Schw}$ is a numerical constant. It was suggested in Ref. \cite{Damour2018} that the same asymptotic behavior
(though with different numerical constants $c^{\chi}_n$) holds for the building blocks $\widehat \chi_{n}^{(k)}(\g)$
introduced above. We shall rediscuss this suggestion below.

%%%%%%
\section{PM-expanded EOB Hamiltonian and EOB radial potential} \label{sec3}

  %%%%%
  \subsection{EOB Hamiltonian in PM gravity}
  
 Refs.  \cite{Damour:2016gwp,Damour2018} introduced a new, PM-based, approach to the conservative  dynamics
 of two-body systems based on  the EOB formalism. This led to simple EOB descriptions of the 1PM  \cite{Damour:2016gwp},
 2PM \cite{Damour2018}, and 3PM \cite{Antonelli:2019ytb} Hamiltonians. Here, we will reconsider the 3PM EOB Hamiltonian
 derived from the quantum-amplitude approach of Refs. \cite{Bern:2019nnu,Bern:2019crd}. Let us start by recalling
 the PM-EOB formalism of Refs.  \cite{Damour:2016gwp,Damour2018}.
 
 The basic feature of the EOB formalism \cite{Buonanno:1998gg,Buonanno:2000ef,Damour:2000we}
 is to describe the two-body dynamics in terms of a 
  one-body Hamiltonian, which describes the dynamics of the {\it relative} two-body motion within the c.m. frame
of the two-body system. The simplest way to define the EOB Hamiltonian is to say that: (i) the (``real") c.m. Hamiltonian
of the two-body system is related to the conserved energy $\Ef$ of the ``effective" dynamics by Eq. \eqref{Ef}, i.e.
\be \label{Heob}
H_{\rm real}({\mathbf R}, {\mathbf P})=M \sqrt{1+2\nu \left(\frac{{\mathcal E}_{\rm eff}}{\mu }-1 \right)}\,;
\ee
and, (ii) the effective energy $\Ef$ is related to the dynamical variables ${\mathbf R}, {\mathbf P}$ describing
the relative c.m. dynamics via a mass-shell condition of the form
\be \label{massshellgen}
0= g_{\rm eff}^{\mu \nu} P_{\mu} P_{\nu} + \mu^2 + Q(X^{\mu}, P_{\mu})\,,
\ee
 where $ g_{\rm eff}^{\mu \nu}$ is (the inverse of) an effective metric of the form
\be \label{geff}
 g^{\rm eff}_{\,\mu \nu} dx^{\mu} dx^{\nu} = - A(R) dT^2 + B(R) dR^2 + C(R) (d \theta^2 + \sin^2 \theta d \varphi^2),
\ee
and where $Q(X^{\mu}, P_{\mu})$ is a Finsler-type additional contribution, which contains higher-than-quadratic  momenta
contributions. The time-invariance, and spherical symmetry, of the effective metric (and of $Q$), implies
(for equatorial motions) the existence of the two conserved quantities $P_0$ and $P_\varphi$,
which are respectively identified with
\be
P_0=- \Ef \; , \; P_\varphi=J \,.
\ee 
For any given additional mass-shell contribution $Q$ expressed as a function of ${\mathbf R}$, ${\mathbf P}$, and $\Ef$,
say $Q=Q({\mathbf R}, {\mathbf P}, \Ef)$,
the effective Hamiltonian $\Ef=H_{\rm eff}({\mathbf R}, {\mathbf P})$ is then obtained by solving
\be \label{massshellgen2}
0=- \frac{\Ef^2}{A} + \frac{P_R^2}{B}+ \frac{P_{\varphi}^2}{C} + \mu^2 + Q({\mathbf R}, {\mathbf P}, \Ef) \,,
\ee
with respect to $\Ef$, and then inserting the result in the real, two-body Hamiltonian \eqref{Heob}.

In a PM framework, i.e. when working perturbatively in $G$, it was shown in \cite{Damour:2016gwp,Damour2018} that:
(i) the effective metric can be taken to be a Schwarzschild metric of mass $M=m_1+m_2$; (ii) the $Q$ term starts at
order $G^2$;  and (iii) one can
(by using some gauge freedom) construct $Q$ so that it depends only on $R = |{\mathbf R}|$ and some
energy-like variable ( ``energy gauge"). There are two simple choices for defining such an energy-gauge. 
Using the shorthand notation
\be
u \equiv \frac{G M}{R}\; , %\; {\widehat Q} \equiv \frac{Q}{\mu^2}\,,
\ee
one can either write ${ Q}$ as a function of $u$ and $\Ef$,
\be \label{QE}
{Q}^E(u, \Ef)= u^2 Q_2(\Ef)+  u^3 Q_3(\Ef)+ u^4 Q^E_4(\Ef)+ O(G^5) ,
\ee
or, one can express $Q$ as a function of position and momenta by writing
\be\label{QH}
{Q}^H(u, H_S)= u^2 Q_2(H_S)+  u^3 Q_3(H_S)+ u^4 Q^H_4(H_S)+ O(G^5),
\ee
where $H_S$ denotes the Schwarzschild Hamiltonian, i.e.
\be
H_S(u,P_R,P_\varphi) =\sqrt{A(R) \left(  \frac{P_R^2}{B(R)}+ \frac{P_{\varphi}^2}{C(R)} + \mu^2 \right)}.
\ee
The second form was initially advocated in \cite{Damour2018} because it allows one to explicitly solve
 the mass shell condition \eqref{massshellgen2} for $\Ef$ as a function of position and momenta,
namely
\bea
&&\Ef = H_{\rm eff}({\mathbf R}, {\mathbf P}) \nonumber \\
&=& \sqrt{A \left(  \frac{P_R^2}{B}+ \frac{P_{\varphi}^2}{C} + \mu^2 + Q^H[u, H_S(u,P_R,P_\varphi)] \right)}.\nonumber \\
\eea
However, Ref. \cite{Damour2018} also used the first form \eqref{QE} because of its usefulness in getting an explicit
energy-dependent potential that can be easily quantized.
As indicated by the notation used in Eqs. \eqref{QE}, \eqref{QH}, the difference between the expansion coefficients 
$Q_n$ entering these two perturbative expansions
starts at order $G^4$. This follows from the fact that $Q$ itself starts at order $G^2$.  

In the following we will mostly work with the first, E-form of the energy gauge. It will also be convenient to work with dimensionless, rescaled quantities, say
\be
 {\widehat Q} \equiv \frac{Q}{\mu^2}\,, \, {\mathbf p} \equiv \frac{{\mathbf P}}{\mu}\,, {\widehat H}_{\rm eff} \equiv \frac{H_{\rm eff}}{\mu},
\ee
and to denote the PM expansion coefficients of ${\widehat Q}$ simply as $q_n \equiv Q_n/\mu^2$, {\it e.g.},
\be \label{QEresc}
{\widehat Q}^E(u, \g)= u^2 q_2(\g)+  u^3 q_3(\g)+ u^4 q^E_4(\g)+ O(G^5) ,
\ee
where we used Eq. \eqref{g=e} to write $\e \equiv \Ef/\mu$ simply as $\g$.

%%%%%%%
\subsection{Energy-dependent, radial scattering potential within the EOB framework} 

In the previous subsection we recalled how PM gravity can be encoded, within the EOB formalism, by means of
a PM-expanded mass-shell function $Q({\mathbf R}, {\mathbf P}, \Ef)$. When discussing the quantum
scattering amplitude corresponding to a given PM-expanded $Q$, it was found convenient in \cite{Damour2018}
to transform $Q$ into an equivalent PM-expanded, energy-dependent radial potential $W({\bar R}, \Ef)$. Let us recall
this transformation.

Most of the past work in EOB dynamics has found it convenient to  represent the EOB effective metric \eqref{geff}
by using a Schwarzschild-like radial coordinate, i.e. by choosing a coordinate $R$
such that the coefficient $C(R)$ of $d \theta^2 + \sin^2 \theta d \varphi^2$ is equal to $R^2$. In keeping with the latter usage,
we shall denote simply by $R$ such a Schwarzschild-like radial coordinate, and by $u$ the corresponding quantity $GM/R$. 
On the other hand, when discussing the effective potential describing the scattering dynamics, it is convenient 
(following the 2PM-level treatment of Sec. X of Ref. \cite{Damour2018}) to use isotropic coordinates, i.e. 
a new radial coordinate, say $\bR$, such that $C(\bR)=\bR^2 B(\bR)$) for the Schwarzschild metric entering the
EOB mass shell condition \eqref{massshellgen2}. The link between $R$ and $\bR$ is
\be
R = \bR \left(1+ \frac{GM}{2 \bR} \right)^2\,,
\ee
or
\be \label{uvsbu}
u= \bu \left(1+ \frac{\bu}{2 } \right)^{-2}\,.
\ee
In these coordinates, the usual formulas $A(u)=1-2u=1/B(u)$ transform into
\be
 {\bar A}(\bu)= \left(\frac{1- \frac12 \bu }{1+ \frac12 \bu}\right)^2 \, ;\,  {\bar B}(\bu)=\left(1+ \frac12 \bu   \right)^4 ,
 \ee
 where we added a bar on $A$, and $B$ (and on the argument $u$), to recall the use of isotropic coordinates. 
  
 We shall denote the Cartesian coordinates  linked in the usual way to $\bR, \theta, \varphi$ as $X^i ={\bf X} $, and the corresponding
 (covariant)  momenta $P_i$ as $ {\mathbf P}$ (for simplicity we do not put bars on ${\bf X}$ and $ {\mathbf P}$).
The E-type mass shell condition then directly leads to an 
 energy-dependent quadratic constraint on the momenta of the form
 \be
 \bP^2= P_{\infty}^2+ W(\bu, P_{\infty})\,,
 \ee
 where
 \be
 P_{\infty}^2 \equiv \Ef^2- \mu^2 = \mu^2 (\g^2-1)\,,
 \ee
 and where the energy-dependent ``potential" $W$ is defined by
 \be \label{Wexact1}
 P_{\infty}^2+ W(\bu, P_{\infty})\equiv {\bar B}(\bu) \left( \frac{\Ef^2}{{\bar A}(\bu)}- \mu^2 - Q(\bu, \Ef) \right)\,.
 \ee
 The radial potential $W(\bu, P_{\infty})$ tends to zero at large distances (i.e. when $\bu=GM/\bR \to 0$)
 and can be rewritten as
 \bea\label{Wexact2}
 W(\bu, P_{\infty})&=& \Ef^2 \left(\frac{{\bar B}(\bu)}{{\bar A}(\bu)}-1 \right) \nonumber \\
 &-& \mu^2 \left({\bar B}(\bu)-1 \right) - {\bar B}(\bu) Q^E(\bu, \Ef).\nonumber\\
 \eea
 Its PM expansion directly follows by combining the $\bu$ expansion of the metric functions ${\bar A}(\bu), {\bar B}(\bu)$,
 with the PM expansion of $Q^E(\bu, \Ef)$. It reads
 \bea \label{Wexp1}
 &&W(\bu, P_{\infty})= W_1 \bu+  W_2 \bu^2+W_3 \bu^3+W_4 \bu^4+ \cdots \nonumber\\
 &=&  \frac{GM W_1}{\bR} + \frac{G^2 M^2 W_2}{\bR^2} +  \frac{G^3 M^3 W_3}{\bR^3} + \frac{G^4 M^4 W_4}{\bR^4} + \cdots\nonumber \\
 \eea
 It is often more convenient to work with a rescaled version of these results in which one
 uses the dimensionless variables
 \be
 \br= \frac{\bR}{GM},  \p=  \frac{\bP}{\mu},  p_{\infty}= \frac{P_{\infty}}{\mu}= \sqrt{\g^2-1}\,.
 \ee
 One then has
 \be
 \p^2= p_{\infty}^2+ w(\bu, p_{\infty})\,,
 \ee
 where
 \be
 w(\bu, p_{\infty})=  \frac{W(\bu, p_{\infty})}{\mu^2}\,,
 \ee
 i.e.
 \bea
 w(\bu, p_{\infty})&=& \g^2 \left(\frac{{\bar B}(\bu)}{{\bar A}(\bu)}-1 \right) \nonumber \\
 &-& \left({\bar B}(\bu)-1 \right) - {\bar B}(\bu) \hQ^E(\bu, \g)\,.
 \eea
 The rescaled potential $w(\bu, p_{\infty})$ has the following PM expansion
 \bea \label{wexp1}
 w(\bu, p_{\infty})&=& w_1(\g) \bu+  w_2(\g) \bu^2+w_3(\g) \bu^3+w_4(\g) \bu^4+ \cdots \nonumber\\
 &=&  \frac{w_1(\g)}{\br} + \frac{w_2(\g)}{\br^2} +  \frac{w_3(\g)}{\br^3} + \frac{w_4(\g)}{\br^4} + \cdots\nonumber \\
 \eea
 where
 \be
 w_n(\g)=  \frac{W_n(\g)}{\mu^2}\,.
 \ee
 Note that these results mean that the {\it relativistic} (scattering) dynamics of a two-body system can be mapped (by using
 the EOB framework) onto  the {\it nonrelativistic} dynamics of one particle of mass $\mu$ in an energy-dependent radial potential. 
 
 We can now use Eq. \eqref{wexp1} to compute the link between the (rescaled) coefficients $w_n(\g)$ 
 entering the PM expansion of the (rescaled) potential $w(\bu,\g)$, and the coefficients $q^E_n(\g)$ entering the 
 PM expansion of the energy-gauge $Q$ function entering the EOB mass shell condition \eqref{massshellgen}. 
 The $Q$ term is numerically independent of the radial gauge used in the EOB effective metric \eqref{geff}, but
 we must distinguish the functions $u \rightarrow {\widehat Q}^E(u, \g)$ and $\bu \rightarrow {\widehat Q}^E(\bu, \g)$. We shall denote
 their respective PM expansion coefficients as
 \be\label{QEu}
{\widehat Q}^E(u, \g)= u^2 q_2(\g)+  u^3 q_3(\g)+ u^4 q^E_4(\g)+ O(G^5)\,,
\ee
and
\be
{\widehat Q}^E(\bu, \Ef)= \bu^2 \bq_2(\g)+  \bu^3 \bq_3(\g)+ \bu^4 \bq^E_4(\g)+ O(G^5)\,,
\ee
with similar equations for ${\widehat Q}^H(u, {\widehat H}_S)$ and ${\widehat Q}^H(\bu, {\widehat H}_S)$.

The relations between the $q_n$'s and the $\bq_n$'s is easily obtained from Eq. \eqref{uvsbu}. For instance, we have
\bea
\bq_2(\g)= q_2(\g) \,,\nonumber \\
\bq_3(\g)= q_3(\g) - 2 q_2(\g) \,,\nonumber \\
\bq^E_4(\g)= q^E_4(\g) - 3 q_3(\g) + \frac{5}{2} q_2(\g) \,.
\eea

We can then express the expansion coefficients $w_n(\g)$ of the EOB potential either in terms of the $q_n$'s or the $\bq_n$'s.
More precisely, the coefficient of $1/\br$ entirely comes from the linearized Schwarzschild metric and reads \cite{Damour2018}
\be
w_1(\g)= 2 (2 \g^2-1)\,,
\ee
while the coefficients of higher powers of $1/\br$ are related to the $\bq_n$'s via
\bea \label{wnvsbqn}
w_2(\g)&=& \frac{15}{2} \g^2 - \frac{3}{2} - \bq_2(\g)  \,,\nonumber \\
w_3(\g)&=& 9 \g^2 - \frac{1}{2} - \bq_3(\g) -2 \bq_2(\g)  \,,\nonumber \\
w_4(\g)&=& \frac{129}{16} \g^2 - \frac1{16} - \bq^E_4(\g) - 2 \bq_3(\g)  - \frac{3}{2} \bq_2(\g) \,, \nonumber \\
\eea
i.e.
\bea \label{wnvsqn}
w_2(\g)&=& \frac{15}{2} \g^2 - \frac{3}{2} - q_2(\g)  \,,\nonumber \\
w_3(\g)&=& 9 \g^2 - \frac{1}{2} - q_3(\g)  \,,\nonumber \\
w_4(\g)&=&  \frac{129}{16} \g^2 - \frac1{16} - q^E_4(\g) + q_3(\g) \,.
\eea
At the 2PM level, it was shown in \cite{Damour2018} that 
\be
q_2(\g, \nu) =  \frac{3}{2} \left( 5 \, \g^2-1 \right) \left[ 1 -  \frac{1}{h(\g,\nu)}  \right]\, ,
\ee
where we recall that $h(\g,\nu)=\sqrt{1+ 2\nu (\g-1)}$, so that
\be
w_2(\g, \nu) =  \frac{3}{2} \left( 5 \, \g^2-1 \right) \frac{1}{h(\g,\nu)}\,.
\ee
The current knowledge of the values of the 3PM coefficients $q_3(\g, \nu)$ and $w_3(\g, \nu)$
will be assessed below.

%%%%

\subsection{Scattering function and scattering invariants of an energy-dependent radial potential}  \label{subsec5C}

Refs. \cite{Damour:2016gwp,Damour2018} showed how to derive the scattering function $\chi(\Ef, J)$ directly
from the $Q$-form of the EOB PM dynamics. An equivalent, alternative procedure is to derive $\chi(\Ef, J)$
from the EOB radial potential $W(\bu, P_{\infty})$ corresponding to the Schwarzschild-metric-plus-$Q$ formulation.
Actually this link is very general and applies to any dynamical formulation involving a radial potential.

The usual formulas of non relativistic potential scattering (recalled, e.g.,  in \cite{Damour:2016gwp}) yield
\be
\pi+ \chi(\Ef,J)= - \int_{-\infty}^{+\infty}   d\bR \frac{\partial P_R(\bR ; \Ef, J)}{\partial J}  \,,
\ee
where the radial momentum $P_R(\bR ; \Ef, J)$ is obtained by solving the mass-shell condition with respect to $P_R$.
When using an energy gauge, the mass-shell condition reads, 
\be
\bP^2= P_R^2+ \frac{J^2}{\bR^2}= P_{\infty}^2+ W(\bu, P_{\infty})\,,
\ee
so that
\be
P_R(\bR ; \Ef, J)= \pm \sqrt{P_{\infty}^2+ W(\bu, P_{\infty}) - \frac{J^2}{\bR^2}}\,.
\ee
Here the (energy-gauge) potential $W(\bu, P_{\infty}) $ (where we recall that $\bu=GM/\bR$ and $P_{\infty}= \sqrt{\Ef^2-\mu^2}$)
does not depend on the angular momentum $J$. We can then write (as in usual non relativistic potential theory)
\be
\frac{\pi}{2}+ \frac12\chi(\Ef,J)= + \int_{\bR_{\rm min}}^{+\infty}  J \frac{d\bR}{\bR^2} \frac{1}{P_R(\bR ; \Ef, J)}\,,
\ee
where $R_{\rm min}= R_{\rm min}(\Ef,J)$ is the radial turning point defined by the vanishing of $P_R$.

In terms of rescaled variables (including $j=J/(GM\mu)$), this reads
\be
\frac{\pi}{2}+ \frac12\chi(\g, j)= + \int_{\br_{\rm min}}^{+\infty}  j \frac{d\br}{\br^2} \frac{1}{p_r(\br ; \g, j)}\,,
\ee
where
\be
p_r(\br ; \g, j)= + \sqrt{p_{\infty}^2+ w(\bu, p_{\infty}) - \frac{j^2}{\br^2}}\,.
\ee
Indeed, one must use the positive squareroots in the integrals above that have been written from the radial turning points ($\bR_{\rm min}$
or $\br_{\rm min}$) to infinity.

In terms of the variable $\bu=1/\br= GM/\bR$, the above integral reads (with $\bu_{\rm max}\equiv 1/\br_{\rm min}$)
\be
\frac{\pi}{2}+ \frac12\chi(\g, j)= + \int_{0}^{u_{\rm max}(\g,j)}  \frac{j \, d\bu}{\sqrt{p_{\infty}^2+ w(\bu, p_{\infty}) - j^2 \bu^2}}.
\ee
Introducing the integration variable
\be
x \equiv \frac{j \, \bu}{p_{\infty}}\,,
\ee
this reads
\be \label{chix}
\frac{\pi}{2}+ \frac12\chi(\g, j)=  \int_{0}^{x_{\rm max}(\g,j)} \frac{dx}{\sqrt{1-x^2+ {\widetilde w}(\frac{ x}{j} , p_{\infty}) }}\,,
\ee
where 
\be \label{wx}
{\widetilde w}\left(\frac{ x}{j} , p_{\infty}\right) \equiv \frac{1}{p_{\infty}^2} \left[w(\bu, p_{\infty}) \right]_{\bu\mapsto x p_{\infty}/j}\,.
\ee
The PM expansion of $w(\bu)$ yields the following large-$j$ expansion of $ {\widetilde w}(\frac{ x}{j} , p_{\infty})$:
\be \label{wexp}
{\widetilde w} \left(\frac{ x}{j} , p_{\infty}\right)= {\widetilde w}_1 \frac{ x}{j} + {\widetilde w}_2 \frac{ x^2}{j^2} +{\widetilde w}_3 \frac{ x^3}{j^3} +{\widetilde w}_4 \frac{ x^4}{j^4} + \cdots
\ee
where we introduced
\bea \label{tildewn}
{\widetilde w}_1(p_{\infty}) &=& \frac{w_1(p_{\infty})}{p_{\infty}} ,\nonumber \\
{\widetilde w}_2(p_{\infty}) &=& w_2(p_{\infty}) ,\nonumber \\
{\widetilde w}_3(p_{\infty})&=& p_{\infty} w_3(p_{\infty}) ,\nonumber \\
{\widetilde w}_4(p_{\infty})&=& p_{\infty}^2 w_4(p_{\infty})\,.
\eea
Before doing any calculation, we see from the integral expression \eqref{chix}, with the expansion \eqref{wexp}, that
the scattering function $\chi(\g,j)$ will only depend on the coefficients
\be
{\widehat w}_n \equiv \frac{{\widetilde w}_n(p_{\infty})}{j^n}\,,
\ee
entering
\be
{\widetilde w}(\frac{ x}{j} , p_{\infty})= \sum_n {\widehat w}_n x^n\,.
\ee
Moreover, as $1/j=O(G)$, the $n$th order term, $\propto G^n$, in the PM expansion of $\frac12 \chi(\g,j)=\sum_n \chi_n/j^n$ must be a polynomial in the ${\widehat w}_m $'s of total degree $\sum m_i=n$. In other words, the coefficient  $\chi_n$ of 
$1/j^n$ must be a polynomial in the ${\widetilde w}_m $'s of total degree $\sum m_i=n$. This trivial remark suffices to
prove that all the coefficients ${\widetilde w}_n(\g) $ are gauge-invariant functions, independent of any canonical
transformation (reducing to the identity when $G\to0$) acting on the rescaled dynamical variables $\x$ and $\p$ (or on their unrescaled versions $\X$, $\bP$).

To have more information on the physical meaning of the various gauge-invariant coefficients ${\widetilde w}_n(\g)$,
one needs to explicitly compute the PM (or $1/j$) expansion of the integral expression  \eqref{chix}. One a priori
technical difficulty is that if one straightforwardly expands the integral on the right-hand side (rhs) of Eq. \eqref{chix}
in powers of $G$, i.e. in powers of ${\widetilde w}(\frac{ x}{j} , p_{\infty})=O(G)$, one generates formally divergent
integrals. In addition, the upper limit of integration (where the expanded integral diverges) depends also on $G$:
$x_{\rm max}(\g,j)= 1 + O(G)$. However, there is a simple way out. It was indeed shown in Ref. \cite{Damour:1988mr},
that the correct result for such an expanded integral is simply obtained by ignoring the expansion of the upper limit,
and by taking the {\it Hadamard partie finie} (Pf) of the divergent integrals. This yields the expansion
\be
\frac{\chi(\g, j)}{2}= \sum_{n \geq1} {\rm Pf}  \int_{0}^{1} dx  {-\frac{1}{2} \choose  n } (1-x^2)^{-\frac12 -n}\left[{\widetilde w}\left(\frac{ x}{j} \right) \right]^n\,.
\ee
Each integral in this expansion (after reexpanding the $n$th power of ${\widetilde w}(x/j)= {\widetilde w}_1 x/j+ {\widetilde w}_2 x^2/j^2+ \cdots$ in powers of $1/j=O(G)$) is an integral of the type
\be
 {\rm Pf}  \int_{0}^{1} dx \,  (1-x^2)^{-\frac12 -n} x^m\,.
\ee
Replacing, e.g., $x$ by $z^{\frac{1}{2}}$, the latter integral becomes an Euler Beta function (and its Hadamard partie finie
is trivially obtained by analytical continuation in the original power $-\frac{1}{2} \rightarrow -\frac{1}{2} + \epsilon$, taking
finally $\epsilon \to 0$). This yields for the coefficients $\chi_n$ of the expansion of $\chi/2$ in powers of $1/j$
\bea \label{chinvswn}
\chi_1&=& \frac12 {\widetilde w}_1, \nonumber\\
\chi_2&=&\frac{\pi}{4} {\widetilde w}_2, \nonumber\\
\chi_3&=& - \frac{1}{24}{\widetilde w}_1^3+ \frac12 {\widetilde w}_1 {\widetilde w}_2 + {\widetilde w}_3 ,\nonumber\\
\chi_4&=& \frac{3 \pi}{8} \left( \frac12 {\widetilde w}_2^2 + {\widetilde w}_1 {\widetilde w}_3+ {\widetilde w}_4 \right)\,.
\eea
By inserting in Eqs. \eqref{chinvswn} the definitions \eqref{tildewn} of the ${\widetilde w}_n$'s one gets the expressions of
the $\chi_n$'s in terms of the coefficients $w_n$ of the potential $W(\bu)$. Relations equivalent to the latter relations have been 
also written down to 4PM order in Eq. (11.25) of \cite{Bern:2019crd}, and to all orders in \cite{Kalin:2019rwq,Bjerrum-Bohr:2019kec}.

Then, by inserting in the latter expressions the expressions \eqref{wnvsqn} of the $w_n$'s in terms of the $q_n$'s,
we get  the $\chi_n$'s in terms of the coefficients $q_n$ of the EOB $Q$ function. For instance,
we get at the 2PM, 3PM and 4PM levels
\bea \label{chinvsqn}
\chi_2&=&\frac{\pi}{4} \left( \frac{3}{2} (5\g^2 - 1) - q_2(\g) \right),\nonumber\\
\chi_3&=& \frac{-1+ 12 p_{\infty}^2 +72 p_{\infty}^4+ 64 p_{\infty}^6 }{3 p_{\infty}^3}\nonumber\\
& -& p_{\infty}\left( q_3(\g) + \frac{2 \g^2-1}{\g^2-1} q_2(\g) \right), \nonumber\\
\chi_4&=& \frac{105\pi}{128}  \left(16 + 48 p_{\infty}^2 + 33 p_{\infty}^4 \right) \nonumber\\
& -&\frac{3\pi}{16} \left[ 3 (4 + 5  p_{\infty}^2) q_2(\g) - q_2(\g)^2  \right. \nonumber\\ 
&&\left.  + (4 + 6  p_{\infty}^2) q_3(\g) + 2  p_{\infty}^2 q_4(\g) \right],
\eea
where we mixed the use of $\g$ and $p_{\infty}\equiv \sqrt{\g^2-1}$.
The first two links (at the 2PM and 3PM levels) have already been obtained (by using the $Q$ route) in \cite{Damour2018},
see Eqs. (5.5), (5.6) and (5.8) there.  

We recall that the $q_n$'s are functions {\it both} of $\g$ and 
of the symmetric mass ratio $\nu$, and that $q_n \to 0$ as $\nu \to 0$. This implies in particular that
the $q_n \to 0$ limits of the rhs's of the above equations are simply the values $\chi_n^{\rm Schw}$ of the $\chi_n$'s
for a test particle moving in a Schwarzschild background (as given in Eqs. (3.18)--(3.21) of \cite{Damour2018}).
Let us also note in passing that, despite the appearance of denominators blowing up at low velocities (when $\peob^2 \to 0$,
 i.e. $\g^2 \to 1$) in some of the expressions we will give below for them, the functions $q_{n}(\g, \nu)$ are all regular as $\peob^2 \to 0$.

\subsection{Summary of the current knowledge of the PM-expanded dynamics}

The above-derived links between $\chi_n$, $q_n$ and $w_n$ can be used in various ways. In particular, if one has derived
the scattering coefficients $\chi_n$ up to some PM level, one can directly deduce from them the values of the corresponding
$q_n$'s and $w_n$'s. This the way Refs. \cite{Damour:2016gwp,Damour2018} derived the values of the $q_n$'s and $w_n$'s
at the 1PM and 2PM levels. Let us summarize these results here.
\begin{align} 
& \chi_1(\g,\nu) =\frac{2 \, \g^2-1 }{\sqrt{\g^2-1}}=\chi_1^{\rm Schw}(\g),\\
&\chi_2(\g,\nu)=\frac{3 \pi}{8} \frac{(5 \, \g^2-1)}{h(\g,\nu)}= \frac{\chi_2^{\rm Schw}(\g)}{h(\g,\nu)}, 
\end{align}
\begin{align} 
& q_1(\g,\nu) =0 \,,\\
&q_2(\g,\nu)= \frac{3}{2} \left( 5 \, \g^2-1 \right) \left[ 1 -  \frac{1}{h(\g,\nu)}  \right]\, ,
\end{align}
\begin{align} 
& w_1(\g,\nu) =2 (2 \g^2-1),\\
&w_2(\g, \nu) =  \frac{3}{2} \frac{\left( 5 \, \g^2-1 \right) }{h(\g,\nu)}\,.
\end{align}
Concerning the 3PM level, we have seen above that it depends on the knowledge of a {\it single} function of $\g$,
entering as the coefficient of $1/h^2(\g,\nu)$ in $\chi_3(\g,\nu)-\chi_3(\g,0)$.
Let us define the auxiliary function $B(\g)$ as 
\be \label{B}
 B(\g) \equiv\frac32 \frac{(2 \g^2-1)(5 \g^2-1)}{\g^2-1}\,,
\ee
and introduce two other functions of $\g$, $A(\g)$ and $C(\g)$, constrained to identically satisfy
\be \label{ABCeq0}
A(\g)+B(\g)+C(\g)\equiv 0\,.
\ee
With this notation (and $p_{\infty}\equiv \peob \equiv \sqrt{\g^2-1}$), 
our results above give the following structural information at the 3PM level
\be\label{chi3vsAB}
\chi_3(\g,\nu)= \chi_3^{\rm Schw}(\g)    - p_{\infty} \left(A(\g)+B(\g)\right) \left( 1-\frac1{h^2(\g,\nu)} \right)\,,
\ee
\be \label{q3vsABC}
q_{3}(\g, \nu)= A(\g) + \frac{B(\g)}{h(\g,\nu)}+ \frac{C(\g)}{h^2(\g,\nu)} \,,
\ee
\be\label{w3vsq3}
w_3(\g,\nu)= 9 \g^2 - \frac{1}{2} - q_3(\g,\nu) \,.
\ee
If we further introduce the notation
\be
{\overline C}(\g) \equiv  (\g-1) \left(A(\g) + B(\g) \right) = - (\g-1) C(\g)\,,
\ee
we can rewrite Eq. \eqref{chi3vsAB} as
\be \label{chi3vsbarC}
\chi_3(\g,\nu)= \chi_3^{\rm Schw}(\g)    - \frac{ 2 \,\nu \, p_{\infty} }{h^2(\g,\nu)}  {\overline C}(\g)\,,
\ee
and  Eq. \eqref{q3vsABC} as
\be \label{q3vsbarC}
q_{3}(\g, \nu)=  B(\g) \left( \frac{1}{h(\g,\nu)} -1\right)+ \frac{2 \,\nu \,{\overline C}(\g)}{h^2(\g,\nu)} \,.
\ee
This shows that the univariate function ${\overline C}(\g)$ directly parametrizes the bivariate 3PM scattering coefficient $\chi_3(\g,\nu)$ via
the expression
\be \label{barCvschi3}
2 \,\nu \, p_{\infty} {\overline C}(\g) = - h^2(\g,\nu) \left( \chi_3(\g,\nu)- \chi_3^{\rm Schw}(\g) \right)\,.
\ee

Let us now discuss  what is our current secure ({\it i.e.}, cross-checked by at least two independent calculations) 
knowledge of $\chi_3(\g,\nu)$, and therefore of the function  ${\overline C}(\g)$.
From the $O(G^3)$ term in the 4PN-accurate expression of the scattering angle
derived in   Ref. \cite{Bini:2017wfr}, one can straightforwardly
derive the following 4PN-accurate value of the function ${\overline C}(\g)$ (expanded
in powers of $\pinf= \peob$):
\be\label{C4PN}
{\overline C}^{4 \rm PN}(\peob)= 4 + 18 \pinf^2 + \frac{91}{10} \pinf^4 + O(\pinf^6)\,.
\ee
Recently, a new (purely classical) method \cite{Bini:2019nra} allowed one to compute the 
5PN-level term in the $O(G^3)$ scattering angle, with the result
\be\label{C5PN}
{\overline C}^{5 \rm PN}(\peob)= 4 + 18 \pinf^2 + \frac{91}{10} \pinf^4 - \frac{69}{140} \pinf^6 + O(\pinf^8)\,.
\ee
On the other hand,
the quantum-amplitude approach of Refs. \cite{Bern:2019nnu,Bern:2019crd} resulted in the computation of a classical value
for $\chi_3(\g,\nu)$ (see Eq. (11.32) of Ref. \cite{Bern:2019crd}, and Ref. \cite{Antonelli:2019ytb}), from which one can derive   
the following value of the function ${\overline C}(\g)$:
\bea\label{bCB}
{\overline C}^B(\g) &=& \frac{2}{3} \g  (14 \g^2+25) \nonumber\\
&+& 4 (4 \g^4 - 12 \g^2 -3) \frac{{\rm as}(\g)}{\sqrt{\g^2-1}}\,,
\eea
where we used the shorthand notation
\be
{\rm as}(\g) \equiv \arcsinh \sqrt{ \frac{\g-1}{2}}\,.
\ee
Note in passing that the expression obtained by inserting Eq. \eqref{bCB} in the above formula for $\chi_3$ is simpler
than (though equivalent to) Eq. (11.32) of Ref. \cite{Bern:2019crd}. In particular, the $a+b/h^2$ structure of $\chi_3$
is present (though somewhat hidden) in their Eq. (11.32).

Let us also note, for future use, other (simpler) forms of the arcsinh function, namely
\be
{\rm as}(\g)= \frac12 \ln \left( \g + p_{\infty} \right)  = - \frac12 \ln \left( \g - p_{\infty} \right)\,,
\ee
where we recall that $p_{\infty} \equiv \sqrt{\g^2-1}$, and
\be
{\rm as}(\g)= \frac14 \ln \frac{ \g + p_{\infty}}{\g - p_{\infty}}=  \frac14 \ln \frac{ 1 + v_{\infty}}{1 - v_{\infty}}\,.
\ee
Here $v_{\infty}$ denotes the (Lorentz-invariant) asymptotic relative velocity between the two bodies
\be
v_{\infty} \equiv \frac{p_{\infty}}{\g} \equiv \sqrt{1- \frac1{\g^2}} \; {\rm  such \; that} \; \g=\frac1{\sqrt{1-v_{\infty}^2}}.
\ee
Note that in the slow-velocity limit ($\g \to 1$, or $p_{\infty} \to 0$)
\be
{\rm as}(\g)= \frac12 p_{\infty} - \frac1{12}  p_{\infty}^3 + \frac{3}{80}  p_{\infty}^5  - \frac{5}{224} p_{\infty}^7 + \ldots
\ee
so that the ratio ${\rm as}({\g})/\sqrt{\g^2-1} ={\rm as}({\g})/p_{\infty}$ entering ${\overline C}^B(\g)$ has a smooth slow-velocity limit
\be
\frac{{\rm as}(\g)}{p_{\infty}}= \frac12  - \frac1{12}  p_{\infty}^2 + \frac{3}{80}  p_{\infty}^4 - \frac{5}{224} p_{\infty}^6 + \ldots
\ee
and is an even function of $p_{\infty}$.

As we shall discuss below, the high-energy ($\g \to \infty$) behavior of the expression \eqref{bCB} seems, at face
value, to be in contradiction with the high-energy behavior found in the SF computation of Ref. \cite{Akcay:2012ea}.
The origin of this tension lies in the fact that the high-energy (HE) behavior of the ${\rm as}(\g)$ function is
\be
{\rm as}(\g) \overset{\rm HE}{=} \frac12 \ln (2 \g)\,,
\ee
so that the leading-order term in the high-energy behavior of the corresponding $q_3$ potential is
\be \label{HEq3B}
q_{3}^B(\g, \nu) \overset{\rm HE}{=} 8 \g^2 \ln (2 \g)\,.
\ee
By contrast, Ref.~\cite{Damour2018} (see Eq.~(6.8) there) had suggested that all EOB coefficients
$q_{n}(\g, \nu)$ should have a  logarithm-free  high-energy behavior of the type
\be
q_{n}(\g, \nu) \overset{\rm HE}{=} c_n^{(q)}\g^2\,,
\ee 
with a $\nu$-independent coefficient $c_n^{(q)}$. The latter high-energy behavior was suggested by several
independent arguments, and notably 
because of its direct compatibility with the high-energy behavior of the SF-expanded EOB Hamiltonian
found in Ref. \cite{Akcay:2012ea}.
We shall further discuss below the relation between the high-energy behavior of $q_{3}^B(\g, \nu) $ and that of the
SF-expanded EOB Hamiltonian and suggest several ways of relieving the tension between the result \eqref{HEq3B},
derived from Refs. \cite{Bern:2019nnu,Bern:2019crd}, and the result of Ref. \cite{Akcay:2012ea}.
We shall also emphasize the importance of 6PN-accurate $O(G^3)$ computations to discriminate between 
various possible ways of relieving the latter tension.

%%%%%%%%%%%
 \section{Map between the 3PM EOB potential and the quantum scattering amplitude} \label{sec4}
 
\subsection{Prelude: quasi-classical scattering amplitude associated with the classical scattering function}

As a prelude to our discussion of the link between the quantum scattering amplitude and the classical dynamics,
let us mention a direct way of using the scattering function $\frac12 \chi(\e, j)$ for constructing the quasi-classical 
(Wentzel-Kramers-Brillouin) approximation to the quantum scattering amplitude. 

 Let us start by clarifying the notation we shall use for the scattering amplitude $\cM$.
  The  Lorentz-invariant amplitude $\cM$ is  defined from the two-body scattering matrix by
\be
\langle p'_1p'_2|  S | p_1p_2\rangle={\rm Identity}+ i (2\pi)^4 \delta^4(p_1+p_2-p'_1-p'_2) \frac{\cM}{N},
\ee 
with the normalization factor  $N= (2E_1)^{1/2} (2E_2)^{1/2} (2E'_1)^{1/2} (2E'_2)^{1/2}$ when using the state
normalization $\langle p'|p\rangle = (2\pi)^3 \delta^3(\p-\p')$. With this definition, $\cM$ is dimensionless.

 Starting from the  dimensionless Lorentz-invariant amplitude $\cM(s,t)$,  it is convenient to
 introduce the associated  amplitude $f_R(\theta)$ defined as 
 \be \label{MvsfR}
 \cM \equiv 8 \pi \frac{s^{1/2}}{\hbar} f_R(\theta)\,.
 \ee
 The amplitude $f_R(\theta)$ has the dimension of a length, and is related to the differential c.m. cross-section via 
 $d\sigma = |f_R(\theta)|^2 d\Omega_{\rm c.m.}$. Let us then consider the partial-wave expansion of the amplitude, written as
 \be \label{partialwave}
 f_R(\theta)= \frac{\hbar}{P_{\rm c.m.}} \sum_{l=0}^{\infty} (2 l+1) \frac{e^{2 i \delta_l} -1}{2i} P_l(\cos \theta)\,.
 \ee
 Here $\theta$ denotes the c.m. scattering angle, and $P_{\rm c.m.}$ the c.m. momentum, related to the Mandelstam
 invariant $s= (E^{\rm tot}_{\rm c.m.})^2= (E_1^{\rm c.m.}+E_2^{\rm c.m.})^2$,
 with $E_1^{\rm c.m.}=\sqrt{m_1^2 + P_{\rm c.m.}^2}$, $E_2^{\rm c.m.}=\sqrt{m_2^2 + P_{\rm c.m.}^2}$. 
 The angle $\theta$ is related to the second Mandelstam invariant $t = - Q_{\rm c.m.}^2$ via
\be
\sqrt{-t} = Q_{\rm c.m.}=  2 \sin \frac{\theta}{2} P_{\rm c.m.}\,.
\ee
 In the expansion \eqref{partialwave}, $\delta_l$ denotes the (dimensionless)
 phase shift of the partial wave corresponding to the c.m. angular momentum $L = \hbar l$, where $l= 0,1,2,\ldots$.
 In the classical limit we can identify the quantized total c.m. angular momentum $L = \hbar l$ with $J$.
 In terms of the dimensionless quantities $l$ and $\delta_l$ entering the expansion \eqref{partialwave}, a quasi-classical
 description of the dynamics {\it a priori} corresponds to a case where both of them are large: $l \gg 1$ and $\delta_l \gg 1$.
 This is formally clear because $ l = L/\hbar$, while, for potential scattering, the quasi-classical (Wentzel-Kramers-Brillouin) approximation
 to the phase shift is $\delta_l \approx \Delta S_L/\hbar$ where $\Delta S_L$ is the (subtracted) half-radial action
 along a classical motion with angular momentum $L$ \cite{LandauQM,FordWheeler1959}. Most useful for our present purpose
 is the fact that the phase-shift $\delta_l $ is linked,
 in the classical limit, to the scattering angle $\chi$ by
 \be
 \frac12 \chi = - \frac{\partial \delta_l}{\partial l}\,.
 \ee
 When expressing $l \equiv L/\hbar \equiv J/\hbar$ in terms of the classical dimensionless angular momentum $j\equiv J/(G m_1 m_2)$,
 the latter result reads
 \be \label{chideltal}
 \frac12 \chi(\e,j) = -  \hh \frac{\partial \delta_l}{\partial j}\,,
 \ee
 where we defined (as in \cite{Damour2018}) the following dimensionless version of $\hbar$
 \be \label{hh}
\hh \equiv \frac{\hbar}{G m_1 m_2} = \frac{\hbar}{GM\mu}  \,.
\ee
 Equation \eqref{chideltal} shows that $\delta_l$ can be obtained (in the classical limit) by integrating over $j$ the classical
 scattering function $\frac12 \chi(\e,j)$. Using the PM-expansion \eqref{chiPM} of $\frac12 \chi(j)$ (and $\e=\g$), 
 then yields the following expansion for $\delta_l$
 \be \label{wkbdeltal}
 \delta_l= \frac1{\hh} \left(  \chi_1(\g, \nu) \ln \left( \frac{j_0}{j}\right) + \frac{ \chi_{2}(\g, \nu)}{j} + \frac12 \frac{\chi_{3}(\g, \nu)}{j^2} + \cdots \right),
 \ee
 where $j_0$ is linked to the IR cutoff needed when evaluating the corresponding IR-divergent Coulomb phase.

 %%%%%%%%
 \subsection{Computation of the quantum scattering amplitude derived from the 3PM EOB potential} 
 
Ref. \cite{Damour2018} had shown how to map the simple 2PM-accurate, energy-gauge EOB description  of the
two-body dynamics onto a corresponding quantum scattering amplitude, say $\cM_{\rm eob}^{\rm 2PM}$, and had checked that 
$\cM_{\rm eob}^{\rm 2PM}$ agreed with what Refs. \cite{Cachazo:2017jef,Guevara:2017csg} (later followed by Refs. 
\cite{Bjerrum-Bohr:2018xdl,KoemansCollado:2019ggb}) had computed as being the ``classical part'' of the $G^2$-accurate
quantum scattering amplitude. In this section we extend this result to the 3PM level. More precisely, we shall show that
the extension of the map defined in Ref. \cite{Damour2018} leads to a 3PM-accurate amplitude, $\cM_{\rm eob}^{\rm 3PM}$,
that coincides with what Refs. \cite{Bern:2019nnu,Bern:2019crd} computed as being the classical part of the $G^3$-accurate
quantum scattering amplitude. 

 Let us  start by recalling that the approach of Ref. \cite{Damour2018} is simply to quantize the
 classical, energy-gauge EOB mass-shell condition, {\it i.e.} to quantize the motion of a particle of mass $\mu$
 moving in a nonrelativisticlike radial potential. Indeed, the  energy-gauge EOB mass-shell condition has the form
 \be
 \bP^2= P_{\infty}^2+ W(R, P_{\infty})\,,
 \ee
  where
 \be
 P_{\infty}^2 \equiv \Ef^2- \mu^2 = \mu^2 (\g^2-1)\,,
 \ee
 and where, to ease the notation, we henceforth suppress
the bar over the  isotropic EOB radial coordinate $R = | \X|$ (and its rescaled avatar $r=R/(GM)=\br$).

The canonical quantization of $\X$ and $\bP$, i.e. 
\be
[X^i , P_j] = i \hbar \,\delta^i_j \,,
\ee
is equivalent to solving the fixed-energy Schr\"odinger equation in the energy-dependent radial potential $W(R, P_{\infty})$.
As in the classical problem, it is convenient to replace the canonically conjugated variables  $\X$, $\bP$ by their (dimensionless) rescaled 
avatars $\x \equiv \X/(G M)$ and $\p \equiv \bP/\mu$ (with $r \equiv |\x|$), satisfying the following rescaled commutation relation:
 \be
[x^i , p_j] = i \hh \,\delta^i_j \;.
\ee
Here (following \cite{Damour2018}) $\hh$ denotes the  (dimensionless) rescaled version of $\hbar$ 
defined in Eq. \eqref{hh}.
In terms of these rescaled variables the mass-shell condition determining $\p$ reads
  \be \label{p2eq}
 \p^2= p_{\infty}^2+ w(r, p_{\infty})\,,
 \ee
where, as we have seen, the PM-expansion of the rescaled radial potential $w\equiv W/\mu^2$ reads
\be \label{wexp2}
 w(r, p_{\infty})
 =  \frac{w_1(\g)}{r} + \frac{w_2(\g)}{r^2} +  \frac{w_3(\g)}{r^3} + \frac{w_4(\g)}{r^4} + \cdots
\ee
One should keep in mind that, as $\frac1{r}= \frac{GM}{{ R}}$, a contribution to the potential  $\propto 1/r^n$ is of order $O(G^n)$.

The quantization of the EOB mass-shell condition \eqref{p2eq} 
 yields the following time-independent Schr\"odinger equation (here truncated at the 3PM level)
\be \label{schro}
- \hh^2 \Delta_{\bf x} \psi({\bf x}) = \left[  p_{\infty}^2 + \frac{w_1}{r} + \frac{w_2}{r^2} +   \frac{w_3}{r^3}+O\left(\frac1{r^4}\right)    \right]\psi({\bf x}) \,.
\ee
In other words (as was already pointed out in \cite{Damour2018,Damourtalks}), the quantization of the isotropic-coordinates formulation of the EOB dynamics of
two spinless particles leads to a potential scattering, with an energy-dependent potential which is a deformation of
a Coulomb potential $ \frac{w_1}{r}$ by higher inverse powers of $r\equiv\br$: $\frac{w_2}{r^2} +   \frac{w_3}{r^3} + \cdots$.

Given an incoming state $|\vk_a\rangle=\varphi_a = e^{i  \vk_a \cdot \x }$ in the infinite past, impinging
on this EOB-potential $w$, 
the scattering amplitude $f_{\rm eob}(\widehat \vk_b)$ (where $\widehat \vk_b=  \vk_b/ | \vk_b|$)
from $|\vk_a\rangle$ to some
outgoing state $|\vk_b\rangle=\varphi_b = e^{i  \vk_b \cdot \x }$ is given by
\be \label{fvsw}
f_{\rm eob}(\widehat \vk_b) = +\frac1{4\pi \hh^2} \langle\varphi_b| w |\psi^+_a \rangle\,.
\ee
Here $\psi^+_a$ is the stationary  retarded-type solution of the scattering equation \eqref{schro} describing
the incoming state $|\vk_a\rangle=\varphi_a = e^{i  \vk_a \cdot \x }$ in the infinite past, and having the
following asymptotic structure at large distances
\be
\psi^+_a  \underset{r \to \infty}{\approx} e^{i  \vk_a \cdot \x } + f_{\rm eob}(\Omega) \frac{e^{i k r}}{r}\,,
\ee
where $\Omega$ denotes the polar coordinates of $\x$ on the sphere of scattering directions.

The crucial point of Ref. \cite{Damour2018} was that, modulo a simple rescaling, namely (see below)
\be \label{cMvsf}
\cM_{\rm eob} =  \frac{8\pi G}{\hbar} \left( E_{\rm real}^{\rm c.m.} \right)^2 f_{\rm eob} = \frac{8\pi G \, s }{\hbar} f_{\rm eob}\,,
\ee
the EOB scattering amplitude  could be identified, at the then existing $O(G^2)$ approximation, with the 
so-called classical part \cite{Cachazo:2017jef,Guevara:2017csg} of the quantum gravity amplitude $\cM$. 
When rewriting Eq. \eqref{cMvsf} in terms of  the corresponding ``non-relativistically-normalized" 
amplitude, say $M^{\rm NR}$, as used in Refs. \cite{Cheung:2018wkq,Bern:2019nnu,Bern:2019crd}, we have
\be \label{MNRvsf}
M^{\rm NR}_{\rm eob} \equiv \frac{\cM_{\rm eob}}{4 E^{\infty}_1 E^{\infty}_2}=  \frac{2\pi G}{\hbar \, \xi_{\infty}}  f_{\rm eob}\,,
\ee
where $\xi_{\infty}= E^{\infty}_1 E^{\infty}_2/(E^{\infty}_1+E^{\infty}_2)^2$ is the asymptotic value
of the symmetric energy ratio $\xi$ defined in \cite{Cheung:2018wkq} (see also Eq.\eqref{defxi} below). 

In the dictionary of Ref. \cite{Damour2018},
the EOB scattering angle $\theta$ between $\widehat \vk_a$ and $\widehat \vk_b$ is directly equal to the physical c.m.
scattering angle, as it enters the physical c.m. momentum transfer
\be
\sqrt{-t} = Q^{\rm c.m.}=  2 \sin \frac{\theta}{2} P_{\rm c.m.}\,.
\ee
This is the quantum version of the fact, proven in Ref. \cite{Damour:2016gwp}, that the classical EOB scattering angle 
coincides with the corresponding c.m. scattering angle. 
 On the other hand, one must remember that the various momenta and wave vectors, $ p_{\infty} =\sqrt{\g^2-1}$,
 $\vk_a$,  $\vk_b$, ${\bf q}$, entering the EOB description differ by some rescaling factors from the corresponding
 physical c.m. ones. First, the link between $ p_{\rm eob} \equiv p_{\infty} =\sqrt{\g^2-1}$ and the
 physical c.m. momentum is
  \be
P_\infty^{\rm EOB} \equiv \mu \, p_\infty = \frac{\E}{M} P_{\rm c.m.}= h(\g) P_{\rm c.m.}\,.
 \ee
 In addition, the conserved norm of the (rescaled) wave vector,  $k= |\vk_a| = |\vk_b|$, is related
to $ p_{\infty} =\sqrt{\g^2-1}$ via
 \be \label{pinfvsk}
p_{\infty} = \hh \, k \,,
\ee
so that the rescaled momentum transfer reads
 \be \label{qvsk}
{\bf q} = \vk_b - \vk_a \; ; \; q = |{\bf q}| = 2 k \sin \frac{\theta}{2} .
\ee
As a consequence of these relations, we have the link
\bea \label{qvsQ}
q  &=& 2 \sin \frac{\theta}{2} \frac{p_{\infty}}{\hh} = 2 \sin \frac{\theta}{2}  \frac{h(\g) P_{\rm c.m.}}{\mu \,\hh}\nonumber \\
&=&   \frac{G M}{\hbar} h(\g) Q^{\rm c.m.} \,.
\eea

Rewriting the link  \eqref{cMvsf} in terms of the relativistic (partial-wave) amplitude $f_R$, defined by Eq. \eqref{MvsfR}, 
leads to the following relation between $f_R$ and $f_{\rm eob}$:
\be
f_R= G \sqrt{s} f_{\rm eob}\,.
\ee
Note that while $f_R$ has the dimension of a length, $f_{\rm eob}$ is dimensionless. The partial-wave expansion of $f_{\rm eob}$
is, in close parallel to Eq. \eqref{partialwave},
\be \label{partialwaveeob}
 f_{\rm eob}(\theta)= \frac{\hh}{p_{\infty}} \sum_{l=0}^{\infty} (2 l+1) \frac{e^{2 i \delta_l} -1}{2i} P_l(\cos \theta)\,,
 \ee
with the same phase shifts, but a prefactor $\frac{\hh}{p_{\infty}}= \frac1k$ which is dimensionless, because of our
various rescalings. At the conceptual level, the relative normalization factor given in Eq. \eqref{cMvsf} is most
clearly understood by saying that the pure phase-shift dimensionless factor of the real amplitude $\cM$, say 
\be
\widehat f(\theta) \equiv \sum_{l=0}^{\infty} (2 l+1) \frac{e^{2 i \delta_l} -1}{2i} P_l(\cos \theta)\,,
\ee
 coincides with the corresponding EOB one. An alternative way \cite{Damour2018} to derive the relative normalization
between $\cM$ and $f_{\rm eob}$ is to compare the LO value, \eqref{born1}, of $\cM$ to the corresponding LO value,
$w_1/(\hh^2q^2)$, of $f_{\rm eob}$, as given in Eq. (10.23) of \cite{Damour2018}, and below.

Let us now derive the 3PM-accurate value of the EOB scattering amplitude $f_{\rm eob}$, and compare it to the result
of Refs. \cite{Bern:2019nnu,Bern:2019crd}. It can be written as
\be
\cM_{\rm eob}^{\rm 3PM} ={\cM}_{\rm eob}^{\prime} + {\cM}_{\rm eob}^{\prime \prime}\,,
\ee
where 
\be
{\cM}_{\rm eob}^{\prime} \equiv \frac{8\pi G \, s }{\hbar} f^w_{\rm eob}\,,
\ee
denotes the first Born approximation to $f_{\rm eob}$ (which is {\it linear} in the potential $w$), while 
\be
{\cM}_{\rm eob}^{\prime \prime} \equiv \frac{8\pi G \, s }{\hbar} f^{w^2 + w^3 +\ldots}_{\rm eob}\,,
\ee
denotes the sum of the terms coming from higher order Born iterations (which are {\it nonlinear} in the potential $w$).

The explicit form of the first Born approximation to $f_{\rm eob}$ is defined by  replacing in Eq. \eqref{fvsw} $\psi^+_a $ 
by the unperturbed state $\varphi_a = e^{i  \vk_a \cdot \x }$:
\bea \label{B1}
f^{w}_{\rm eob}({\bf q}) &=&  +\frac1{4\pi \hh^2} \langle\varphi_b| w(r) |\varphi_a \rangle \nonumber \\
 &=&  +\frac1{4\pi \hh^2} \int d^3\x e^{- i {\bf q} \cdot \x} w(r) \,.
\eea
We recall that the EOB potential, $w(r)$, Eq. \eqref{wexp2},
is a sum of contributions $\sum_n w_n/r^n$ coming from successive PM approximations, {\it i.e.}  $w_n/r^n= O(G^n)$. 
This generates a corresponding sum of contributions in the
first Born approximation \eqref{B1}, namely
\be
f^{w}_{\rm eob}({\bf q})= \sum_n f^{w_n}_{\rm eob}({\bf q})\,,
\ee
with
\bea \label{fwn}
f^{ w_n}_{\rm eob}({\bf q}) &=&  +\frac1{4\pi \hh^2} \langle\varphi_b| \frac{w_n}{r^n} |\varphi_a \rangle \nonumber \\
 &=&  +\frac1{4\pi \hh^2} \int d^3\x e^{- i {\bf q} \cdot \x} \frac{w_n}{r^n}\,.
\eea
This is easily computed from the value of the Fourier transform of $1/r^n$, which is (in space dimension $d$)
\be
{\cal F}^{(d)}\!\left[ \frac1{r^n}\right] \equiv \int d^d \x e^{- i {\bf q} \cdot \x} \frac1{r^n} = \frac{C_n^{(d)}}{q^{d-n}}\; ,
\ee
 where
 \be
 C_n^{(d)} =\pi^{\frac{d}{2}} \frac{2^{\bar n} \Gamma(\frac12 {\bar n})}{\Gamma(\frac12 n)}\; ; \; {\rm with} \; {\bar n} \equiv d -n \,.
 \ee
 The Fourier transforms of the $1/r$ (1PM) and $1/r^2$ (2PM) potentials are convergent in dimension $d=3$,
 \be
 {\cal F}^{(3)}\!\left[ \frac1{r}\right] =\frac{4 \pi}{q^2} \; ;  \; {\cal F}^{(3)}\!\left[ \frac1{r^2}\right]=\frac{2 \pi^2}{q} \,,
 \ee
 while the 3PM-level $1/r^3$ potential leads to a UV ($r \to 0$) divergence whose dimensional regularization ($d=3+\epsilon$)
 yields the result:
 \bea
 {\cal F}^{(3 + \epsilon)}\!\left[ \frac1{r^3}\right]&=& 4 \pi \left[ \frac1{\epsilon} - \ln q  + \frac12 \ln (4\pi) - \frac12 \gamma_E +O(\epsilon) \right] \nonumber \\
 &\equiv& - 4 \pi \ln \frac{q}{\widehat \Lambda}\,,
 \eea
 where, in the last line, we denoted by $\widehat \Lambda$ a UV cutoff (in its EOB-rescaled version).
 This yields
 \be \label{feobw}
 \hh^2 f^w_{\rm eob}=\frac{w_1}{q^2}+ \frac{\pi}{2}  \frac{w_2}{q}  - w_3  \ln \frac{q}{\widehat \Lambda}\,.
 \ee
 When inserting in this result the values of $w_1$ and $w_2$ derived in \cite{Damour2018}, and the value of $w_3$ obtained
 by inserting   \eqref{bCB} in Eqs. \eqref{q3vsABC}, \eqref{w3vsq3},  and using the above-defined rescalings,
 it is straightforwardly checked that this yields
 \be \label{fwvsM'}
 {\cM}_{\rm eob}^{\prime} \equiv \frac{8\pi G \, s }{\hbar} f^w_{\rm eob}  = {\cM}_1^{\prime} + {\cM}_2^{\prime}+{\cM}_3^{\prime}\,,
 \ee
 where, following the notation used in Refs. \cite{Bern:2019nnu,Bern:2019crd}, $  {\cM}_i^{\prime}$, $i=1,2,3$, denote
 the IR-finite parts of the classical part of the amplitude $\cM$ derived there (written in Eqs. (13) and the first three lines
 of Eq. (8) in  \cite{Bern:2019nnu}). We work here with the Lorentz-invariant amplitude $\cM$, {\it i.e.} we do
 not include the factor $(4 E^{\infty}_1 E^{\infty}_2)^{-1}$. [At the technical level, Eq. \eqref{fwvsM'}
 means  that, at the 3PM level, the EOB potential coefficient
$ w_3$ can be simply identified with $- 1/\left(6 h^2(\g,\nu)\right)$ times the bracket 
$\left[ 3 - 6 \nu + 206 \nu \sigma +\cdots\right]$ multiplying
$\log {\mathbf q}^2$ in  Eq. (8) of \cite{Bern:2019crd}.]

The latter simple link between the Fourier transform of the EOB energy potential and the IR-finite part of the classical part of 
the amplitude $\cM$ of Refs. \cite{Bern:2019nnu,Bern:2019crd} has also been pointed out in  recent works 
\cite{Kalin:2019rwq,Bjerrum-Bohr:2019kec}, however, we wish to emphasize that it is in great part tautological 
(in the sense that it follows from definitions).
Indeed, on the one hand (as clearly recognized in Ref. \cite{Bjerrum-Bohr:2019kec}) the EOB formulation \cite{Damour2018} of the 
map between the classical dynamics and the amplitude $\cM$ trivially shows that the linear-in-potential part of $\cM$ is
simply given by the Fourier transform of the EOB energy-gauge potential (as was explicitly explained in several talks \cite{Damourtalks}),
 and, on the other hand Refs. \cite{Bern:2019nnu,Bern:2019crd} are actually {\it defining} $  {\cM}_i^{\prime}$ by selecting
 the parts of the total two-loop amplitude which satisfy two criteria: (i) to correspond to the $\sim G/q^2$, $\sim G^2/q$ and $G^3 \ln q$
 terms that are precisely corresponding to the classical dynamics; and, (ii) to have been amputated of the extra contributions coming from
 iterated Born approximations of the type denoted ${\cM}_{\rm eob}^{\prime \prime} \equiv \frac{8\pi G \, s }{\hbar} f^{w^2 + w^3 +\ldots}_{\rm eob}$ above. Indeed, as is stated in Ref. \cite{Bern:2019crd}, and as we shall now check,
  the latter terms are precisely the IR-divergent contributions left in the
 form of integrals in Eq. (9.3) of Ref. \cite{Bern:2019crd}. In other words, given the simple EOB map of Ref. \cite{Damour2018}, 
 and given the methodology of extracting the so-called classical part of $\cM$ proposed in \cite{Cheung:2018wkq}, and implemented
 in \cite{Bern:2019nnu,Bern:2019crd}, the apparently striking result \eqref{fwvsM'} is a tautology.

Let us now discuss the detailed structure of the iterated Born approximations
${\cM}_{\rm eob}^{\prime \prime} \equiv \frac{8\pi G \, s }{\hbar} f^{w^2 + w^3 +\ldots}_{\rm eob}$
that must be added to the linear-in-potential contribution ${\cM}_{\rm eob}^{\prime} \equiv \frac{8\pi G \, s }{\hbar} f^w_{\rm eob} $.
 As $w_n = O(G^n)$, the 3PM ($O(G^3)$) accuracy necessitates to consider both the second iteration
(with contributions proportional to $ w_1^2$ and $w_1 w_2$), and the third iteration (with contributions proportional to $ w_1^3$).
[The 3PM-level contribution coming from $w_3/r^3$ is included in the first Born approximation, and does not need
to be iterated.]
The iterations of the Coulomb-type $w_1/r$ potential can actually be deduced from the known, exact Coulomb scattering
amplitude \cite{LandauQM}.  
Alternatively, one can extract both the first two iterations of the  $w_1/r$ potential ($O(w_1^2) +  O(w_1^3)$) and the mixed iteration of 
the $w_1/r$ and $w_2/r^2$ potentials ($O(w_1 w_2)$) from an old result of  Kang and Brown  \cite{KangBrown}. Indeed,
the latter reference computed the higher-Born approximations for the Coulomb scattering amplitude of a Klein-Gordon
particle, i.e. for the wave equation
\be \label{KG}
- \hbar^2 \Delta \psi=\left[ \left(E- \frac{Z e^2}{r}\right)^2 - \mu^2 \right] \psi\,,
\ee
whose potential involves both a $w^{KG}_1/r= - 2 E Z e^2/r$ potential and a $w^{KG}_2/r^2=+ (Z e^2)^2/r^2$ one.
 
Transcribing the results of Ref. \cite{KangBrown} in terms of our scattering equation \eqref{schro} yields
explicit forms for the various Born-iterated contributions. We introduce the notation
\be
\delta_1= \frac{i}{2} \frac{w_1}{\hh^2 \, k} \ln \frac{q^2}{{\widehat \lambda}^2}= \frac{i}{2} \frac{w_1}{\hh \, p_{\infty}} \ln \frac{q^2}{{\widehat \lambda}^2}\,,
\ee
for the  IR-divergent Coulomb phase ($\widehat \lambda$ being an IR cutoff introduced by the replacement 
$Z e^2/r \to  Z e^2 e^{- {\widehat \lambda} r}/r $ in the Klein-Gordon equation \eqref{KG}).
[Note the fact that  $\delta_1$ contains a factor $1/\hbar$. This crucial property of the Born
expansion will be discussed at length in the following subsection, starting with Eq. \eqref{quantum}.]

\be\label{fw1^2}
\hh^2 f_{\rm eob}^{w_1^2} = \delta_1 \frac{w_1}{q^2}\,,
\ee
\be \label{fw1^3}
\hh^2 f_{\rm eob}^{w_1^3} =  \frac12 \delta_1^2 \frac{w_1}{q^2}\,,
\ee
\be \label{fw1w2}
\hh^2 f_{\rm eob}^{w_1 w_2}= \delta_1 \frac{\pi}{2}  \frac{w_2}{q} + \frac{w_1 w_2}{\hh^2 q^2} x B_{29}(x)\,.
\ee
Here the variable $x$ denotes
\be
x \equiv \sin \frac{\theta}{2} = \frac{q}{2 k}\,,
\ee
and the function
$B_{29}(x)$ denotes (see the last bracket in Eq. (29) of \cite{KangBrown})
\be
B_{29}(x)= i \pi \ln \frac{4}{(1+ x)}+ \ln x \ln \frac{1-x}{1+x}+ L_2(x) - L_2(-x),
\ee
where 
\be
L_2(x)= x+ \frac{x^2}{2^2}+ \frac{x^3}{3^2}+  \frac{x^4}{4^2}+ \cdots
\ee
is the dilogarithm function. All the above iterated contributions are clearly IR divergent because they all
contain a term proportional to the IR-divergent Coulomblike phase $\delta_1$.

Adding all those iterated Born contributions to the first-Born approximation $\hh^2 f_{\rm eob}^{w} $, 
Eq. \eqref{feobw}, yields the complete 3PM-accurate EOB amplitude
\bea \label{h2f}
\hh^2 f_{\rm eob}&=&(1+\delta_1 + \frac12 \delta_1^2)\frac{w_1}{q^2}+ (1+\delta_1) \frac{\pi}{2}  \frac{w_2}{q} \nonumber\\
&+& \frac{w_1 w_2}{\hh^2 q^2} x B_{29}(x) - w_3  \ln \frac{q}{\widehat \Lambda}\,.
\eea

Let us note in passing that the 3PM-expanded amplitude \eqref{h2f} is compatible with the fact (proven by Weinberg \cite{Weinberg:1965nx})
that the (gravitational) IR-divergent Coulomb phase $\delta_1$ exponentiates, {\it i.e.} that one can factorize $f_{\rm eob}$
as
\bea \label{factorizedh2f}
\hh^2 f_{\rm eob} && = e^{\delta_1} \left[   \frac{w_1}{q^2}+  \frac{\pi}{2}  \frac{w_2}{q} \right. \nonumber \\
&& \left. +   \frac{w_1 w_2}{\hh^2 q^2} x B_{29}(x) - w_3  \ln \frac{q}{\widehat \Lambda} \right] + O(G^4)
\eea
where the terms within the square brackets are IR-finite. 

As already explained, the methodology used in \cite{Bern:2019nnu,Bern:2019crd} consists of setting aside the 
various IR-divergent (Born-iterated) contributions \eqref{fw1^2}, \eqref{fw1^3}, \eqref{fw1w2}, in \eqref{h2f},
thereby retaining only the linear-in-$w$ ones. This means in particular that Refs. \cite{Bern:2019nnu,Bern:2019crd}
set aside not only the IR-divergent term proportional to $\delta_1 w_2$, but also its Born-iterated partner  $\propto w_1 w_2$ (recall that $\delta_1 \propto w_1$). They then considered as only IR-finite $O(G^3)$ contribution the last term (proportional to $\ln q$)
in Eq. \eqref{h2f}, namely
\be\label{fw3}
 - w_3  \ln \frac{q}{\widehat \Lambda}\,.
\ee
As we shall discuss next, a different IR-finite result would have been obtained if one had (following Weinberg) first
factored $e^{\delta_1}$, and then taken the small-$q$ limit.

Let us, indeed, discuss the small-angle limit,  $q \to 0$, and therefore $x \to 0$, of the complete 3PM EOB amplitude \eqref{h2f}.
We have the expansion
\bea
x \, B_{29}(x)&=& i \pi (x \ln 4 - x^2 + O(x^3))  
\nonumber \\ &+& \ln x ( -2 x^2 + O(x^4))  + 2 x^2 + O(x^4) \,.\nonumber \\
\eea
Here, the leading term $O(x)$ in the imaginary part modifies the Coulomb phase factor  $(1+\delta_1) $
in front of the $w_2/q \propto w_2/x$ term.  The terms  $O(x^2)$ (both in the imaginary part
and in the real part)  yield (after division by the $q^2$ prefactor)  contributions $\propto q^0$, which are the 
Fourier transforms of  contact terms. 

Of most interest for our discussion of the non-analytic-in-$q$ contributions in the $q\to 0$ limit,
is the fact that the $O(x^2 \ln x)$ term in the small-$x$ expansion of the function $x \, B_{29}(x)$
yields the following additional contribution to the amplitude
\be
\hh^2 f^{w_1 w_2}_{\rm eob}= -  \frac12  \frac{w_1 w_2}{p_{\infty}^2}  \ln \frac{q}{2 k}\,.
\ee
This contribution has the same $\ln q$ structure as the linear-in-$w$ contribution coming from $w_3/r^3$.

Summarizing: the real part of the 3PM, $O(G^3)$, amplitude contains the following contributions (where we recall
that $k=p_{\infty}/\hh$)
\bea \label{f3PMlogs}
\hh^2 {\rm Re}\left[ f^{3PM}_{\rm eob}\right]&=& - \frac12 \frac{w_1^3}{\hh^2 p_{\infty}^2 } \frac1{q^2}\left( \ln \frac{q}{\widehat \lambda} \right)^2 \nonumber\\
&-& \frac12  \frac{w_1 w_2}{p_{\infty}^2}  \ln \frac{q}{2 k}
-  w_3  \ln \frac{q}{\widehat \Lambda} \,.
\eea

\subsection{General concern about the link between a quantum scattering amplitude and  classical dynamics }  \label{concern}
 
 Several recent works have discussed the issue of the relation between $\cM$ and classical dynamics,
 see Refs.\cite{Donoghue:1993eb,Donoghue:1994dn,BjerrumBohr:2002kt,Bjerrum-Bohr:2013bxa,Neill:2013wsa,Cachazo:2017jef,Guevara:2017csg,Damour2018,Cheung:2018wkq,Bern:2019nnu,Bern:2019crd,Kosower:2018adc,Maybee:2019jus,Kalin:2019rwq}.  In particular, some {\it one-way} maps between (EOB or EFT) Hamiltonians describing the classical dynamics and the
 scattering amplitude have been defined, and implemented at both the 2PM \cite{Damour2018,Cheung:2018wkq} and 3PM levels 
 \cite{Bern:2019nnu,Bern:2019crd}. However, we wish here to express a general concern (which has been already
 raised in \cite{Damourtalks}) about  
 the program of transferring information between a quantum scattering amplitude and classical dynamics. As far as we know,
 this concern has not been explicitly addressed in the recent literature.

The basic idea of extracting classical information from an amplitude
 is simply that a same theory (namely GR) is underlying both the classical and the quantum
dynamics, so that there should exist some ``classical limit'' under which it should be possible to extract the classical
dynamics from a quantum scattering amplitude. [This idea was already the one of Refs. 
\cite{Corinaldesi:1956,Barker:1966zz,Barker:1970zr,Corinaldesi:1971sz,Iwasaki:1971vb,Okamura:1973my}.]
 It seems that many recent papers simply assumed the existence of a `` precise
demarcation between classical and quantum contributions to the scattering amplitude'' (as formulated in the Introduction of
 \cite{Bern:2019crd}). We wish to stress that the existence of such a demarcation is {\it a priori} unclear to us for a variety of related issues.
 
 First, let us recall the basic fact 
 that the domain of validity of the standard quantum scattering perturbation expansion (Born-Feynman
 expansion) {\it does not overlap} with the domain of validity of the standard classical scattering perturbation expansion when
 considering a Coulomblike potential $V=  Z_1 Z_2 e^2/r + O(1/r^2)$, or  $V=  - G E_1 E_2/r + O(1/r^2)$ in the gravitational
 case.
 Here, $ E_1$ and $ E_2$ denote, say, the c.m. energies of two colliding particles  (we set $c=1$).
 This fact was eloquently expressed in the classic 1948 paper of Niels Bohr on the penetration of charged
 quantum particles in matter \cite{Bohr1948}, and is also stressed in the treatise of
 Landau and Lifshitz \cite{LandauQM,Landau4}. 
 The basic point is that the quantum expansion is {\it a priori} valid
 when the dimensionless ratios ($v$ denoting the relative velocity)
 \be \label{quantum}
 \frac{ Z_1 Z_2 e^2}{ \hbar \, v} \ll 1 \; {\rm or} \;  \frac{ G E_1 E_2 }{\hbar \, v} \ll 1 \; ({\rm quantum})\,,
 \ee
 while the domain of validity for a quasi-classical description of the scattering is just the opposite, namely
  \be\label{classical}
 \frac{ Z_1 Z_2 e^2}{ \hbar \, v} \gg 1 \; {\rm or} \;  \frac{ G E_1 E_2}{\hbar \, v} \gg 1 \; ({\rm classical})\,.
 \ee
 When a precise definition of the relative velocity $v$ is needed, we shall define it as
\be
v_{\infty} \equiv \sqrt{1- \frac1{\g^2}} \; {\rm  such \; that} \; \g=\frac1{\sqrt{1-v_{\infty}^2}}.
\ee
 At the formal level of considering limits for $\hbar$, the classical domain of validity \eqref{classical} does correspond
 to the expected limit $\hbar \to 0$, while the quantum domain of validity \eqref{quantum} corresponds
 to the less usually considered formal limit $\hbar \to \infty$.

 The necessity of the inequalities \eqref{classical} and \eqref{quantum} can be seen in various ways.
At the conceptual level, Bohr points out (see subsection 1.3 of Ref. \cite{Bohr1948}) that  the condition
\eqref{classical}  is {\it necessary and sufficient} for being able ``to construct wave packets which, to a high degree
of approximation, follow the classical orbits'' during the entire scattering process. Bohr only discusses nonrelativistic 
Coulomb-like scattering. Let us show how it works in the relativistic case, and in the c.m. frame. Each particle is described
by  an incoming relativistic wavepacket having a relatively small
transversal size $d$, {\it e.g.} realized (says Bohr) by a hole of radius $d$ in a screen. The quantum diffraction angle $\phi$
caused by the hole is of order $\phi \sim \lambda/d$ where $\lambda = \hbar/P_{\rm c.m.}$ is the (reduced) de Broglie wavelength 
of each particle. In other words, $\phi \sim \hbar/(d P_{\rm c.m.})$ measures the
angular spreading of the quantum wave packets. To be able to measure the classical scattering angle $\chi$ in spite of the
quantum spread, one must have the inequality $\phi \ll |\chi|$. In addition, the transverse size must be small compared
to the impact parameter: $ d \ll b$. The leading-order (half) scattering angle is of the form 
\be
\frac12 \chi = \frac{a_s}{b},
\ee
where the length $a_s$ depends on the spin of the (massless) exchanged particle (scalar, vector or tensor). More precisely, one has
\be
a_s=  G_s \frac{Q_1 Q_2}{\mu} \frac{h(\g,\nu)}{\pinf^2}f_s( u_1 \cdot u_2)\,,
\ee
where $G_s$ is a coupling constant, $Q_a$ a (scalar, electric or gravitational) charge, and where the factor
$f_s( u_1 \cdot u_2)$ comes from the current-current interaction between the two worldlines, so that,
for the  scalar, electromagnetic and gravitational cases, respectively, one has
\be
f_0=1 \; , \; f_1=  u_1 \cdot u_2\; , \;   f_2= 2( u_1 \cdot u_2)^2 -1\,.
\ee 
Combining the inequalities $\phi \ll |\chi|$ and  $ d \ll b$ then leads to the inequality
\be
\frac{|a_s| P_{\rm c.m.}}{\hbar} \gg1\,,
\ee
where
\be
\frac{a_s P_{\rm c.m.}}{\hbar} = \frac{G_s Q_1 Q_2}{\hbar} \frac{f_s(\g)}{\sqrt{\g^2-1}}\,.
\ee
For instance, in the gravitational case, we have $G_2=G$, $Q_a=m_a$, so that the necessary inequality 
for quasi-classicality reads
\be \label{classical2}
 \frac{G m_1 m_2}{\hbar} \frac{2 \g^2-1}{\sqrt{\g^2-1}} =  \frac{G}{\hbar} \frac{2 (p_1\cdot p_2)^2 -p_1^2p_2^2}{ \sqrt{(p_1\cdot p_2)^2 -p_1^2p_2^2}}\gg1\,.
\ee
This is easily seen to be (approximately) equivalent to the second condition \eqref{classical},
for all values of the relative velocity.

An important point for our discussion is that this inequality
must be satisfied even when considering very large impact parameters, corresponding to {\it a priori} quasi-classical
very large angular momenta (and very small scattering angles). 

Another way of seeing the necessity of the inequality \eqref{classical2} comes from considering
 the LO contribution to the phase shift $\delta_l$, namely
\bea \label{deltalLO}
\delta_l^{\rm LO}&=&\frac{G m_1 m_2}{\hbar}\frac{2 \g^2-1}{ \sqrt{\g^2-1}}  \ln \left( \frac{j_0}{j}\right) \nonumber \\
&=& \frac{G }{\hbar} \frac{2 (p_1\cdot p_2)^2- m_1^2m_2^2}{ \sqrt{(p_1\cdot p_2)^2- m_1^2m_2^2}}  \ln \left( \frac{j_0}{j}\right).
\eea
This directly confirms that the classicality condition, Eqs. \eqref{classical}, \eqref{classical2}, 
corresponds to large phase shifts $\delta_l \gg 1$,
 which is one of the standard conditions for the validity of the classical limit \cite{LandauQM}.
 
In addition, let us  recall the basic structure of the perturbative expansion of the quantum scattering amplitude $\cM$.
The  LO ($O(G/\hbar)$) contribution to $\cM$ 
 coming from a one-graviton exchange  in the $t$-channel (discarding the $u$- and $s$-channel contributions), reads (see, e.g., Refs. \cite{DeWitt:1967uc,Bjerrum-Bohr:2018xdl})
\be \label{born1}
 \cM^{\left( \frac{G}{\hbar}\right)}(s,t) = 16 \pi \frac{G}{\hbar} \,\frac{2 \, (p_1\cdot p_2)^2 - p_1^2 \, p_2^2 + (p_1\cdot p_2) Q^2}{Q^2}\,,
\ee
 where $Q= p'_1-p_1= - (p'_2-p_2)$, so that $Q^2 = -t$. When considering, for orientation, a generic relativistic collision, 
 with large velocities $v\sim 1$, and significant momentum transfers, $ Q^2 = -t \sim s $, the order of magnitude of the LO
 contribution \eqref{born1} is
 \be
  \cM^{\left( \frac{G}{\hbar}\right)} \sim  \frac{G s}{\hbar}  \sim \alpha_g.
 \ee
 Here, we introduced the gravitational analog of the quantum electrodynamics coupling constant $\alpha= e^2/\hbar$ (or,
 more generally, $Z_1 Z_2 e^2/\hbar$), say
 \be
 \alpha_g \equiv \frac{G E_1 E_2}{\hbar}\,.
 \ee
 Dimensional analysis (in the simple one-scale regime where $ s \sim -t \gtrsim m_1^2 \sim m_2^2$)
 then shows that the Born-Feynman expansion (or loop-expansion)  of $\cM$ has the rough structure
 \bea \label{bornexp}
 \cM  &\sim&   \frac{G s}{\hbar}  +  \left(\frac{G s}{\hbar}\right)^2 +   \left(\frac{G s}{\hbar}\right)^3 +\ldots \nonumber\\
 &\sim& \alpha_g +  \alpha_g^2 +  \alpha_g^3 +\ldots
 \eea
 This exhibits the {\it a priori} necessity of the quantum condition \eqref{quantum} (which implies $\alpha_g \ll 1$) for a reliable
 use of the Born-Feynman expansion of $\cM$. [Let us note in passing that the systematic use of the small-velocity limit $ v\to 0$
 in Refs. \cite{Neill:2013wsa,Cheung:2018wkq,Bern:2019nnu,Bern:2019crd} might exacerbate the classical-quantum conflict 
 by making  the usual, non relativistic Coulomb coupling constant $\frac{ G E_1 E_2 }{\hbar \, v}$ parametrically larger than the natural dimensionless quantum coupling
 constant $ \alpha_g=\frac{ G E_1 E_2 }{\hbar }$ entering the loop expansion of $\cM$.]

 How can one hope to bridge the gap between the classical domain \eqref{classical}, and the quantum one \eqref{quantum} ?
 If we could control the exact dependence of the function $\cM(s,t,\alpha_g)$ for all values of $\alpha_g$ (both small and large),
 it would be straightforward to read off the classical dynamics (say via the use of the quasi-classical phase shifts \eqref{wkbdeltal}).
 However, we often have only knowledge of the first few terms in the Born-Feynmann (small $\alpha_g$) expansion of $\cM(s,t,\alpha_g)$.
 Several suggestions have been made in the recent literature for extracting classical information from $\cM$.
 
 On the one hand, Refs. \cite{Neill:2013wsa,Cachazo:2017jef,Guevara:2017csg,Cheung:2018wkq,Bern:2019nnu,Bern:2019crd}
  have emphasized that a crucial tool for retrieving classical information from $\cM$ is 
 to focus, at each order in the formal Born-Feynman expansion in powers of $ \alpha_g =\frac{ G E_1 E_2 }{\hbar }$
 on a secondary expansion in $ Q^{\rm c.m.} $. As the corresponding small dimensionless
 parameter is $Q^{\rm c.m.}/P_{\rm c.m.}= 2 \sin \frac{\theta}{2}$, this corresponds to a small-scattering-angle expansion.
 The  idea is here related to the fact that the classical PM expansion is a large-impact-parameter limit,
 corresponding to a small-scattering-angle limit. This intuitive idea is certainly appealing, but the point,
 recalled above, made by Bohr \cite{Bohr1948} that sufficiently slowly-spreading wave packets can
 only be constructed when the classicality condition \eqref{classical} (which implies $\alpha_g \gg 1$) is satisfied
 makes it unclear (at least to the author) that focussing on a  secondary expansion in $ Q^{\rm c.m.} $ is 
 sufficient for correctly extracting, at all orders, the classical dynamical information.
 It would be interesting to examine in detail whether this conflicts with the approach pursued in 
 Refs. \cite{Kosower:2018adc,Maybee:2019jus} for extracting classical results from $\cM$. Indeed, it seems that
 the latter approach assumes the existence of  wave packets staying well-localized during the entire scattering process,
 but also uses the Born-Feynman perturbative expansion of $\cM$ in powers of $\frac{ Z_1 Z_2 e^2}{ \hbar}$
 or $ \frac{ G E_1 E_2 }{\hbar}$.

 On the other hand, Refs. \cite{tHooft:1987vrq,Amati:1990xe,Kabat:1992tb,Saotome:2012vy,Akhoury:2013yua,Bjerrum-Bohr:2018xdl,KoemansCollado:2019ggb,DiVecchia:2019myk,DiVecchia:2019kta} have emphasized
 the usefulness of focussing on the so-called eikonal approximation, under which one can hopefully prove that part of the
 perturbative expansion of $\cM$ can be resummed by exponentiating a suitably defined ``eikonal phase''.  
 The idea here is that perturbative theory can correctly compute some of the first few diagrams, and therefore
 their associated exponentiated version.
 However, this program can reliably give the (large) quasi-classical exponentiated phase (as in Eq. \eqref{wkbdeltal})
 only if one {\it proves}  which perturbative diagrams do exponentiate and which do not. This is a non trivial task,
 as shown, for instance, at the one-loop level in Ref. \cite{Akhoury:2013yua}. [The first and second versions
 of Ref. \cite{Akhoury:2013yua} differed in their conclusion of which perturbative contributions do exponentiate.]
 For further discussion of the subtleties of the eikonal approach and of the exponentiating
 contributions, see Refs. \cite{DiVecchia:2019myk,DiVecchia:2019kta,Krachkov:2015uva}.

Let us just mention a specific example  suggesting (without, however, proving) that, even when focussing on the small 
$ Q^{\rm c.m.} $ limit,  it is delicate to try to unambiguously extract from the perturbative expansion of the amplitude
the corresponding classical PM-expanded information. At the one-loop level (second order in $\alpha_g$), there
appears, when considering the $t/s \ll 1$ limit (or $q \to 0$), a non-analytic $\ln q$ term \cite{Donoghue:1993eb,Donoghue:1994dn,BjerrumBohr:2002kt,Bjerrum-Bohr:2013bxa,KoemansCollado:2019ggb}. This term corresponds to a quantum modification
of the LO gravitational potential $-G m_1 m_2 (2 \g^2-1)/R$ (in physical units) by an additional term of the type
($L_P^2 \equiv \hbar G$ denoting the squared Planck length)
\be\label{oneloop}
-\frac{G m_1 m_2(2 \g^2-1) }{R} \left[ 1+  A(\g, \nu) \frac{L_P^2}{R^2} \right]\,,
\ee
which corresponds, in the rescaled EOB units, to a correction of the potential $w(r)=w_1/r + \ldots$ of the type
\be
\delta w(r)= \nu \hh  A(\g, \nu) \frac{w_1}{r^3}\,,
\ee
{\it i.e.} a modification of the 3PM coefficient $w_3$ of the type
\be \label{oneloopw3}
\delta w_3= \nu \hh  A(\g, \nu) {w_1}\,.
\ee
Here the dimensionless coefficient $A(\g, \nu)$ has a
finite limit at low velocities ($\g \to 1$) \cite{Donoghue:1993eb,Donoghue:1994dn,BjerrumBohr:2002kt,Bjerrum-Bohr:2013bxa}, 
but was recently found \cite{KoemansCollado:2019ggb} to grow logarithmically at high energies ($\g \to \infty$). More precisely, 
Ref. \cite{KoemansCollado:2019ggb} (see Eq. (2.25) there) found that the logarithmically
growing part of $A(\g, \nu)$ comes from a factor proportional to the same arcsinh function entering the result of Ref. \cite{Bern:2019nnu},
denoted ${\rm as}(\g)$ above. We note that, in the domain of validity of the perturbative regime $\alpha_g \to 0$, 
{\it i.e.} $\hh \to \infty$,
the one-loop  contribution \eqref{oneloopw3} to $w_3$ is {\it parametrically larger} than the (3PM-level) value $w_3^B$ derived from 
the two-loop amplitude of Ref. \cite{Bern:2019nnu}. This makes it unclear to us that a formal analytic continuation (in $\alpha_g$)
of the perturbative two-loop computation to the classically-relevant domain where $\alpha_g \gg 1$, {\it i.e.} $\hh \ll1$ can
unambiguously read off the needed classical contribution to $w_3$. We hope that our remarks will prompt some clarification
of these subtle issues.

 %%%%%%
 \section{Self-force (SF) theory and PM dynamics}
 
  Before explaining in detail why the result of Ref. \cite{Akcay:2012ea} seems to be in conflict with the logarithmic growth
  \eqref{HEq3B}, derived from Refs. \cite{Bern:2019nnu,Bern:2019crd}, let us point out a potentially
 interesting new  use of SF theory for deriving {\it exact} PM dynamical results.

 \subsection{On the use of self-force (SF) theory to derive {\it exact} PM dynamics}

Let us start by recalling that the discussion in Section \ref{sec2} above allowed one to give a stringent
upper bound on the number of unknown functions of $\g$ entering each PM order.
In particular, we found that, both at the 3PM and the 4PM levels, there 
was {\it only one} a priori unknown function of $\g$. Namely, in the parametrization of Eqs. \eqref{chi3pm} and \eqref{chi4pm},
the function $\widehat \chi_{3}^{(2)}(\g)$ at the 3PM level, and the  function $\widehat \chi_{4}^{(3)}(\g)$ at the 4PM level.
We wish to point out here the rather remarkable fact that  SF theory 
(which, in the framework of EOB theory means expanding the EOB dynamics to linear order in $\nu$),
can, in principle,  be used to derive in an {\it exact} manner the 3PM
and 4PM dynamics. The main point is that the first-order SF (1SF) expansions of the 3PM and 4PM scattering functions
$\chi_{3}(\g, \nu)$ and $\chi_{4}(\g, \nu)$, i.e their expansions in powers of $\nu$, keeping only the term linear in $\nu$,
contain enough information to compute the exact functions  $\chi_{3}(\g, \nu)$ and $\chi_{4}(\g, \nu)$.
Indeed,  using the fact that 
\be
 h(\g, \nu) =\sqrt{1+ 2\nu (\g-1)} = 1 + \nu (\g-1) + O(\nu^2),
 \ee
 and considering first the 3PM level, the 1SF expansion of $\chi_{3}(\g, \nu)$ reads, from  Eq. \eqref{chi3pm},
\be \label{chi3sf}
\chi_{3}(\g, \nu)= \chi_{3}^{\rm Schw}(\g) - 2 \nu (\g-1) \widehat \chi_{3}^{(2)}(\g) + O(\nu^2).
\ee
Therefore the linear-in-$\nu$, or 1SF contribution,  to $\chi_{3}(\g, \nu)$ is proportional to the function $(\g-1) \widehat \chi_{3}^{(2)}(\g)$,
so that an analytical knowledge of  $\chi_{3}^{1 \rm SF}$ yields enough knowledge to compute $\widehat \chi_{3}^{(2)}(\g)$,
and thereby the exact, non-SF-expanded value  Eq. \eqref{chi3pm} of $\chi_{3}(\g, \nu)$.

The same result holds at the 4PM level. Namely, starting from Eq. \eqref{chi4pm}, the 1SF expansion of $\chi_{4}(\g, \nu)$ reads
\be \label{chi4sf}
\chi_{4}(\g, \nu)=(1- \nu (\g-1) )\chi_{4}^{\rm Schw}(\g) - 2 \nu (\g-1) \widehat \chi_{4}^{(3)}(\g) + O(\nu^2).
\ee
Using the exact value of $\chi_{4}^{\rm Schw}(\g)$, Eqs. \eqref{chischw2}, we see that 
an analytical knowledge of  $\chi_{4}^{1 \rm SF}$ yields enough information to compute $\widehat \chi_{4}^{(3)}(\g)$,
and thereby the exact, non-SF-expanded value  Eq. \eqref{chi4pm} of $\chi_{4}(\g, \nu)$.

One does not have today general enough 1SF results allowing one to extract $\widehat \chi_{3}^{(2)}(\g)$, $\widehat \chi_{4}^{(3)}(\g)$,
and their higher-order analogs. Actually, the SF theory of scattering motions is still in its developing stages.
Some years ago Ref. \cite{Damour:2009sm} had  pointed out the interest
of extending the SF approach (which is usually applied only to circular, or near-circular, states) to scattering states,
and showed what information it could give.
Due to technical issues, it is only very recently \cite{Barack:2019agd} that  a
numerical implementation of one of the scattering-type SF computations proposed in Ref. \cite{Damour:2009sm} has been accomplished.
Here, we are suggesting to develop an analytical, PM-expanded SF framework, e.g. based on the $G-$expansion
of the Mano-Suzuki-Takasugi formalism, for computing the $G$-expansion of the scattering angle in
large-mass-ratio binary systems. When a second-order SF formalism becomes available, the same idea will
allow one to compute the exact 5PM and 6PM (conservative) dynamics. Indeed, a look at Eqs. \eqref{chi3456} shows
that, after using the test-mass knowledge  ($\chi_5^{\rm Schw}, \chi_6^{\rm Schw}$), one has two unknown
functions of $\g$ at 5PM and at 6PM, so that it is enough to know  the 1SF ($O(\nu)$) and the 2SF ($O(\nu^2)$)
contributions to the SF expansions of $\chi_{5}(\g, \nu)$ and  $ \chi_{6}(\g, \nu)$ to reconstruct their
exact expressions for any mass ratio.

In Appendix C we discuss the high-energy limit of SF scattering theory, and the information it could bring on 
the structure of the PM expansion.
 
 \subsection{Tension between the 3PM dynamics of Refs. \cite{Bern:2019nnu,Bern:2019crd} 
 and the HE behavior of the SF Hamiltonian of an extreme mass-ratio two-body system}

Let us show in what technical sense the (numerical)  circular-orbit SF computation of  Ref. \cite{Akcay:2012ea} provides 
a direct handle on the high-energy (HE) limit of the 
1SF-expanded\footnote{We recall that ``1SF'' means ``first order in the symmetric mass-ratio $\nu$''.} two-body dynamics. 
To be concrete, and explicitly display how the 3PM-level result of \cite{Bern:2019nnu,Bern:2019crd} seems to conflict, in the HE limit, with the  
1SF HE result of \cite{Akcay:2012ea}, let us  consider the 1SF expansion of the 3PM-accurate EOB Hamiltonian derived
in \cite{Antonelli:2019ytb} from the results of \cite{Bern:2019nnu,Bern:2019crd}. We recall that the two-body
Hamiltonian is expressed by the general formula \eqref{Heob} in terms of the effective Hamiltonian  
$\Ef=H_{\rm eff}({\mathbf R}, {\mathbf P})$. In turn, the effective Hamiltonian is obtained by solving
the EOB mass-shell condition \eqref{massshellgen2} for $\Ef$. In the $H$-type energy gauge this yields
a squared effective Hamiltonian of the form (in rescaled variables)
\be
{\widehat H}_{\rm eff}^2({\mathbf r}, {\mathbf p})=  {\widehat H}_S^2+  (1-2u) {\widehat Q}^H(u, {\widehat H}_S)\,,
\ee
where
\be
{\widehat H}_S^2({\mathbf r}, {\mathbf p})=  (1-2u) \left(1+ (1-2u) p_r^2 + u^2 p_\varphi^2  \right)\,,
\ee
and
\be \label{QHresc}
{\widehat Q}^H(u, \g,\nu)= u^2 q_2(\g,\nu)+  u^3 q_3(\g,\nu) + O(G^4) \,.
\ee
The 2PM coefficient $q_2(\g,\nu)$ is given by \cite{Damour2018}
\be
q_2(\g, \nu) =  \frac{3}{2} \left( 5 \, \g^2-1 \right) \left[ 1 -  \frac{1}{h(\g,\nu)}  \right]\, ,
\ee
while the 3PM coefficient derived in \cite{Antonelli:2019ytb} by combining the results of \cite{Bern:2019nnu,Bern:2019crd}
and \cite{Damour2018} reads
\bea \label{q3vsBC}
q_{3}^B(\g, \nu)&=& B(\g)\left(\frac1{h(\g,\nu)}-1 \right)+ C^B(\g) \left(\frac1{h^2(\g,\nu)}-1 \right) \nonumber\\
&=&  B(\g)\left(\frac1{h(\g,\nu)}-1 \right) + 2 \nu \frac{{\overline C}^B(\g)}{h^2(\g,\nu)}\,,
\eea
where
\be
 B(\g) \equiv\frac32 \frac{(2 \g^2-1)(5 \g^2-1)}{\g^2-1}\,,
\ee
and where
\be
C^B(\g)= - \frac{  {\overline C}^B(\g) }{  \g -1 }\,,
\ee
with the explicit value of $ {\overline C}^B(\g) $  witten in Eq. \eqref{bCB} above.

A crucial point is that the HE limit $\g \to \infty$ and the SF limit $\nu \to 0$ do not commute
because of the denominators involving powers of $h(\g,\nu)= \sqrt{1 + 2 \nu (\g-1)}$.
When discussing SF results we are interested in performing first a linear expansion in $\nu$,
and in then taking the HE limit of this linear expansion.
Let us denote, for simplicity, by $F^{1 \rm SF} $ the coefficient of $\nu$ in the linear-in-$\nu$, or 1SF, expansion
of any EOB function, $F$, considered
as a function of the EOB phase-space variables ${\mathbf r}, {\mathbf p}$, and of $\nu$: 
$F({\mathbf r}, {\mathbf p},\nu) = F({\mathbf r}, {\mathbf p},0)+ \nu F^{1 \rm SF}({\mathbf r}, {\mathbf p}) + O(\nu^2)$.

Applied to $q_2(\g, \nu)$ this yields first 
\be
q_2^{1 \rm SF} =  \frac{3}{2}  (\g -1) \left( 5 \, \g^2-1 \right) \,,
\ee
which becomes in the HE limit $\g \to \infty$
\be
q_2^{1 \rm SF} \overset{\rm HE}{=}  \frac{15}{2}  \g^3 \,.
\ee
Applying the same (non commuting) successive limits to $q_3^B(\g, \nu)$ yields
\be
q_3^{B\,1 \rm SF} \overset{\rm HE}{=}  \frac{11}{3}  \g^3 + 16  \g^3 \ln (2 \g) \,.
\ee
 Let us consider
\be
{\widehat Q}^{1 \rm SF} = \frac{ \left[{\widehat H}_{\rm eff}^2\right]^{1 \rm SF} }{1-2u}\,.
\ee
We have
\be
 {\widehat Q}^{1 \rm SF}_B = u^2 q_2^{1 \rm SF}+ u^3 q_3^{B\,1 \rm SF} + O(u^4)\,.
\ee
Its HE limit reads
\be \label{Q1SFBHE}
{\widehat Q}^{1 \rm SF}_B \overset{\rm HE}{=} \frac{15}{2}  \g^3  u^2 +  \frac{11}{3}  \g^3 u^3 + 16  \g^3 \ln (2 \g) u^3 + O(u^4).
\ee
The crucial point to note here is that the $ \ln (2 \g)$ contribution coming from the arcsinh term implies that the ratio
${\widehat Q}^{1 \rm SF}_B /\g^3$ does not have a finite HE limit, when considered at the 3PM level, namely
\be \label{Q1SFBbyg3}
\frac{  {\widehat Q}^{1 \rm SF}_B }{\g^3} \overset{\rm HE}{=}   \frac{15}{2}   u^2 +  \frac{11}{3}  u^3 + 16   \ln (2 \g) u^3 + O(u^4)\,.
\ee
 In other words, if we truncate the PM expansion at the 3PM level included, and use
 \be
 {\widehat Q}^{\leq 3 \rm PM}_B= u^2 q_2(\g,\nu) + q_{3}^B(\g, \nu) u^3\,,
 \ee
 to define some exact dynamics, the latter dynamics implies a logarithmic growth of the ratio  ${\widehat Q}^{1 \rm SF} /\g^3$
 in the HE limit.

Such a logarithmic growth is in conflict with a result of  Akcay {\it et\,al.} \cite{Akcay:2012ea}.
Indeed,  Ref. \cite{Akcay:2012ea} has (numerically) computed a 1SF-accurate gauge-invariant function which can be directly related to 
$ {\widehat Q}^{1 \rm SF}$. More precisely, Ref. \cite{Akcay:2012ea} considered the sequence of circular orbits of a small
black hole (of mass $m_1$) around a large black hole (of mass $m_2$) and computed a function $a_E^{1 \rm SF}(u) $
which (using results
from Refs. \cite{LeTiec:2011ab,LeTiec:2011dp,Barausse:2011dq}) can be related to $ {\widehat Q}^{1 \rm SF}$ in the
following (gauge-invariant) way (see \cite{Akcay:2012ea} for details)
\be
\frac{a_E^{1 \rm SF}(u) }{(1-2u)^2}
 =\left[\frac{  {\widehat Q}^{1 \rm SF} }{{\widehat H}_S^3}\right]^{\rm circ}  \,.
\ee
The superscript circ on the right-hand side means that the arguments of the EOB function $ {\widehat Q}^{1 \rm SF} /{\widehat H}_S^3$
must be evaluated along the sequence of circular orbits around a Schwarzschild black hole of mass $M$, {\it i.e.} that we have
the relation 
\be
\g^{\rm circ}={\widehat H}_S^{\rm circ}= \frac{1-2u}{\sqrt{1-3u} } \,.
\ee
Rigorously speaking, only the  part of the sequence of circular orbits describing the
unstable orbits below $R=4GM$, {\it i.e.} $ \frac14 < u < \frac13$,  leads to a value of $\g^{\rm circ}>1$
that can be directly inserted in the formulas above. However, one can formally consider the analytic continuation
of the formulas above for smaller values of $u$. In particular, we could satisfactorily check that, in the PN limit $u \to 0$,
 $\left[  {\widehat Q}^{1 \rm SF} /{\widehat H}_S^3\right]^{\rm circ}   = 2 u^3 + O(u^4)$, which agrees with the LO
 PN term in $ a_E^{1 \rm SF}(u) /(1-2u)^2$.

The tension with the result above then comes when focussing on the limit $u \to \left(\frac13\right)^-$. This
limit, which physically corresponds to considering HE circular orbits near the light ring of the large-mass black hole, realizes
the above-considered HE limit $\g\to \infty$. The crucial point is that Ref. \cite{Akcay:2012ea} could numerically
study with high accuracy the behavior of the $a_E^{1 \rm SF}(u)$ in this limit, and found that it admitted a {\it finite}
limit yielding
\be \label{akcay}
\lim_{\g\to \infty}\left[\frac{  {\widehat Q}^{1 \rm SF} }{{\widehat H}_S^3}\right]^{\rm circ}= \frac{27}{4} \zeta\,,
\ee
where $\zeta$ is a finite number equal to 1 to good accuracy.
In particular, the study of the behavior of $a_E^{1 \rm SF}(u)$ in the close vicinity of $u=\frac13$ definitely excluded
the presence of a LO logarithmic singularity $\propto \ln (1-3u)$, {\it i.e.} $\propto \ln \g$. On the other hand, the
numerical results of \cite{Akcay:2012ea} were compatible with the additional presence of a {\it subleading} logarithmic singularity, {\it i.e.}
a behavior of $  {\widehat Q}^{1 \rm SF} /\g^3 - 27 \zeta/4$  of the form  $\propto (1-3u) \ln (1-3u)$, {\it i.e.} $\propto \g^{-2} \ln \g$.

How can we reconcile the (apparently) conflicting HE behaviors \eqref{Q1SFBbyg3} and \eqref{akcay} ?
Barring some hidden numerical flaw in the work of \cite{Akcay:2012ea}, several possibilities come to mind.
We wish here to propose two different possibilities for relieving the tension between \eqref{Q1SFBbyg3} and \eqref{akcay}.

The first possibility was suggested to the author by a statement made in the second sentence below Eq. (9.5) of \cite{Bern:2019crd} to the
effect that their general ansatz for their $O(G^3)$, 3PM amplitude ${\cal M}_3$ was uniquely fixed only by the knowledge 
of the PN expansion of ${\cal M}_3$ at the 6PN-level included. However, as we recalled above, at the time of writing of (the preprint version of) this
paper (November 2019), there existed no classical computation having confirmed the 3PM dynamics of \cite{Bern:2019crd} at the 6PN level. 
The highest PN level which had been independently checked was the 5PN level, as  obtained in Ref. \cite{Bini:2019nra}. 
This then suggested exploring the conjecture that some error might have crept at the 6PN level
in the computations of Ref. \cite{Bern:2019crd}
(which rely in great part on working with  the PN-expansion of the  two-loop integrand), and in looking for a modified version
of the 3PM dynamics exhibiting a softer, logarithmic-free HE behavior. This possibility is briefly discussed in the following section.

A second possibility relies on the fact that there might exist correlations between the various PM contributions to 
${\widehat Q}(u, \g,\nu)$, 
\be \label{QEresc2}
{\widehat Q}^E(u, \g,\nu)= u^2 q_2(\g,\nu)+  u^3 q_3(\g,\nu) +  u^4 q_4(\g,\nu) + \ldots \,.
\ee
leading to a cancellation\footnote{This possibility was briefly alluded to in the preprint version of this work,
but not pursued there because of its apparently fine-tuned nature.} of the problematic logarithmic term in Eq. \eqref{Q1SFBbyg3}.
This second possibility relies on making another type of conjecture on the structure of 
the EOB potential ${\widehat Q}(u, \g,\nu)$. It is also explored in the following section.

%%%%%%%%%
\section{Different conjectures on the HE behavior of PM gravity and their consequences} \label{HEPM}

\subsection{Conjecture on the HE behavior of PM gravity} \label{simpleHEconj}

A striking feature of 2PM-level gravity, which is especially clear in its  EOB formulation \cite{Damour2018},
is that it has a remarkably simple HE limit. Specifically, the (energy-gauge) EOB mass-shell condition \eqref{massshellgen}
(which, for general energies and momenta, is a complicated, nonlinear function of energies and momenta) drastically simplifies
in the HE limit and becomes quadratic in $P_\mu$. Moreover, in this limit the dependence on the mass ratio $\nu$ completely disappears.
Indeed, when $\g \to \infty$, the $O(G^2)$ $Q$ term ${\widehat Q}^{2 \rm PM}(u, \g,\nu)= u^2 q_2(\g,\nu)$, where
\be
q_2(\g, \nu) =  \frac{3}{2} \left( 5 \, \g^2-1 \right) \left[ 1 -  \frac{1}{h(\g,\nu)}  \right]\, ,
\ee
 reduces to
\be
{\widehat Q}^{2 \rm PM}(u, \g,\nu) \overset{\rm HE}{=} \frac{15}{2}  u^2 \,  \g^2 \,,
\ee
where we recall that $\g \equiv -P_{0}/\mu \equiv \Ef/\mu$,
so that the  mass-shell condition \eqref{massshellgen} simplifies to the following quadratic constraint
\be 
0= g_{\rm Schw}^{\mu \nu} P_{\mu} P_{\nu} + \frac{15}{2} u^2 \, P_{0}^2\,,
\ee
or, explicitly,
\be \label{massshell2PMHE}
0=- \frac{\Ef^2}{A_{\rm HE \, 2PM}} + \frac{P_R^2}{B_{\rm HE\, 2PM}}+ \frac{P_{\varphi}^2}{C_{\rm HE\, 2PM}} \,,
\ee
where $\frac1{A_{\rm HE \, 2PM}}= \frac1{A_{\rm Schw}} - \frac{15}{2} u^2$, {\it i.e.}, inserting the
values of the Schwarzschild-metric coefficients (with $u\equiv GM/R$),
\bea
A_{\rm HE\, 2PM}(u)&=& \frac{A_{\rm Schw}(u)}{1- \frac{15}{2} u^2 A_{\rm Schw}(u)}= \frac{1-2u}{1- \frac{15}{2} u^2 (1-2u)} \,,\nonumber\\
B_{\rm HE\, 2PM}(u)&=& B_{\rm Schw}(u)= \frac1{1-2u}\,,\nonumber\\
C_{\rm HE\, 2PM}(u)&=& C_{\rm Schw}= R^2 \,.
\eea
In other words, at the 2PM-level, and in the HE limit $\g \to \infty$ (which is equivalent to taking the massless limit
$m_1 \to 0$, $m_2 \to 0$), the c.m. scattering angle of a two-body system becomes
blind to the values of the masses and can be obtained from the null geodesic motion in the effective HE metric
\bea \label{g2PMHE}
 g^{\rm HE\, 2PM}_{\,\mu \nu} dx^{\mu} dx^{\nu} &=& - A_{\rm HE\, 2PM}(R) dT^2 + B_{\rm HE\, 2PM}(R) dR^2 \nonumber\\
 &+& C_{\rm HE\, 2PM}(R) (d \theta^2 + \sin^2 \theta d \varphi^2)\,.
\eea
Such a simple conclusion (equivalent\footnote{Modulo a different parametrization leading there to $A_{\rm HE\, 2PM}(u)= (1-2u)(1+ f(u))$.} to the discussion in section VII of \cite{Damour2018}), seems to be physically quite satisfactory.
 Indeed, as (classical and quantum) gravity couples to energy, rather than to rest-mass, one would a priori
 expect that a limit  where the two masses $m_1, m_2$
tend towards zero, while keeping fixed the energies $E_1= \sqrt{m_1^2+ p_1^2}$, $E_2= \sqrt{m_2^2+ p_2^2}$, 
should exist, and be describable by the interaction of two (classical or quantum) massless particles.
We sketch in appendix B how a classical PM scattering computation might prove that such a limit exists.

 Ref. \cite{Damour2018} assumed that such a HE limit exists not only at the 2PM level, but also at any higher PM
 order. Let us recall at this point that, contrary to the PN expansion which can (and does) involve logarithms of $\frac1c$, 
 there seems to be no way in which the
PM expansion (when considered at finite $\g$) can involve logarithms of the gravitational coupling constant $G$. 
Indeed, as was indicated in Section \ref{sec2},
the PM expansion of the classical scattering angle (equivalent to the knowledge of $ \widehat Q$) must, at each PM order,
 be a polynomial in the masses, and therefore in $Gm_1/b$ and $Gm_2/b$. Therefore, when considering the generic case of finite masses and arbitrary (but finite) values of $\g$, we must have an expansion in powers of $u= GM/R$ of the type 
 \be \label{PMexpQ0}
 \widehat Q= \sum_{n\geq2} q_n(p_\lambda,\nu) u^n\,.
\ee
 Ref. \cite{Damour2018} then assumed that the HE limit of each PM coefficient $ q_n(p_\lambda,\nu)$ 
 would become  a 
$\nu$-independent quadratic form in $p_\lambda$. [This is equivalent to saying that the corresponding coefficient in the
unrescaled $Q$ becomes a $\nu$-independent quadratic form in $P_\lambda$.]

The precise expression for the limiting behavior of $q_n(p_\lambda,\nu)$
depends on the gauge chosen to write it. In the first form \eqref{QE} of the energy gauge (where $q_n(p_\lambda,\nu)$ is only a function of 
$p_0= - \Ef/\mu = \e$), one would have
 \be \label{qnHE1}
\lim_{\e \to \infty} q^E_n(\e , \nu) \approx c^{(qE)}_n \e^2 \,.
\ee
This is what we used in our 2PM-level discussion above.
In the second (Hamiltonian) form \eqref{QH} of the energy gauge one would have
\be \label{qnHE2}
\lim_{\hHfS \to \infty} q_n(\hHfS , \nu) \approx c^{(qH)}_n \hHfS^2 \,.
\ee
One can easily see that the two conditions are equivalent to each other, with some transformation between
the coefficients corresponding to rewriting the higher-PM version of $ A_{\rm HE\, 2PM}(u)$
either as
\be
 A_{\rm HE}(u)= (1-2u)\left(1-  c^{(qE)}_2 u^2 -\ldots -c^{(qE)}_n u^n  -\ldots \right)^{-1},
 \ee
 or as
 \be
 A_{\rm HE}(u)= (1-2u)\left(1+ c^{(qH)}_2 u^2 +\ldots +c^{(qH)}_n u^n  +\ldots \right).
 \ee

\subsection{Uniqueness of a  conjectured 3PM dynamics compatible with the simple HE behavior \eqref{qnHE1}}

The 3PM-level EOB $Q$ potential derived from the 3PM result of \cite{Bern:2019nnu,Bern:2019crd} is given by Eq. \eqref{q3vsBC}.
We discussed above why its 1SF expansion is in tension with the SF result \eqref{akcay}. In addition, its HE limit (without doing any SF expansion)
does not respect the expected HE behavior \eqref{qnHE1}. Indeed, we have
\be
\frac{\widehat Q^{\leq \rm 3 PM \, B}(u,\g,\nu)}{\g^2} \overset{\rm HE}{=} \frac{15}{2} u^2  + \left(8  \ln (2 \g) - \frac{17}{3}\right) u^3\,,
\ee
where the logarithmically growing term $  8 u^3  \ln (2 \g)$ comes from the term  $16 \g^4  {\rm as}({\g})/\sqrt{\g^2-1}$
in the function ${\overline C}^B(\g)$, Eq. \eqref{bCB}.

In  the present subsection we propose a conjectured modification of the function  ${\overline C}^B(\g)$ that has the property
of being compatible at once with four different constraints: (i) the same restricted analytic structure as ${\overline C}^B(\g)$;
(ii) the SF result \eqref{akcay}; (iii) the independently confirmed 5PN-level expansion of the 3PM dynamics;
and (iv) the simple HE behavior \eqref{qnHE1}.
Moreover, these  properties uniquely determine our conjectured modified function  ${\overline C}(\g)$.

The general ansatz\footnote{If we knew sufficiently many terms in the PN expansion of ${\overline C}(\g)$
the method of Ref. \cite{Blumlein:2019bqq} would allow us to derive its exact form without assuming such a restricted form.}
 made (and motivated by several arguments) in Ref. \cite{Bern:2019crd} is (when transcribed in terms of 
${\overline C}(\g)$) that
\be\label{bCc0}
{\overline C}^c(\g)=c_1\g + c_3 \g^3 + (d_0+d_2 \g^2+d_4 \g^4) \frac{\arcsinh \sqrt{ \frac{\g-1}{2}}}{\sqrt{\g^2-1}}\,,
\ee
with some numerical coefficients $c_1, c_3, d_0,d_2,d_4$. This general structure corresponds to the structure of the
coefficients $\tau_1$ and $\tau_3$ in Eq. (9.5) of \cite{Bern:2019crd}, as determined by the requirement
indicated  just below Eq. (9.5) there that (after completing it by the overall factor $ m_1^3 m_2^3$) ${\overline C}(\g)$ must
be a polynomial in\footnote{I thank Mikhail Solon for clarifying  the precise meaning of the statement written
below Eq. (9.5)  of \cite{Bern:2019crd}.}
$m_1^2$, $m_2^2$ and $(p_1\cdot p_2)$. 

When redoing the computation of the HE limit of the quantity $\frac{  {\widehat Q}^{1 \rm SF} }{\g^3}$
considered in Eq. \eqref{Q1SFBbyg3} above for the general ansatz \eqref{bCc0}, one finds that 
this ratio now takes the general form
 \be
 \frac{  {\widehat Q}^{1 \rm SF} }{\g^3} \overset{\rm HE}{=}   \frac{15}{2}   u^2 +  (c_3-15)  u^3 + \frac{d_4}{2}   \ln (2 \g) u^3 + O(u^4)\, .
 \ee 
Barring the possibility (separately explored below) that the  $O(u^4)$ 4PM remainder term in this result cancells the $O(\ln (2 \g) u^3)$ term,
the compatibility with the SF result \eqref{akcay}, together with the general requirement \eqref{qnHE1}, then determines
that the coefficient $d_4$ in Eq. \eqref{bCc0} should vanish:
\be \label{d4eq0}
d_4=0\,.
\ee
We note  in passing that the term proportional to $d_2$ in Eq. \eqref{bCc0}
will generate a subleading logarithmic term in the SF quantity computed in \cite{Akcay:2012ea} that is
compatible with the best fits obtained there.

This leaves only {\it four} unknown parameters in the so-restricted ansatz \eqref{bCc0}, namely 
$c_1, c_3, d_0, d_2$. 
If we now use the independently derived  (by using purely classical methods)
5PN-level value of ${\overline C}(\g)$ \cite{Bini:2019nra},
as written in Eq. \eqref{C5PN} above, we have in hands four equations for the four unknowns $c_1, c_3, d_0, d_2$.
By solving these four equations, we have found that they uniquely determine $c_1, c_3, d_0, d_2$, thereby
uniquely determining a  3PM dynamics with softer HE behavior from the sole use of 5PN-level information\footnote{This result is
compatible with the statement made in the second sentence below Eq. (9.5) of \cite{Bern:2019crd} to the
effect that their more general ansatz (involving an extra term $d_4 \g^4$ in the coefficient of the arcsinh) is uniquely
fixed by the 6PN-level $O(G^3)$ amplitude.}. The resulting unique value
of ${\overline C}(\g)$ is found to be
\be\label{bCTDbis}
{\overline C}^c(\g)=  \g (35+ 26 \g^2) -( 18 + 96\g^2) \frac{{\rm as}(\g)}{\sqrt{\g^2-1}}\,.
\ee
Let us briefly contrast the predictions following from the conjectured 3PM dynamics defined by Eq. \eqref{bCTDbis}
to those  following from the result \eqref{bCB} of Refs.~\cite{Bern:2019nnu,Bern:2019crd}. First, the corresponding 
3PM-level contribution to the scattering angle, namely
\be\label{chi3vsABTD}
\chi^c_3(\g,\nu)= \chi_3^{\rm Schw}(\g)    - p_{\infty} \frac{{\overline C}^c(\g)}{\g-1} \left( 1-\frac1{h^2(\g,\nu)} \right)\,,
\ee
has a HE limit (equivalent to the massless limit at fixed momenta) equal to\footnote{The negative coefficient $ - \frac{14}{3}$
comes from combining the positive Schwarzschild contribution $+ \frac{64}{3}$  with the contribution
$- c_3 = - 26$ from the first term in ${\overline C}^c(\g)$.}
\be
\chi_3(\g,\nu) \overset{\rm HE}{=} - \frac{14}{3} \g^3\,.
\ee
Using the notation (following Ref. \cite{Damour2018})
\be \label{defalpha}
\alpha \equiv \frac{\g}{j} \equiv \frac{G M \Ef}{J}=\frac{G}{2} \frac{s - m_1^2  -m_2^2 }{J}\,,
\ee
and adding the HE limits of the 1PM and 2PM scattering angles, we get as conjectured 3PM-accurate prediction for
the HE-limit of the  scattering angle the following {\it finite} result
\be \label{chicHE}
\frac12 \chi^c \overset{\rm HE}{=}  2 \alpha -   \frac{14}{3} \alpha^3\,.
\ee
By contrast, if one formally computes the 
HE limit of the scattering angle derived from the 3PM dynamics of  Refs. \cite{Bern:2019nnu,Bern:2019crd}
one gets a logarithmically divergent 3PM-level contribution, namely (at the leading-logarithm accuracy),
\be \label{chiBHE}
\frac12 \chi^B \overset{\rm HE}{=}  2 \alpha -   8 \ln (2 \g)  \alpha^3\,.
\ee
We note in passing that the sign of the logarithmically divergent coefficient $ -   8 \ln (2 \g)$ of $\alpha^3$ is {\it negative}.
This agrees with the sign of the corresponding (finite) term in Eq. \eqref{chicHE}. By contrast, 
the eikonal-approximation two-loop result of Amati, Ciafaloni and Veneziano \cite{Amati:1990xe}
(which has been recently checked to hold also in several supergravity theories \cite{DiVecchia:2019kta,Bern2020}, and 
confirmed  in the pure gravity case \cite{Bern2020}) gives the result (after using a correction suggested by 
Ciafaloni and Colferai \cite{Ciafaloni:2014esa} and confirmed in \cite{Bern2020})
\be \label{chiACV}
 \frac12 \chi^{\rm eikonal} \overset{\rm HE}{=}  2 \alpha + \frac{16}{3} \alpha^3\,,
\ee
where the sign of $\alpha^3$ is {\it positive}. Independently of the consideration of the HE-softer conjecture Eq. \eqref{bCTDbis},
we note that the HE limit of the result of Refs. \cite{Bern:2019nnu,Bern:2019crd} disagrees with the HE eikonal result
of Refs. \cite{Amati:1990xe,Ciafaloni:2014esa,DiVecchia:2019kta,Bern2020}.

As a second type of predictions from Eq. \eqref{bCTDbis}, let us note that it leads to a specific 3PM-accurate EOB $Q$
potential of the form
\be\label{QEu3PM}
{\widehat Q}^{3 \rm PM \, c}(u, \g,\nu)= u^2 q_2(\g,\nu)+  u^3 q^c_3(\g,\nu)\,,
\ee
where $q^c_3(\g,\nu)$ is obtained by replacing in Eq. \eqref{q3vsABC} the function $C^c(\g)$ given in Eq. \eqref{bCTDbis}
(using also  Eqs. \eqref{ABCeq0} and \eqref{B}). Let us now consider the 1SF-accurate value of ${\widehat Q}^{c\,3 \rm PM}(u, \g,\nu)$, {\it i.e.}, 
the coefficient of $\nu$ in the $\nu$-expansion of the full function ${\widehat Q}^{c\,3 \rm PM}(u, \g,\nu)$.
A straightforward calculation  yields for the HE behavior of ${\widehat Q}^{c\,3 \rm PM \, 1SF}(u, \g)$, {\it i.e.}, its
asymptotic behavior as $\g \to \infty$, the value
\be \label{Q1SFTDHE}
\frac{  {\widehat Q}^{c\, 3 \rm PM \,1SF} }{\g^3} \overset{\rm HE}{=}   \frac{15}{2}   u^2 +  37  u^3 + O(u^4)\,.
\ee
Contrary to the corresponding result following from \cite{Bern:2019crd} that led to the logarithmically divergent
result Eq. \eqref{Q1SFBHE}, we now get a finite limit when inserting the value $u=\frac13$ corresponding to the lightring, namely
\bea
\lim_{\g\to \infty}\left[\frac{  {\widehat Q}^{c\, 3 \rm PM \,1SF} }{\g^3}\right]^{\rm lightring}&=& \frac{5}{6} + \frac{37}{27}\nonumber\\
&=&\frac{119}{54} \approx 2.2037
\eea
The corresponding numerical result of \cite{Akcay:2012ea}, Eq. \eqref{akcay}, was $\approx \frac{27}{4} = 6.75$.
We should not expect a close numerical agreement because we have used in our analytical estimate only the
first two terms (2PM and 3PM) in the (visibly badly convergent) infinite PM expansion of this ratio. However, 
the 3PM conjectural expression Eq. \eqref{bCTDbis} is (contrary to the 3PM result of Bern {\it et al.})
qualitatively compatible in sign and in order of magnitude (and in its finiteness!) with the numerical SF result of \cite{Akcay:2012ea}.

On the other hand, the  conjectured, HE-softer, 3PM dynamics starts differing from the result of Refs.~\cite{Bern:2019nnu,Bern:2019crd} at the 6PN level.
Indeed, the 6PN-accurate expansion of \eqref{bCB} reads
\bea\label{CB6PN}
{\overline C}^{B \, 6 \rm PN}(\peob)&=& 4 + 18 \pinf^2 + \frac{91}{10} \pinf^4  - \frac{69}{140} \pinf^6  \nonumber\\
&-& \frac{1447}{10080} \pinf^8 + O(\pinf^{10})\,,
\eea
while that of \eqref{bCTDbis} reads
\bea \label{CTD6PN}
{\overline C}^{c \, 6 \rm PN}(\peob)&=& 4 + 18 \pinf^2 + \frac{91}{10} \pinf^4  - \frac{69}{140} \pinf^6  \nonumber\\
&-& \frac{233}{672} \pinf^8 + O(\pinf^{10})\,.
\eea
Several independent groups have very recently performed  6PN-accurate $O(G^3)$ computations 
\cite{Blumlein:2020znm,Cheung:2020gyp,Bini:2020wpo}. All those
calculations agree among themselves and have directly
confirmed the 6PN-accurate expansion  \eqref{CB6PN}, thereby disproving the  (HE-softer) conjectured  3PM
dynamics \eqref{bCTDbis}, leading to Eq. \eqref{CTD6PN}.

We must therefore discard the  possibility, explored above, of relieving the tension between the high-energy behavior 
\eqref{HEq3B} derived from Eq. \eqref{bCB} and the high-energy behavior found in Ref. \cite{Akcay:2012ea} by
softening the HE behavior of the 3PM dynamics in the simple-minded way\footnote{We shall not explore here the
more far-fetched possibility that the 3PM dynamics involve
nonperturbative factors, say $\propto 1-\exp\left({- \frac1{\pinf}}\right)$, that would not be detectable at any finite PN approximation,
but that would soften the HE behavior of the arcsinh term.} \eqref{d4eq0}.
Let us, however, emphasize again that our search for some type of resolution of the tension between 
the result of Refs.~\cite{Bern:2019nnu,Bern:2019crd} and the HE result of Ref. \cite{Akcay:2012ea} should continue.
In addition, we have emphasized the presence of another tension between the HE limit of the result of 
Refs.~\cite{Bern:2019nnu,Bern:2019crd} and the (now confirmed) HE eikonal result of  ACV.
Before continuing our effort towards understanding how to reconcile these contrasting results, let us put forward
what we consider to be minimal requirements concerning the HE behavior of PM gravity.

%%%%%%%%%
\subsection{Minimal requirement on the HE behavior of PM gravity}

We recalled above the arguments suggesting that the HE (or massless) limit of the EOB mass-shell constraint \eqref{massshellgen}
should yield a (mass-independent) massless quadratic constraint of the type
\be 
0= g_{\rm HE}^{\mu \nu} P_{\mu} P_{\nu} \,.
\ee
This constraint is equivalent to requiring that the HE limit of the exact unrescaled $Q(u,P_\mu)$ term be quadratic in 
the unrescaled effective momentum $P_\mu$, or that the exact rescaled $\widehat Q(u,p_\mu) \equiv Q/\mu^2$ term be quadratic in 
the rescaled effective momentum  $p_\lambda \equiv P_\lambda/\mu$:
\be \label{QHEp}
\widehat Q(u,p_\mu) \overset{\rm HE}{=} q^{\mu \nu}_{\rm HE}(u) p_\mu p_\nu\,,
\ee
with a mass-independent tensor $q^{\mu \nu}(u)$. In the energy gauge, this requirement reads
\be \label{QHEg}
\widehat Q(u,\g,\nu) \overset{\rm HE}{=} q_{\rm HE}(u) \g^2\,.
\ee
Above, we implicitly assumed that the limiting HE behavior of Eqs. \eqref{QHEp}, \eqref{QHEg} {\it separately} applies
at each PM order. In other words, we assumed that the two limits $G \to 0$ and $\g \to \infty$ commuted.
However, another possibility is that these two limits do not commute,
and that though individual PM contributions  $ q_n(\g,\nu) u^n$ in
Eq. \eqref{PMexpQ0} do not separately exhibit
the expected HE quadratic behavior, the sum of all the PM contributions does lead to a nice quadratic mass-shell condition 
\eqref{QHEp},\eqref{QHEg} in the HE limit. A structure allowing  such a mechanism is presented in the next subsections.

%%%%%
\subsection{Transmutation of post-Minkowskian order in the radiative corrections to the dynamics} \label{gruzinov}

We shall present below a mechanism  able to reconcile
the 3PM dynamics derived in \cite{Bern:2019nnu,Bern:2019crd}, with the 1SF, HE behavior found in \cite{Akcay:2012ea}.
The basic idea of this mechanism is a particular type of non commutativity of the two limits $G \to 0$ and $\g \to \infty$
by which the HE (or massless) limit of  the $O(G^{\geq 4})$  dynamics trickles down to the $O(G^3 \ln G)$ level. Before presenting a
specific conjecture exhibiting such an effect and thereby reconciling Refs. \cite{Bern:2019nnu,Bern:2019crd} and \cite{Akcay:2012ea},
 let us show that such effects are indeed present in PM gravity, when considering
the conservative part of classical radiative corrections.

We recall that it was pointed out long ago \cite{Blanchet:1987wq} that classical radiative effects start having a non purely dissipative
dynamical effect at the 4PN level, via the so-called hereditary tail. At the 4PN level, a part of the near-zone gravitational field
becomes a nonlocal functional of the two worldlines that cannot be simply obtained by a usual, small-retardation PN expansion.
The conservative part of the corresponding non-local-in-time dynamics can be described by a nonlocal action, either of the
Schwinger-Keldysh type  \cite{Foffa:2011np}, or of the Fokker type \cite{Damour:2014jta}. The latter conservative radiative correction
is the source of the first logarithm entering the PN
expansion of the two-body dynamics. This logarithm\footnote{As shown in Ref. \cite{Bini:2017wfr} the 4PN, 
$O(G^4)$, logarithmic contribution to the scattering angle involves the
logarithm of a dimensionless velocity $\ln \pinf$, but {\it does not involve} the logarithm of $G$
({\it e.g.} through the form $\ln j$, with $j=J/(G m_1 m_2)$).} arises at the $O(G^4)$ (and 4PN) level \cite{Damour:2009sm,Blanchet:2010zd}.
 This might suggest that delicate physical effects linked to time-nonlocality
start occurring at the $O(G^4)$ level, and have no effect on the $O(G^3)$ level. 
This is likely to be true when considering non-zero masses and a
finite value of $\g$. However, the following argument shows that this is not true when considering the HE limit where $\g \to \infty$
(with the masses $m_i \to 0$, keeping fixed the c.m. energy $E$).

Let us start from a simple formula obtained\footnote{It was written down there at the leading PN order, but, in view of 
Refs. \cite{Blanchet:1987wq,Foffa:2011np,Damour:2015isa,Galley:2015kus,Foffa:2019eeb}
it clearly has a general validity.} in Ref. \cite{Bini:2017wfr} for the value of
the scattering angle associated with the conservative effect of the radiative correction, namely
\be \label{chirads}
\chi^{\rm rad}_s(E,J)=\frac{\partial}{\partial J} W^{\rm rad}_s(E,J)\,,
\ee
where
\be
W^{\rm rad}_s= 2 \,G \,H\int d\omega \frac{dE^{\rm gw}}{ d\omega} \ln\left(2 e^{\gamma_E}|\omega| s \right)\,.
\ee
Here, $E=H$ is the total c.m. energy of the binary system, $\frac{dE^{\rm gw}}{ d\omega}$ is the spectrum of the energy that
would be emitted in gravitational waves if one would use a retarded Green's function (rather than a time-symmetric one), 
and $s$ is a length scale to be chosen (after differentiation) of order of the size of the system. [The results of Ref. \cite{Bini:2017wfr}
show, for scattering motions, that taking $s=b$ allows one to capture all the relevant nonlocal effects.]

The crucial point is that $W^{\rm rad}_b \equiv W^{\rm rad}_{s=b}$, and the corresponding $\chi^{\rm rad}_{s=b}$,
is of order $G^4$ in the case of the scattering of
massive particles at finite $\g$ (see \cite{Bini:2017wfr}), but becomes of order $G^3 \ln(1/G)$ in the case of the classical scattering 
of massless particles. Indeed,  following the results of Gruzinov and Veneziano \cite{Gruzinov:2014moa} on the gravitational radiation from
{\it classical  massless} particle collisions\footnote{The results of Ref.~\cite{Gruzinov:2014moa} have been confirmed by a quantum-amplitude derivation \cite{Ciafaloni:2015xsr}.}, we see that, while in the
finite-$\g$ (and finite masses) scattering case  $\frac{dE^{\rm gw}}{ d\omega} $ (which is $O(G^3)$) starts decaying exponentially above a frequency of order 
$v/b$ \cite{Bini:2017wfr}, in the massless case  $\frac{dE^{\rm gw}}{ d\omega} $ decays only very slowly ($\propto \ln(1/(GE\omega)$)
above $1/b$. Using the approximate expression of $\frac{dE^{\rm gw}}{ d\omega} $ (when $1/b \lesssim \omega \lesssim 1/(GE)$)
derived in Ref. \cite{Gruzinov:2014moa}, and neglecting the contribution\footnote{The $\sim 1/\omega$ decay of the latter 
contribution might generate an additional logarithmic factor.} from $\omega \gtrsim 1/(GE)$, yields, for $W^{\rm rad}_b$,
an integral proportional to 
\be
\int_{1/b}^{1/GE} d\omega \ln (\omega b) \ln \left(\frac{1}{GE\omega} \right) \approx \frac{1}{GE} \ln \left(\frac{b}{GE} \right).
\ee
The crucial point to note is that this integral generates a factor $\frac{1}{GE}$ due to the slow decay of the HE gravitational-wave spectrum
between $1/b$ and $1/GE \gg 1/b$. [We are considering the small scattering angle case, $GE/b \ll1$.] Adding the factor
corresponding to the zero-frequency limit of $\frac{dE^{\rm gw}}{ d\omega} $ \cite{Weinberg:1965nx}, and the characteristic
tail prefactor $2GH =2 GE$, leads to
the following estimate
\be
W^{{\rm rad}\, \g \to \infty}_b \sim + \frac{G^3 E^4}{b^2} \ln \left(\frac{b}{GE} \right).
\ee
Differentiating with respect to $J \approx Eb/2$, finally leads to a scattering angle for massless particles of order
\be \label{chirad}
\chi^{{\rm rad}\, \g \to \infty}_b \sim - \left(\frac{GE}{b} \right)^3 \ln \left(\frac{b}{GE} \right) \sim - \chi^3 \ln \frac1{\chi},
\ee
where $\chi\sim GE/b$, on the rhs, denotes the leading-order scattering angle.
By contrast, the radiative contribution to $\chi$ in the finite-masses, finite-$\g$ case is 
\be
\chi^{{\rm rad} \,\g \,{\rm finite} }_b\sim \left(\frac{GE}{b} \right)^4 \sim \chi^4 =O(G^4).
\ee
As announced, we have here a conservative  dynamical effect, the radiative contribution to the scattering angle of two classical particles,
which is $O(G^4)$ when $\g$ is finite, but becomes $O(G^3 \ln 1/G)$ in the  $\g \to \infty$ limit.
Note that our estimates only concern the non-local (tail-transported \cite{Blanchet:1987wq}) contribution 
to the conservative dynamics. However, this is a clear proof that   the 4PM-level ($O(G^4)$)
conservative dynamics undergoes a transmutation of PM order (down to the $O(G^3 \ln 1/G)$ level)
in  the $\g \to \infty$ limit.

We also note that our reasoning 
indicates that, at the leading-log approximation, the  sign of $\chi^{{\rm rad}\, \g \to \infty}_b $ is {\it negative}.
Indeed, both $\ln (\omega b)$ and $\frac{dE^{\rm gw}}{ d\omega} $ are positive in the relevant interval; and
the differentiation with respect to $J$, {\it i.e.} $b$, changes the sign. We will come back below to this point.

%%%%%%%
\subsection{Second conjecture to reconcile the 3PM result of Refs.~\cite{Bern:2019nnu,Bern:2019crd}, with the 1SF, HE 
behavior of Ref.~\cite{Akcay:2012ea}} \label{transmutation}

Let us recap  the conundrum we are trying to solve. The  two-loop result of Refs.~\cite{Bern:2019nnu,Bern:2019crd}
leads to the following 3PM-accurate EOB $Q$ potential
 \be \label{Q3PMB}
\widehat Q^{\leq \rm 3 PM \, B}(u,\g,\nu) =  q_2(\g,\nu) u^2 + q_3^B(\g,\nu) u^3\,,
\ee
where the 3PM-level coefficient $q_3^B(\g,\nu)$ reads
\be \label{q3vsBC2}
q_{3}^B(\g, \nu)
=  B(\g)\left(\frac1{h(\g,\nu)}-1 \right) + 2 \nu \frac{{\overline C}^B(\g)}{h^2(\g,\nu)}\,,
\ee
with
\be
 B(\g) \equiv\frac32 \frac{(2 \g^2-1)(5 \g^2-1)}{\g^2-1}\,,
\ee
and 
\bea\label{bCB2}
{\overline C}^B(\g) &=& \frac{2}{3} \g  (14 \g^2+25) \nonumber\\
&+& 4 (4 \g^4 - 12 \g^2 -3) \frac{{\rm as}(\g)}{\sqrt{\g^2-1}}\,.
\eea
The crucial contribution in  ${\overline C}^B(\g)$ is the term\footnote{As it is the large $\g$ behavior that is of concern here, 
we could rephrase our discussion below  by replacing everywhere the factor $16 \g^4$ by its gravitational-coupling origin 
$w_1^2=4( 2\g^2-1)^2$,  as per the penultimate equation (9.2) of Ref. \cite{Bern:2019crd}.} 
$16 \g^4 {\rm as}(\g)/ \sqrt{\g^2-1}$, where we recall that the arcsinh function  can be written as
\bea \label{asvarious}
{\rm as}(\g)&=& \frac12 \ln \left( \g + p_{\infty} \right)  = - \frac12 \ln \left( \g - p_{\infty} \right)\nonumber\\
&=& \frac14 \ln \frac{ \g + p_{\infty}}{\g - p_{\infty}}=  \frac14 \ln \frac{ 1 + v_{\infty}}{1 - v_{\infty}}\,,
\eea
where  $v_{\infty} \equiv \frac{p_{\infty}}{\g} \equiv \sqrt{1- \frac1{\g^2}}$. 

Indeed, this contributes to $q_{3}^B(\g, \nu)$ the term
\be
q_{3}^{\rm log}(\g, \nu)=\frac{16 \nu \g^4}{h^2(\g,\nu) p_{\infty}} \ln \left( \g + p_{\infty} \right) \,.
\ee
The latter term is the source of the various logarithmic divergences entailed by the result of Refs.~\cite{Bern:2019nnu,Bern:2019crd}.
First, it causes the HE ($\g \to \infty$) behavior of $q_{3}^B(\g, \nu)$ to contain a $\ln \g$ enhancement of the $\sim \g^2$
behavior ensuring a well-defined massless limit (see Eq. \eqref{QHEg} above), indeed 
\be \label{HEQB}
\widehat Q^{\leq \rm 3 PM \, B}(u,\g,\nu) \overset{\rm HE}{=} \frac{15}{2} u^2  \g^2+  \left( 8\ln (2 \g)- \frac{17}{3}\right)u^3\g^2 \,.
\ee
Second, it is the source of the tension with the result of Ref.~\cite{Akcay:2012ea}. Indeed, it generates
a logarithmically divergent contribution to the 1SF quantity ${\widehat Q}^{1 \rm SF}_B /\g^3$:
\be \label{Q1SFBbyg3bis}
\frac{  {\widehat Q}^{1 \rm SF}_B }{\g^3} \overset{\rm HE}{=}   \frac{15}{2}   u^2 +  \frac{11}{3}  u^3 + 16   \ln (2 \g) u^3 \,,
\ee
while $ {\widehat Q}^{1 \rm SF}/\g^3$ was found in Ref.~\cite{Akcay:2012ea} to have a finite HE limit  (see Eq. \eqref{akcay}).
And third, it also leads to a logarithmic divergence when considering the HE limit $\g \to \infty$ 
(letting the masses $m_i \to 0$, and keeping fixed the c.m. energy $E$) of the  two-particle scattering
angle, namely (at the leading-logarithm accuracy),
\be \label{chiBHE2}
\frac12 \chi^B \overset{\rm HE}{=}  2 \alpha -   8 \ln (2 \g)  \alpha^3\,,
\ee
with $\alpha \overset{\rm HE}{=} G E^2/(2 J)$, as defined in Eq. \eqref{defalpha}. The latter result is in tension with the
eikonal computations of the gravitational scattering angle of (quantum) massless particles \cite{Amati:1990xe,Bern2020}.

These tensions have motivated us to propose above a modification (having a softer HE behavior) of the 3PM dynamics of 
Refs.~\cite{Bern:2019nnu,Bern:2019crd}. However, the recently performed  6PN-accurate $O(G^3)$ computations 
\cite{Blumlein:2020znm,Cheung:2020gyp,Bini:2020wpo} have disproved our softer-HE conjecture Eq. \eqref{bCTDbis}.

We wish now to present an alternative conjectural mechanism for cancelling the three related logarithmic divergences,
Eqs. \eqref{HEQB}, \eqref{Q1SFBbyg3bis}, and \eqref{chiBHE2}. We have seen
 in the previous subsection, that the $\g \to \infty$ limit of the radiative contribution entering the 4PM-level ($O(G^4)$)
scattering angle $\chi$ had the remarkable property of descending from the $O(G^4)$ level to the $O(G^3 \ln G)$ one.
In a similar manner, our proposed mechanism invokes the presence of a structure in 
 (part of) the 4PM-level contribution to the EOB $Q$ potential whose  HE limit trickles down
to the 3PM level  and  tames the three problematic 3PM-level logarithmic growths linked
to the presence of the arcsinh function  in Eq. \eqref{bCB}. To motivate the possibility of this mechanism, let us
start by noticing that one way to understand the technical origin of these various logarithmic growths is to view them 
(when considering the various rewritings of the arcsinh function exhibited in Eq. \eqref{asvarious}) as due to
the vanishing of the quantity $\g - \pinf$, or, equivalently, of $1-v_\infty$, as $\g\to \infty$. 
[In the HE limit, $\g \to \infty$,  $ \g - p_{\infty} =( \g + p_{\infty})^{-1}$ tends to zero  like $O(\frac1{\g})$,
while $1 - v_{\infty}$ tends to zero  like $O(\frac1{\g^2})$.]
Both these quantities make use (in their construction) of the flat Minkowski metric, $\eta_{\mu \nu}$. E.g., 
$1-v_\infty = (1-v_\infty^2)/(1 +v_\infty)$ crucially involves $1-v_\infty^2=-\eta_{\mu \nu}dx^\mu dx^\nu/(dx^0)^2$.
Now, the crucial contribution \eqref{bCB2} comes from the  $O(G^3)$ ``H-diagram'' ${\bf 7}$ in Fig. 14 of  
Ref. \cite{Bern:2019crd}. At the next PM levels, $O(G^{\geq 4})$, there will appear (among other diagrams) 
modifications of the H-diagram comprising extra graviton exchanges between one of the external massive particle lines
and, either the other massive particle, or one of the internal graviton lines. From the classical point of view, such
 modifications are related to some extra coupling to the metric field $h_{\mu \nu}=O(G)$, and can therefore be viewed as
modifying some of the occurrences of the flat metric $\eta_{\mu \nu}$ within the $O(G^3)$ diagrams.
This intuitive argument suggests the possibility of an effective  blurring of the light-cone-related
quantity $1-v_\infty^2=-\eta_{\mu \nu}dx^\mu dx^\nu/(dx^0)^2$ that is at the root of the logarithmic
blow-up of the $O(G^3)$ H-diagram. In other words, it is conceivable that some $O(G^{\geq 4})$ corrections
will soften the $\g \to \infty$ logarithmic blow-up contained in the $O(G^3)$ arcsinh function. Such a possibility is connected
with the known fact (discussed next) that the classical PM expansion is {\it not valid} for all Lorentz factors $\g$, but makes
sense  only if $\g$ is {\it smaller} than some $G-$dependent upper limit. 

The issue of the domain of physical validity of the PM expansion  has been
discussed in the literature on relativistic gravitational bremsstrahlung 
\cite{Peters:1970mx,DEath:1976bbo,Kovacs:1977uw,Kovacs:1978eu}, though with unclear or conflicting answers.
Peters \cite{Peters:1970mx} concludes (in the small mass-ratio case, $\nu \to 0$), that the PM expansion
is valid only if 
\be \label{peters}
\g^2 \frac{GM}{b} \ll 1 \,; 
\ee
while D'Eath (see p. 1016 in \cite{DEath:1976bbo}), cited by Kov\'acs and Thorne \cite{Kovacs:1978eu}, concludes that,
for comparable masses ($\nu =O(1)$), the PM expansion is valid only if
\be \label{death}
 h^2 \frac{GM}{b} \sim \g \frac{GM}{b} \ll 1 \,.
\ee
To illustrate one of the technical origins of the limit \eqref{peters}, let us consider  the scalar 
$h_{2 \, \mu\nu} u_1^{ \mu} u_1^{ \nu}$, where (see, {\it e.g.}, \cite{Bini:2017xzy})
\be \label{huu}
h_{2 \, \mu\nu}(x)=2 \frac{G m_2}{R_2} (2 u_{2 \, \mu} u_{2 \, \nu} + \eta_{\mu \nu})\,,
\ee
is the value, along the worldline of the first particle $m_1$,
of the harmonic-gauge linearized gravitational field generated by the second particle $m_2$. During a small-angle hyperbolic encounter,
the scalar \eqref{huu} reaches the maximum value
$(h_{2 \, \mu\nu} u_1^{ \mu} u_1^{ \nu})^{\rm max}= 2 \frac{G m_2}{b} (2 \g^2-1)=w_1  \frac{G m_2}{b}$. It seems then
natural to require for the validity of the PM expansion that $w_1  \frac{G m_2}{b} \sim \g^2  \frac{G M}{b} \ll 1$ (in agreement with Eq.~\eqref{peters}) .

If we reexpress the various possible limits of validity of the PM expansion, Eqs. \eqref{peters} or \eqref{death}, in terms of the scattering angle
\be
\frac12\chi= \frac{GM h(\g,\nu)}{b} \frac{2 \g^2-1}{\pinf^2}+O(G^2)\,,
\ee
we get  limits of validity  of the general type 
\be \label{limitchi}
 \chi \g^n \ll 1\,,
\ee
with some (strictly) positive $n$ ($n=\frac32$ according to Peters, and our argument, and $n=\frac12$ according to D'Eath).

Independently of the differences\footnote{These differences could due to a gauge dependence, 
and could also refer to the domains of validity of different observables.}
 between these various validity constraints, the general requirement \eqref{limitchi} (with any positive exponent $n$) is saying that one
cannot trust taking the HE limit $\g \to \infty$ independently of the $\chi \to 0$ (or of the $G \to 0$) limit. 
 This points out towards  a possible non commutativity of the two limits $\g \to \infty$ and $G \to 0$.

We are interested in transcribing the validity limit \eqref{limitchi} 
 in terms of the EOB gravitational potential $u=GM/R$, which enters the $Q$ potential.
When considering a small-angle scattering, the maximum value of $u$ is defined by inserting $p_r=0$
in the  free-motion ($G\to 0$) EOB mass-shell condition, $ \e^2=1 + p_r^2+ j^2 u^2$, so that
\be
u_{\rm max} = \frac{\pinf}{j}= \frac{GM h(\g,\nu)}{b}\,.
\ee
We thereby see that in the HE limit $u_{\rm max} \sim \chi$. Therefore the general limit \eqref{limitchi} is equivalent to
\be \label{limitu}
 \g^{n} u_{\rm max} \ll 1\, {\rm with} \; n >0 \; ({\rm and \, probably} \;\frac12 \leq n \leq \frac32).
 \ee
Combining this information about the limit of validity of the PM expansion, both with the reasoning above concerning
$O(G^{\geq 4})$  corrections to the crucial $O(G^3)$ H-diagram, and with the proof given in the previous
subsection of a $O(G^4) \mapsto O(G^3 \ln G)$ transmutation of PM order in the radiative part of the
scattering angle, leads us to conjecture that the higher-PM contribution to the 3PM-accurate EOB $Q$ potential
will contain a term $\Delta \widehat Q(\g, u,\nu)$ which is of order $u^4 =O(G^{ 4})$ when $\g^{n} u \ll 1$,
but which becomes of order $u^3 \ln (\g^n u)$ when $\g^{n} u \gg 1$, and which  cancells the HE logarithmic
blow-up, Eq. \eqref{HEQB}, of the 3PM potential $\widehat Q^{\leq \rm 3 PM \, B}(u,\g,\nu)$, Eq. \eqref{Q3PMB}.

Such a general requirement about the nature of the non commutativity of the two limits $G \to 0$
and $\g \to \infty$ might be realized in many different ways. Let us illustrate 
the possibility of such a mechanism by a specific example of a  $O(G^{\geq 4})$ term $\Delta \widehat Q(\g, u,\nu)$.
We are not claiming here that our example must be exactly the one that will enter the $O(G^{\geq 4})$ dynamics, but
we propose it as an existence proof of a $O(G^{\geq 4})$ modification of the 3PM dynamics  having interesting 
HE properties, and, in particular, reconciling the current 3PM dynamics, Eqs. \eqref{Q3PMB}, \eqref{q3vsBC2},
 with the SF result, Eq. \eqref{akcay}.

Our proposed example consists in modifying the 3PM-accurate $Q$ potential, Eqs. \eqref{Q3PMB},  \eqref{q3vsBC2}, by an 
extra $O(G^{\geq 4})$ contribution of the form
 \be\label{deltaQ0}
\Delta \widehat Q(\g, u,\nu)=-\frac{16 \nu \g^3}{h^2(\g,\nu)}   u^3 \frac{  \ln (1+ \g^n u)}{n}\,,
\ee
where $n>0$ refers to the exponent entering the general limit of validity, Eq. \eqref{limitu}, of the PM expansion.
[For simplicity, we did not include in the illustrative model \eqref{deltaQ0} various possible modifications, such as
a prefactor containing lower powers of $\g$ ({\it e.g.}, $w_1^2/\g$ in lieu of $16 \g^3$), and a numerical 
coefficient in front of $\g^n u$ in the argument of the logarithm.]
The only crucial elements (for our discussion below) entering this illustrative definition of $\Delta \widehat Q(\g, u,\nu)$ are the following:
(i) the factor 16 in front of $\g^3$; (ii) the fact that the mass ratio $\nu$ only enters via the overall factor $\nu/h^2(\g,\nu)$;
and (iii) the fact that the function $u^3   \ln (1+ \g^n u)/n$ is of order $O(u^4)$ as $u \to 0$, and 
$\approx u^3 (\ln \g + \frac1n \ln u)$ as $\g \to \infty$.
[Evidently, many other functions could realize such requirements, or suitable variants of them.]

The   dynamics  defined by the {\it modified} $Q$ potential
\be \label{newQ3PM}
 \widehat Q^{\rm mod}(u,\g,\nu) \equiv \widehat Q^{\leq \rm 3 PM\, B}(u,\g,\nu)+ \Delta \widehat Q(u,\g,\nu) \,,
\ee
has the following properties.

First, it relieves  the tension between Refs. \cite{Bern:2019nnu,Bern:2019crd} and  Ref. \cite{Akcay:2012ea}. Let us take the 1SF
(linear in $\nu$) contribution to the modified EOB $Q$ potential \eqref{newQ3PM}, and then consider its HE limit.
As our modification enters the EOB Hamiltonian multiplied by the overall factor $2\nu/h^2(\g,\nu)$, the 1SF piece in the new EOB $Q$ potential \eqref{newQ3PM} is  given by:
\be \label{Q1SFBbyg3new}
\frac{  {\widehat Q}^{\rm mod \,1 SF } }{\g^3} \overset{\rm HE}{=}   \frac{15}{2}   u^2 + \left(\frac{11}{3}+ 16 \ln(2) - \frac{16}{n}   \ln (  u) \right) u^3 \,.
\ee
The major difference with the previous result, Eq.  \eqref{Q1SFBbyg3}, is that the divergent logarithm $ +\ln ( \g)$ has
been now replaced by $ - \frac1n \ln ( u)$. When evaluated at the lightring $u=\frac13$, we thereby get a finite
contribution involving $ - \frac1n \ln (\frac{1}{3})$ instead of the divergent $ \ln ( \g)$, in qualitative agreement with the finite
result found in \cite{Akcay:2012ea}. 

 Second,  let us consider the HE limit, $\g\to \infty$, of the modified EOB $Q$ potential \eqref{newQ3PM}.
 Contrary to the  HE limit  of $\widehat Q^{\leq \rm 3 PM \, B}(u,\g,\nu)$, displayed in Eq.  \eqref{HEQB} above,
 which did not define  a good, quadratic-in-$\g$ HE limit, the $\g\to \infty$ limit  of
$ \widehat Q^{\rm mod}$  now leads to a well-defined quadratic-in-$\g$ HE limit, namely
\be \label{Q3pmnewHE}
\widehat Q^{ \rm mod}(\g, \nu,u)  \overset{\rm HE}{=}  q^{\rm HE \, new}(u) \g^2 \,,
\ee
where
\be \label{q3pmnewHE}
q^{\rm HE \, new}(u) \overset{\rm HE}{=} \frac{15}{2} u^2 + \left(- \frac{17}{3} + 8 \ln(2) - \frac{8}{n} \ln( u) \right) u^3\,.
\ee
As announced, the latter HE limit has featured a  phenomenon of transmutation of PM order. The HE limit of the  $O(u^4)=O(G^4)$ additional contribution $\Delta \widehat Q$  has been transmuted into a contribution of order $u^3 \ln u =O(G^3 \ln G)$. This property is intimately linked with the fact that
the additional contribution $\Delta \widehat Q$ was devised so as to cancell the $\g^2 \ln \g$ contribution present in the
HE limit of $\widehat Q^{\leq \rm 3 PM \, B}(u,\g,\nu)$. 

 Third, if we consider a fixed,  finite value of $\g$, and take the PM expansion of $\Delta \widehat Q$, {\it i.e.}, its expansion in powers of $G$,
we find that its 4PM-level, $O(u^4)=O(G^4)$, structure reads 
\be\label{deltaQ1}
\Delta \widehat Q^{\rm PM-expanded}=-\frac{16 \nu \g^4 }{n h^2(\g,\nu)}  \g^{n-1} u^4 + O(u^5)\,.
\ee
Taking the HE limit of this PM-expanded contribution yields
\be \label{delQPMHE}
\Delta \widehat Q^{\rm PM-expanded}\overset{\rm HE}{=}  -\frac{8}{n} \g^{2+n} u^4 + O(u^5)\,.
\ee
As we had assumed $n > 0$ (and probably  $n\geq \frac12$), 
we see that this contribution violates (in a {\it power-law} fashion) the expected quadratic-in-$\g$ HE behavior.
This violation at the level of the Hamiltonian entails a corresponding power-law violation of the naively expected behavior
of scattering observables (at the 4PM level). Namely, instead of having a 4PM-level contribution $\chi_4(\g,\nu)$
behaving, when $\g \to \infty$, $\propto \g^4$ (like the test-particle one), the term \eqref{delQPMHE} would
yield a contribution $\propto \g^{4+n}$. This apparent fast growth as $\g \to \infty$ would, however, be an
effect of having PM-expanded the factor $\ln (1+\g^n u)$ and is absent in the exact, non-PM-expanded
scattering angle $\chi(\g, j, \nu)$.

 Indeed, the real value of the HE scattering angle predicted by Eq. \eqref{newQ3PM} is obtained from  the HE limit of the modified
$Q$ potential, {\it i.e.}, from the HE quadratic mass-shell constraint Eq. \eqref{Q3pmnewHE}, with Eq. \eqref{q3pmnewHE}. 
Similarly to what happened
for the 1SF-level contribution ${\widehat Q}^{1 \rm SF \, new}$, one finds that that this now predicts a {\it finite} 3PM-level
massless scattering angle wich differs from the previous logarithmically divergent one, Eq. \eqref{chiBHE}, by the replacement
of  $ \ln ( \g)$ by  $ - \frac1n \ln ( \alpha)$. Namely, at the leading-logarithm accuracy ({\it i.e.} modulo some $\propto \alpha^3$
contribution), one finds
\be \label{chiHEnew}
\frac12 \chi^{\rm mod} \overset{\rm HE}{=}  2 \alpha -   \frac8{n} \ln \left(\frac1{\alpha}\right)  \alpha^3\,.
\ee
We have written it here in terms of $\ln \left(\frac1{\alpha}\right)$ to emphasize that the sign of the finite 
logarithmic contribution is the same (namely negative) as the sign of the previously divergent contribution 
$ -   8 \ln (2 \g)  \alpha^3$. On the one hand, this sign differs from the corresponding positive 3PM contribution $\sim + \alpha^3$
found by ACV, and recently confirmed in Ref. \cite{Bern2020}. On the other hand, we note that the estimate \eqref{chiHEnew}
agrees in magnitude and sign with the contribution $O(\alpha^3 \ln1/\alpha)$ in Eq. \eqref{chirad} derived
above from considering the radiative correction to the classical scattering angle. 

Let us finally note that the structure we used in our illustrative model \eqref{newQ3PM} is not, by itself,
leading to a scattering angle satisfying the general mass-ratio-dependence properties discussed in section \ref{sec2}.
There are, however, ways to design a modified version of  $\Delta \widehat Q(\g, u,\nu)$ that would incorporate the latter
expected mass-ratio-dependence. We found that such better (but more complicated) models predict the same general
features we just discussed. For simplicity, and in absence of precise guidelines for choosing among such models,
we do not feel useful to complicate our discussion by indicating the construction of such models.

In conclusion: our  second (illustrative) ansatz \eqref{newQ3PM} relieves
the tension between  Refs. \cite{Bern:2019nnu,Bern:2019crd} and  Ref. \cite{Akcay:2012ea}, and leads
to some generic predictions for the HE behavior of, both, the full dynamics, and its 4PM-truncated version.
There remains (as was the case with the first conjecture, Eq. \eqref{bCTDbis})  a tension between 
the massless limit of the {\it classical} scattering, and the 
{\it quantum}, eikonal-based massless scattering angle of Refs. \cite{Amati:1990xe,Bern2020}. 
As we already pointed out, the root of the latter discrepancy might reside in subtleties of the quantum-to-classical
transition (with a possible non commutativity of the two limits $\g \to \infty$, and $\hbar \to 0$),
or in the use of  the quantum-eikonal-approximation.

%%%%%%%%
\section{Summary} 

This paper has derived new general properties of post-Minkowskian (PM) gravity,
notably in its effective one body (EOB) formulation. Our first result has been to prove
general expressions for the dependence of the momentum transfer (during the classical scattering of two masses)
on the two masses, and  thereby on the symmetric mass ratio $\nu$ (see Eqs. \eqref{QPM}, \eqref{Qvsnu}).
This implies  specific constraints on the $\nu$ dependence of the scattering angle considered as a function
of the reduced angular momentum $j\equiv J/(G m_1 m_2)$ (see Eqs. \eqref{chi1pm}, \eqref{chi2pm}, \eqref{chi3pm}).
A useful consequence of these results is that the full knowledge of the 3PM  dynamics is encoded
in  a single function of the single variable $\g = - (p_1 \cdot p_2)/(m_1 m_2)$. Moreover the same property holds 
also at the 4PM level. We pointed out that these properties  allow  {\it first-order self-force} (linear
in mass ratio) computation
of scattering to give access to the {\it exact} 3PM and 4PM dynamics.

We then generalized our previous work \cite{Damour2018} by deriving, up to the 4PM level included,
 the explicit links between the scattering angle and the two types of potentials entering the Hamiltonian
 description of PM dynamics within EOB theory. The first type of potential is the $Q$ potential entering
 the mass-shell condition of EOB dynamics
 \be \label{massshellgenbis}
0= g_{\rm Schwarz}^{\mu \nu} P_{\mu} P_{\nu} + \mu^2 + Q(X,P)\,,
\ee
while the second one is an ordinary, energy-dependent radial potential $W(E, {\bar R})$ entering a
non-relativistic-like quadratic constraint on the EOB momentum,
\be
 \bP^2= P_{\infty}^2+ W(\bu, P_{\infty})\,.
 \ee
The first formulation is usually expressed in terms of a Schwarzschild-like radial coordinate $R$ (with $u=GM/R$),
while the second one uses an isotropic-like radial coordinate ${\bar R}$ (with ${\bar u}=GM/{\bar R}$).
The links between the PM expansion coefficients of both types of formulations, as well as their links with
the PM expansion coefficients of the scattering function, were given in section \ref{sec3}. [See Appendix \ref{A} 
for the link of the EOB potential with the potential used in Refs. \cite{Bern:2019nnu,Bern:2019crd}.]
At the end of section \ref{sec3}, we summarized the current knowledge of the PM-expanded dynamics and emphasized the
apparent incompatibility between the recent classical 3PM-level dynamics derived by Bern {\it et \, al. } \cite{Bern:2019nnu,Bern:2019crd} 
and  the self-force computation of Ref. \cite{Akcay:2012ea}.  We then suggested two different types of resolution of this tension.
The first resolution conjectures that the 3PM dynamics has a softer high-energy (HE) behavior than the one derived 
in Refs. \cite{Bern:2019nnu,Bern:2019crd}. Namely, we conjectured that the function ${\overline C}(\g)$ entering the
3PM dynamics might have a softer HE behavior than Eq. \eqref{bCB} (see Eq. \eqref{bCTDbis}). However, several recent 6PN-accurate
$O(G^3)$ computations \cite{Blumlein:2020znm,Cheung:2020gyp,Bini:2020wpo} have disproved the (HE-softer) conjectured  3PM
dynamics \eqref{bCTDbis}.

In subsection \ref{concern} we recalled a classic argument of Niels Bohr showing the lack of overlap between the
domains of validity of classical and quantum scattering theory. This fact might entail subtleties in the quantum-to-classical 
maps used in several recent works.

We also presented a second type of possible resolution of the tension between Refs. \cite{Bern:2019nnu,Bern:2019crd} 
and  Ref. \cite{Akcay:2012ea}. This second resolution does not call for a modification of the 3PM dynamics of 
Refs. \cite{Bern:2019nnu,Bern:2019crd} when it is considered at a finite value of the Lorentz factor $\g$ (denoted $\sigma$
in Refs. \cite{Bern:2019nnu,Bern:2019crd}), but assumes a particular type of non commutativity of the two
limits  $\g \to \infty$, and $ G \to 0$. We emphasized that the PM expansion is expected to lose its validity when 
$\g$ becomes larger than some inverse power of $GM/b$ (or $GM/r$), see Eqs. \eqref{limitchi}, \eqref{limitu}.
We gave an illustrative model of  higher PM ($O(G^{\geq 4})$) contributions
to the currently known $O(G^{\leq 3})$ dynamics able to reconcile the results of Refs. \cite{Bern:2019nnu,Bern:2019crd} 
and  Ref. \cite{Akcay:2012ea}; see Eq. \eqref{newQ3PM}.  This model makes some generic predictions (explained
in the previous section) and exhibits an interesting phenomenon of HE transmutation of post-Minkowskian order.
Namely the HE limit of a  $O(G^{\geq 4})$ term becomes of order $O(G^3 \ln G)$ when  $\g^n u \gg 1$, for
some positive exponent $n$. Independently of the motivation for our conjecture, we showed (in subsection \ref{gruzinov}) that such 
 a HE transmutation of PM order (from ($O(G^4)$ down to $O(G^3 \ln G)$) does take place in
 the radiative contribution to the scattering angle of classical massless particles.

Section \ref{sec4} presented the 3PM generalization of a result of Ref. \cite{Damour2018},
namely the computation of the scattering amplitude derived from quantizing the 3PM EOB potential. Our computation explicitly
takes into account the IR-divergent contributions coming from the Born iterations of the EOB radial potential. The usual 
potential-scattering amplitude $f_{\rm eob}$ in the EOB radial potential is  linked to a corresponding Lorentz-invariant
amplitude $\cM$ via the simple rescaling
\be \label{cMvsfbis}
\cM_{\rm eob} = \frac{8\pi G \, s }{\hbar} f_{\rm eob}\,.
\ee

Section \ref{HEPM} (as well as Appendices B and C) 
discusses various features of the high-energy (or massless) limit of the PM dynamics.

Note finally that a general theme of the present work has been to highlight some of the subtleties involved when
considering several {\it a priori} non-commuting limits: $\hbar \to 0$ versus $\hbar \to \infty$; $G \to 0$; $\g  \to \infty$;
and $\nu \to 0$. The existing tension between: (i) the (logarithmically divergent) high-energy limit, Eq. \eqref{chiBHE2}, of the scattering angle 
of Refs. \cite{Bern:2019nnu,Bern:2019crd}; (ii) the quantum-eikonal-based  computation \cite{Amati:1990xe,Bern2020} of the 
 scattering angle of {\it quantum} massless particles, Eq. \eqref{chiACV} ; and, 
 (iii) the type of (finite) scattering angle of {\it classical} massless particles
 predicted by both our HE-softer conjectures (see notably Eq. \eqref{chiHEnew}), needs further clarification.

%%%%%%%%%%%%%

%%%%%%%%%%
 \section*{Acknowledgments}
I thank Johannes Bl\"umlein, Marcello Ciafaloni, Alain Connes, Henri Epstein, Stefan Hollands, Alfredo Guevara, Maxim Kontsevich, 
Vincent Rivasseau, Slava Rychkov, Pierre Vanhove, and Stefan Weinzierl for informative exchanges. Special
thanks go to to  Dimitri Colferai, Alexander Milstein, Mikhail Solon, and Gabriele Veneziano  for enlightening discussions.
 %%%%%%%%%%%

 %%%%%%%%%
 \appendix

%%%%%%%%% A
\section{Map between the EOB potential and the potential of Cheung, Rothstein and Solon} \label{A}

Cheung, Rothstein and Solon (CRS) \cite{Cheung:2018wkq} have proposed to describe the classical dynamics
of a two-body system by the same type of Hamiltonian that was considered long ago by Corinaldesi and Iwasaki,
namely
\begin{eqnarray} \label{Husualbis}
H(\x_1,\x_2,\p_1,\p_2) &=& c^2\sqrt{ m_1^2 + \frac{\p_1^2}{c^2}} +
 c^2\sqrt{ m_2^2 + \frac{\p_2^2}{c^2}} \nonumber \\ &+& V(\x_1-\x_2,\p_1,\p_2)\,,
\end{eqnarray}
except that they did not limit themselves to working with the PN-expanded form of such an
Hamiltonian (\`a la Eq. \eqref{HPN}). In addition, when working in the  c.m. frame (with the 
c.m. Hamiltonian reduction $\p_1=- \p_2= \bP$), they required a specific isotropic-like 
gauge-fixing of the c.m. potential $V(\x_1-\x_2,\p_1,\p_2)^{\rm c.m.}=V(\X,\bP)$ such that
$V(\X,\bP)$ depends only on $\bP^2$ and $ R \equiv |\X|$:
\be \label{VPMbis}
V(\bP, \X)= G \frac{c_1(\bP^2)}{|\X|}+ G^2 \frac{c_2(\bP^2)}{|\X|^2}+ G^3 \frac{c_3(\bP^2)}{|\X|^3}+ \cdots 
\ee
Ref. \cite{Cheung:2018wkq} derived a 2PM-accurate potential (from the  quasi-classical one-loop amplitude of Refs. 
\cite{Guevara:2017csg,Bjerrum-Bohr:2018xdl}) without connecting this potential to the previously derived (simpler) 2PM-accurate
EOB potential of Ref. \cite{Damour2018}. To complete our study, let us sketch how the two types of potentials are related
by using the tools we have introduced above. We will be brief because results essentially equivalent (and sometimes to higher-orders)
to the results below
(though formulated differently) have already been displayed in Refs. \cite{Bern:2019crd,Kalin:2019rwq,Bjerrum-Bohr:2019kec}.

The gauge-invariant characterisation of the successive coefficients $w_n$ entering the energy-dependent
version of the EOB potential obtained in  subsection \ref{subsec5C} gives a simple algorithmic procedure for extracting 
the gauge-invariant information from the PM expansion \eqref{VPMbis} of the CRS potential $V(\X,\bP)$. Let us sketch
how this can done.

Starting from
\be \label{H2body}
H(\bP, \X)= \sqrt{ m_1^2 + \bP^2}+ \sqrt{ m_2^2 + \bP^2}+  V(R,\bP^2)\,,
\ee
with
\be \label{VCRS}
V(R,\bP^2)= G \frac{c_1(\bP^2)}{R}+  G^2 \frac{c_2(\bP^2)}{R^2}+  G^3 \frac{c_3(\bP^2)}{R^3}+\cdots
\ee
and denoting as $P_{\infty}$ the (common) magnitude of the c.m. incoming (and outgoing) momenta,
such that the total (conserved) energy $E_{\rm real}= \sqrt{s}$ of the two-body system reads
\be
E_{\rm real}(P_{\infty}^2)=\sqrt{ m_1^2 + P_{\infty}^2}+ \sqrt{ m_2^2 + P_{\infty}^2}\,,
\ee
we can perturbatively solve the energy conservation law $E_{\rm real}=H(\bP, \X)$ for $\bP^2$. 
Beware that, in this appendix, we will use the notation $P_{\infty}$ (without extra label) to denote the magnitude
of the asymptotic physical c.m. three-momentum. This quantity differs from the corresponding 
 EOB incoming momentum, which was also denoted $P_{\infty} = \mu p_{\infty}$ in the main text.
Here, we will denote the latter EOB incoming momentum as $P^{\rm EOB}_{\infty} = \mu p^{\rm eob}_{\infty}$.
The relation between  $P_{\infty} \equiv P^{\rm cm}_{\infty}$ and $P^{\rm EOB}_{\infty}$ will be recalled below.

We look for a PM expansion of the type
\be \label{P2PM}
\bP^2= P_{\infty}^2+ \frac{ W_1(P_{\infty})}{R}+ \frac{ W_2(P_{\infty})}{R^2}+\frac{ W_3(P_{\infty})}{R^3}+\cdots
\ee
where $W_n \propto G^n$, such that the insertion of the expansion \eqref{P2PM} in Eq. \eqref{H2body}, with the
PM-expanded potential \eqref{VCRS} solves the constraint $E_{\rm real}=H(\bP, \X)$. At first order in $G$, this yields
the constraint
\be
 \frac{d E_{\rm real}(P_{\infty}^2)}{d P_{\infty}^2} \frac{ W_1(P_{\infty})}{R}+ G \frac{c_1(P_{\infty}^2)}{R}=0\,,
\ee
which uniquely determines $ W_1(P_{\infty})$ in terms of $c_1(P_{\infty}^2)$, namely
\be
W_1(P_{\infty}) = -  \left(    \frac{d E_{\rm real}(P_{\infty}^2)}{d P_{\infty}^2}\right)^{-1} G \, c_1(P_{\infty}^2)\,.
\ee
At second order in $G$, we similarly get an equation uniquely determining $ W_2(P_{\infty})$ in terms of $c_2(P_{\infty}^2)$,
of the $P_{\infty}^2$ derivative of $c_1(P_{\infty}^2)$,
and of the previously determined $W_1(P_{\infty}^2)$, namely
\bea
W_2(P_{\infty}) &=&  -  \left(    \frac{d E_{\rm real}}{d P_{\infty}^2}\right)^{-1} \left( G^2\,  c_2(P_{\infty}^2) \right.\nonumber\\
 &+& \left. G \frac{d c_1(P_{\infty}^2)}{d P_{\infty}^2} W_1  + \frac12  \frac{d^2 E_{\rm real}}{(d P_{\infty}^2)^2 } W_1^2  \right).
\eea
This algorithmic procedure successively determines the  coefficients $W_n(P_{\infty})$ entering the PM expansion \eqref{P2PM}
in terms of the sequence of functions $c_n(\bP^2)$. The results of this procedure agree with the corresponding results 
in section 11.3.1 of Ref. \cite{Bern:2019crd}, but we will use them here to relate the EOB $Q$ potential to the CRS $V$ potential.

The next step is to transform the coefficients $W_n(P_{\infty})$  into their corresponding gauge-invariant avatars 
${\widetilde W}_n(P_{\infty}) $, defined in the same way as in Eq. \eqref{tildewn} above, namely
\bea \label{tildeWn}
{\widetilde W}_1(P_{\infty}) &=& \frac{W_1(P_{\infty})}{P_{\infty}}, \nonumber \\
{\widetilde W}_2(P_{\infty}) &=& W_2(P_{\infty}), \nonumber \\
{\widetilde W}_3(P_{\infty})&=& P_{\infty} W_3(P_{\infty}),\nonumber \\
{\widetilde W}_4(P_{\infty})&=& P_{\infty}^2 W_4(P_{\infty}).
\eea
Then, applying the reasoning made around Eq. \eqref{tildewn} above, we conclude that the 
${\widetilde W}_n(P_{\infty}) $'s extracted from the sequence of functions $c_n(\bP^2)$'s must be 
numerically identical to the ${\widetilde w}_n(p_{\infty}) $'s entering the EOB potential.
One must simply take care of the presence of a factor $(Gm_1 m_2)^n$ due to
 the rescaling factors, $P= \mu p$, $E= M h$, $J = G M \mu j$, used  above,
 and of the (crucial) fact that the CRS and EOB quantities are expressed as functions of different variables,
 namely  $P_{\infty} \equiv P^{\rm cm}_{\infty}$ versus $P^{\rm EOB}_{\infty} = \mu \peob$.
 At this stage, we need to recall that, according to, {\it e.g.}, Eq. (10.27) of Ref. \cite{Damour2018},
 the (rescaled) EOB incoming momentum $\peob=p^{\rm eob}_{\infty}$ is related
 to the real, c.m. incoming momentum $P_{\infty}$ by
 \be
 E_{\rm real} P^{\rm real}_{\infty}= m_1 m_2 \sqrt{\g^2-1} \equiv m_1 m_2 p^{\rm eob}_{\infty}\,.
 \ee
 Finally, we have the simple relations
 \bea 
{\widetilde W}_1(P_{\infty}) &=& G m_1 m_2  \, {\widetilde w}_1^{\rm eob}(\g), \nonumber \\
{\widetilde W}_2(P_{\infty}) &=&  (G m_1 m_2)^2 \, {\widetilde w}_2^{\rm eob}(\g),\nonumber \\
{\widetilde W}_3(P_{\infty})&=&   (G m_1 m_2)^3 \, {\widetilde w}_3^{\rm eob}(\g),\nonumber \\
{\widetilde W}_4(P_{\infty})&=&  (Gm_1 m_2)^4 \, {\widetilde w}_4^{\rm eob}(\g)\,.
\eea
The first two EOB PM levels have been computed in Ref. \cite{Damour2018} and yielded
the results
\bea
{\widetilde w}_1^{\rm eob}(\g) &=& \frac{2(2\g^2-1)}{\sqrt{\g^2-1}} \,,\nonumber \\
{\widetilde w}_2^{\rm eob}(\g) &=& \frac{3}{2}\frac{(5\g^2-1)}{h(\g,\nu)}\,.
\eea
We have checked that by inserting the latter  simple expressions in the relations written above
gave the (much more intricate) expressions of $c_1$ and $c_2$ derived in \cite{Cheung:2018wkq}.
Note, in particular, that the  asymptotic value $\xi_{\infty}$ of the symmetric energy ratio defined 
in \cite{Cheung:2018wkq}, namely
\be \label{defxi}
\xi(\bP^2) \equiv \frac{ \sqrt{ m_1^2 + \bP^2} \sqrt{ m_2^2 + \bP^2} }{ \left(\sqrt{ m_1^2 + \bP^2}+ \sqrt{ m_2^2 + \bP^2}\right)^2},
\ee
which does not appear in the EOB results, enters in $c_1$ via the derivative
\be
 \frac{d E_{\rm real}(P_{\infty}^2)}{d P_{\infty}^2} =\frac{1}{2 \xi_{\infty} \, E_{\rm real}(P_{\infty}^2)}\,.
\ee
When working at the 3PM-level one can similarly relate the coefficients $c_3$, ${\widetilde W}_3(P_{\infty})$,
${\widetilde w}_3^{\rm eob}(\g)$ and $q_3(\g)$, and explicitly check that the value of $c_3$ given in the last
Eq. (10.10) of \cite{Bern:2019crd} is equivalent to the (much simpler) expression of $q_3$ obtained in the main text 
(and also derived in Ref. \cite{Antonelli:2019ytb} by using the formulas of \cite{Damour2018}). 
Let us finally note that Refs. \cite{Kalin:2019rwq,Bjerrum-Bohr:2019kec} derived all-order expressions for the
links between the quantities $c_n$ and $w_n$ (without considering, however, the more basic EOB coefficients $q_n$).

%%%%%%%%% B
\section{On the structure of the HE limit of PM gravity}

To complete our discussion of  PM gravity, let us briefly discuss some of the  structures that might arise in the HE limit of the classical
momentum transfer $\sQ$, considered as a function of the impact parameter, Eq. \eqref{Qvsnu}. We have discussed above two
different possibilities for reconciling the current quantum-based computation of 3PM dynamics, and older HE SF results.
The first possibility assumes that the HE limit of classical scattering is as tame at the third (and higher) PM order(s) than it is at the 
first and second PM orders. The second possibility allows for violations of the
latter tame HE behavior. We shall contrast the structures corresponding to these two possibilities.

To discuss the HE behavior, let us reformulate the
classical time-symmetric Lorentz-invariant, PM  perturbation-theory computation of the momentum change $\Delta p_{1 \mu}= -\Delta p_{2 \mu}$.
 Above we wrote this PM perturbation theory in terms of two worldlines parametrized by their proper times $s_a$,
so that $u_a^{\mu} = d x_a^{\mu}/ds_a$ were two unit vectors, because we wanted to keep track of the dependence
on the two rest masses $m_a$, entering the stress-energy tensor as multiplicative factors. But we could have, instead,
as was actually done in \cite{Damour:2016gwp,Damour2018}, use worldline parameters $\s_a= s_a/m_a$
such that $d x_a^{\mu}/d\s_a= m_a u_a^{\mu}=  p_a^{\mu}$. In this parametrization the stress-energy tensor does
not involve the masses, but only the momenta, and reads
\bea
T^{\mu \nu}(x) 
&=& \sum_{a=1,2}  \int d\sigma_a   p_a^{\mu} p_a^{\nu} \, \frac{\delta^4(x-x_a(\sigma_a))}{\sqrt{g}}\nonumber\\
&=& \sum_{a=1,2} \int    p_a^{\mu} dx_a^{\nu} \, \frac{\delta^4(x-x_a)}{\sqrt{g}}\,.
\eea
One then checks that the masses will never explicitly occur in this reformulation of PM perturbation theory.
 This reformulation is useful for treating the limiting case where $m_a \to 0$, $u_{a}^{\mu} \to \infty$,
 keeping fixed the values of the momenta $ p_a^{\mu}= m_a u_{a}^{\mu}$. In this limit the two momenta, and the two worldlines,
 become lightlike: $p_a^2= - m_a^2 \to 0$. The expressions written down in Refs. \cite{Damour:2016gwp,Damour2018}
 then define a formal  PM perturbation theory that applies when one or two of the particles are massless.
Let us consider the case where both particles are massless. A difference with the massive case is that the
convolution of the time-symmetric propagator $\propto \delta \left[ (x-y)^2 \right]$ with a $T^{\mu \nu}(y)$
localized along a null geodesic (which is straight at LO) selects a single (advanced or retarded) source point $x_a$
on each worldline. [Indeed, the LO equation to be solved in $\s_a$, for a given field point $x$, namely
$ (x- x^0_a- p^0_a \s_a)^2=0$, is linear, rather than quadratic,
in $\s_a$ because $ (p^0_a)^2=0$.]
The corresponding linearized approximation for the metric (in harmonic gauge) reads
\be \label{AS}
 h_{\mu \nu}^{m_a=0}(x)= \sum_a 4 G \frac{p_{a \mu} p_{a \nu}}{| (x-x_a) \cdot p_a|} + O(G^2)\,.
 \ee
In the presently considered case where the $p_a$'s are null,
the expression \eqref{AS} represents a sum of  Aichelburg-Sexl metrics \cite{Aichelburg:1970dh} associated with each worldline.
Each Aichelburg-Sexl metric is flat (zero curvature) outside of the null hyperplanes $(x-x_a). p_a=0$, but
 has  nonzero curvature concentrated (in a Dirac-delta manner) on these hyperplanes. Correspondingly, the decay at
 large distances of $ h_{\mu \nu}^{m_a=0}(x)$ (in harmonic gauge) is non uniform, and weaker in some directions than
 for its finite-mass analog. This raises 
 delicate issues about the convergence of the integrals appearing at each order of the PM expansion.
 Some of these issues have been discussed by D'Eath \cite{DEath:1976bbo} (who works with large but finite $\gamma$),
 and by Gruzinov and Veneziano \cite{Gruzinov:2014moa} (who argue that the massless limit, $\g \to \infty$, does exist).
 This issue might be alleviated by choosing a suitable (non harmonic) gauge for representing the physical content of the
 metric \eqref{AS}.
 
 We shall assume here that the formal PM perturbation theory for the scattering of two massless
 particles leads to well-defined integral expressions for  the vectorial momentum transfer 
 $\Delta p_{ \mu} \equiv \Delta p_{1 \mu}= p'_{1 \mu} - p_{1 \mu}= - \Delta p_{2 \mu}$.
 
The (incoming) vectorial impact parameter $b_\mu$ (such that $ b \cdot p_{1}(- \infty)=0 =  b \cdot p_{2}(- \infty)$)
is easily seen to be uniquely defined by the geometrical configuration made by the two incoming
(null) worldlines. One can then write $\Delta p_{ \mu}$ as a Poincar\'e-covariant function of $b_\mu$ and of the two
incoming momenta. As before the corresponding scalar  
\be
{\sf Q}(p_1,p_2,b) \equiv \sqrt{-t} \equiv \sqrt{\eta^{\mu \nu}\Delta p_{1 \mu} \Delta p_{1 \nu}}\,,
\ee
must be a Lorentz scalar covariantly constructed from the vectors
 $b_\mu$, and $p_{a \mu}$ (the latter denoting the incoming values of the momenta). 
 As $b_\mu$ is (by definition) orthogonal to the two momenta, and as the momenta have vanishing Lorentz norms,
 the only non-zero scalar product (besides $b^2 \equiv (b \cdot b)$) that can be extracted from the geometrical configuration $p_1,p_2,b$ is
 the scalar product $ |( p_1 \cdot p_2)| =  - (p_1 \cdot p_2)$. 
[We  assume that $p_1$ and $p_2$ are both future-oriented so that $(p_1 \cdot p_2) <0$.]
This technical fact can be geometrically understood as follows.
 After fixing the vectorial impact parameter $b_\mu$,
 the geometrical configuration defined by the two incoming null worldlines admits as symmetry group
 the subgroup of the Lorentz group made of boosts acting in the two-plane 
 spanned by the two null vectors $p_1$ and $p_2$. If we consider a null frame with two
 null vectors $\ell^\mu $, $n^\mu$, respectively parallel to  $p_1$ and $p_2$,
 but normalized so that $\ell \cdot n = -1$, these boosts are parametrized by a scalar $k$
 (equal to $\sqrt{(1-v)/(1+v)}$ in terms of the usual boost velocity $v$) acting on the null
 frame $\ell, n$ as $\ell \to k \ell$, $n \to k^{-1} n$. These boosts change the components
 of $p_1$ and $p_2$ along the null basis vectors $\ell, n$ (say $p_1^\mu =p_{1 \ell} \ell^\mu$
 and $p_2^\mu =p_{2 n} n^\mu$) by factors $k^{-1} $ and $k$, respectively. 
 The Lorentz scalar  ${\sf Q}(p_1,p_2,b)$ must be invariant under these Lorentz frame transformations.
[One could gauge-fix this residual Lorentz symmetry by going to the c.m. frame where the
spatial components of $p_1$ and $p_2$ are opposite, but the idea here is, on the contrary,
to use this symmetry to constrain the expression of ${\sf Q}(p_1,p_2,b)$.]

Summarizing:  The (classical) scalar  momentum transfer ${\sf Q}(p_1,p_2,b)$ can only be a function
of the two scalars $ |( p_1 \cdot p_2)| =  - (p_1 \cdot p_2)$ and $b$. 

The first term in the PM expansion of ${\sf Q}(p_1,p_2,b)$ is obtained by taking the  massless limit $m_a \to 0$, $p_a^2 \to 0$ 
(equivalent to considering the HE limit) of the
beginning of the finite-mass expression of ${\sf Q}(p_1,p_2,b; m_1,m_2)$:
\bea
&&\frac12 {\sf Q}(p_1,p_2,b,m_1,m_2)=  \frac{G}{b} \frac{2 (p_1.p_2)^2- p_1^2 p_2^2}{ \sqrt{(p_1.p_2)^2- p_1^2 p_2^2}} \nonumber\\
&+&  \frac{3\pi}{8}  \frac{G^2}{b^2} \frac{(m_1+m_2) (5 (p_1.p_2)^2- p_1^2 p_2^2)}{\sqrt{(p_1.p_2)^2- p_1^2 p_2^2}} 
+ O(G^3)\,. \nonumber\\
\eea
This yields
\be
\frac12 {\sf Q}(p_1,p_2,b,0,0) =  \frac{2 G |(p_1.p_2)|}{b} + O\left( \frac{G^3}{b^3}\right)\,.
\ee
The structure of PM perturbation theory formally generates, at each PM order $G^N$, an expression for
${\sf Q}(p_1,p_2,b)$ that is a homogeneous polynomial of order $N+1$ in $p_{1 \ell}$ and $p_{2 n}$,
and that is proportional to $1/b^N$.  Using now  dimensional analysis, and looking at the dimension of 
${\sf Q}(p_1,p_2,b) \sim \frac{ G |(p_1.p_2)|}{b}$, 
it is easy to see  that $N+1$ must be an even integer, and that the $O(G^N)$ contribution to
${\sf Q}(p_1,p_2,b)$ must be a polynomial (of order $(N+1)/2$) in the product of
components $p_{1 \ell} p_{2 n}$, i.e. in the scalar product $ |( p_1 \cdot p_2)| =  - (p_1 \cdot p_2)$. 
This leads to a PM expansion for ${\sf Q}(p_1,p_2,b)$ of the form
\bea \label{masslessQexp}
\frac12 {\sf Q}^{\rm massless}(p_1,p_2,b) &=& 2 \frac{ G  |( p_1 \cdot p_2)| }{b} + {\sf Q}_3 \frac{ G^3  |( p_1 \cdot p_2)|^2 }{b^3} \nonumber\\
 &+& {\sf Q}_5 \frac{ G^5  |( p_1 \cdot p_2)|^3 }{b^5} + \cdots
\eea
with some {\it dimensionless} odd-order coefficients $ {\sf Q}_3$, $ {\sf Q}_5$, etc. The corresponding structure for the
scattering angle, considered as a function of 
\be
\alpha \equiv \frac{G |( p_1 \cdot p_2)|}{J} \overset{\rm HE}{=}\frac{\g}{j}  ,
\ee
 is
\be
\frac{\chi}{2} \overset{\rm HE}{=}  2 \alpha + c^{\chi}_3 \alpha^3 + c^{\chi}_5 \alpha^5 +  c^{\chi}_7 \alpha^7+ \cdots 
\ee
with some corresponding  dimensionless coefficients $c^{\chi}_3$,  $c^{\chi}_5$, etc. 

We have thereby recovered, at the classical level, the 
structure that was deduced, in the case of the HE quantum scattering, by Amati, Ciafaloni and Veneziano \cite{Amati:1990xe}
from analyticity requirements in $s$. We see that it follows from the classical symmetry discussed above.

Let us first emphasize that there are two possibilities concerning the dimensionless coefficients 
$ {\sf Q}_3$, $ {\sf Q}_5, \ldots$, or  $c^{\chi}_3$,  $c^{\chi}_5, \ldots$, which can be thought of corresponding
to the two possible conjectures made in the text.
The most conservative scenario is that the latter dimensionless coefficients are pure numbers. 
This would naturally correspond to our first conjecture (of a soft HE behavior). Indeed, the 
conjectured (HE-soft) 3PM dynamics \eqref{bCTDbis} leads to a non-zero $O(G^3/b^3)$ contribution in the HE limit
of the form  ${\sf Q}_3 G^3  |( p_1 \cdot p_2)|^2 / b^3 = {\sf Q}_3 G^3 (m_1 m_2)^2 \g^2/b^3$,
with a finite numerical coefficient ${\sf Q}_3$. 
Let us note that this term is the only term to survive, at $O(G^3)$, in the HE limit of the general finite-mass expression \eqref{QPM} 
because the corresponding coefficient ${\sf Q}_{12}^{3 \rm PM}(\g)$ grows like $\g^2$ when $\g \to \infty$, {\it i.e.}, faster than
${\sf Q}_{11}^{3 \rm PM}(\g)={\sf Q}_{22}^{3 \rm PM}(\g) \sim \g$. [One can check that, at any PM order, all the coefficients 
${\sf Q}_{11\ldots1}$, or  ${\sf Q}_{22\ldots2}$ of the terms involving only one of the two masses, grow like $\g$ when $\g \to \infty$.]
A similar HE dominance $\sim \g^{n+1}$ of the coefficient 
of  $(m_1 m_2)^{n}$ at $(2n+1)$-PM ({\it e.g.} $ {\sf Q}_{1122}^{5 \rm PM}(\g) \sim \g^3$) would ensure that the HE limit
of Eq. \eqref{QPM} yields the form \eqref{masslessQexp}\footnote{I thank Gabriele Veneziano for a useful question concerning
this issue.}. Moreover, in that case, the vanishing of the even coefficients $c_{2n}^{\chi}$ implies a specific HE behavior for the
corresponding coefficients $q_{2n}^E(\g,\nu)$ in the PM expansion of the energy-gauge EOB $Q$ 
potential\footnote{Beware of not confusing the EOB $Q$ potential with the momentum transfer $\sQ$.};
\be
{\widehat Q}^E(u, \g,\nu)= u^2 q_2(\g,\nu)+  u^3 q_3(\g,\nu)+ u^4 q^E_4(\g,\nu)+ O(G^5) .
\ee
Under our present soft-HE-behavior assumption, we would have a {\it quadratic} HE behavior for $q_n^E(\g, \nu)$, 
namely relations of the type
\eqref{qnHE1} or \eqref{qnHE2}, with {\it $\nu$-independent} numerical coefficients $c_n^{qE}$ or $c_n^{qH}$.
Then the vanishing of the even asymptotic coefficients $ c_{2n}^{\chi}$ leads 
to the following links
\be
c_2^{qE}=c_2^{qH}= \frac{15}{2}\,,
\ee
which we already knew, and the new links
\be
c_3^{qE}=c_3^{qH}=-  c_3^{\chi}+\frac{64}{3}- 2 c_2^{qE}= -  c_3^{\chi}+ \frac{19}{3}\,,
\ee
and
\be
c_4^{qE}=-  3 c_3^{qE} + \frac{705}{16}= 3  c_3^{\chi}+  \frac{401}{16}\,.
\ee
For instance, the first conjectured 3PM result \eqref{bCTDbis} implies $c_3^{\chi}= - \frac{14}{3}$,
which would, in turn, imply the following results
\be
c_3^{qE}= + 11\,, \; {\rm and} \; \; c_4^{qE}= \frac{177}{16}\,.
\ee
In other words, the  corresponding HE mass-shell condition would read
\be
- \e^2 \left( \frac{1}{1-2u} - {\bar f}(u)\right) + (1-2u) p_r^2 + j^2 u^2=0\,,
\ee
with ${\bar f}(u) \equiv \frac{15}{2} u^2 + 11 u^3 +  \frac{177}{16} u^4 + O(u^5)$.
Though the specific soft-HE conjecture \eqref{bCTDbis} is now disproved, we mention these facts here 
to emphasize that, under the present (soft HE) assumption, one can derive 4PM-level
information from a 3PM-level one (similarly to  Eq. (7.14) of \cite{Damour2018} which used the ACV result
as input information).

Let us now discuss the impact of our alternative conjecture, exemplified above by 
the additional piece \eqref{deltaQ0}. In such a scenario, the result \eqref{chiHEnew} shows that
the dimensionless coefficient $ {\sf Q}_3$ is no longer a pure number but involves 
the logarithm of the dimensionless quantity 
\be
\delta = \frac{G  E_{\rm c.m.}}{b}=\frac{G  \sqrt{s}}{b} \overset{\rm HE}{=}\frac{G  \sqrt{2 |(p_1.p_2)|}}{b}\,.
\ee
As we indicated in subsection \ref{transmutation} above, the corresponding ($\propto G^3 \ln G$) $\ln \delta$ contribution 
to $ {\sf Q}_3$ has descended from a $O(G^4)$ contribution in the usual finite-mass PM expansion. We can similarly expect
that the higher odd-order coefficients ${\sf Q}_{2n+1}$ will also involve the logarithm of $\delta$.
Note that we are talking here about logarithmic contributions that might occur in the  scattering angle of {\it classical} massless particles.
The analytic structure of the scattering angle of {\it quantum} massless particle might be different, notably if analyticity
requirements forbid the presence of $\ln s$ (and therefore $\ln \alpha$) in $\chi$.

%%%%%%%%% C

\section{On the interplay between the SF expansion, the HE behavior and the PM expansion}

Let us show how SF theory gives us access to some structural information about the HE limit of the scattering angle.
We can use a reasoning which generalizes the one used in Ref. \cite{Akcay:2012ea} to understand the  HE behavior
found there when considering 1SF expanded quantities near the light ring.  
 
Let us imagine analytically computing the SF
expansion for the total change of momentum of a small-mass particle (say of mass $m_1$) scattering (at some given
impact parameter, or with some given angular momentum)
on a large-mass black hole (say of mass $m_2 \gg m_1$).  It can be formally obtained by replacing
on the right-hand side of
\be \label{deltapmu3}
\Delta u_{1 \mu}
= \frac{1}{2}\int_{- \infty}^{+\infty} \,\D_\mu g_{\alpha \beta}(x_a) \, u_a^{ \alpha} dx_a^{ \beta}\,,
\ee
$g_{\alpha \beta}$ by $g^{(0)}_{\alpha \beta}(m_2) +  h_{\alpha \beta} $ (and
correlated $O(\nu)$ changes in $u_a^{ \alpha}$ and the worldline). Here, the perturbation $h_{\alpha \beta}$ of the
metric must be determined by solving the linearized perturbed Einstein equations (around $g^{(0)}_{\alpha \beta}(m_2)$), say
\be
\frac{\delta G^{\mu \nu}}{\delta g_{\alpha \beta}}\left[ h_{\alpha \beta}\right]=  8\pi G m_1 \int    u_1^{\mu} dx_1^{\nu} \, \frac{\delta^4(x-x_1(s_1))}{\sqrt{g}}\,.
\ee
If we consider an {\it ultra-relativistic motion} ($ u_1^{\mu} \gg1$, keeping the product $ m_1 u_1^{\mu}$ small)
of the small particle, the perturbation $h_{\alpha \beta}$ of the metric (which is sourced by $ m_1 u_1^{\mu}$)
will be proportional to, say, the conserved energy ${\mathcal E}_1= - m_1 u_1^{\mu} \xi_\mu$ (where $\xi^\mu$
is the time-translation Killing vector of the background $g^{(0)}_{\alpha \beta}(m_2)$). A direct consequence of this simple remark is that 
the fractional 1SF change in the scattering angle will be of order $O( {\mathcal E}_1/m_2)$, rather than
the naive estimate $O(m_1/m_2)$ that holds for particles with velocities small or comparable to the velocity of light.
In the EOB formalism, the 1SF effects are described by the linear-in-$\nu$ piece in the mass-shell term $Q$.
The previous reasoning shows that, when considering the small back-reaction ultrarelativistic double limit where
$-  u_1^{\mu} \xi_\mu \to \infty$, $m_1 \to 0$ with  ${\mathcal E}_1= - m_1 u_1^{\mu} \xi_\mu$ fixed
but much smaller than $m_2$, i.e. a limit where one {\it first} expands to linear order in $\nu$, and {\it then}
formally considers the limit where $\g = - (p_1 \cdot p_2)/(m_1 m_2) \approx {\mathcal E}_1/m_1 \to \infty$,
one will have fractional corrections to $\chi$ of order $\nu \g$. In other words, if we define the 1SF contribution
to the scattering function $\chi(\g,j;\nu)$ by writing 
\be \label{chi1SF}
\chi(\g, j,\nu) =\chi^{\rm Schw}(\g,j)+  \nu \, \chi^{1 \rm SF}(\g,j) +O(\nu^2)\,,
\ee
we expect the ratio $\chi^{1 \rm SF}(\g,j)/\g$ to have a finite limit as $\g \to \infty$, when keeping fixed
the impact parameter, and therefore the ratio $\alpha \equiv \frac{\g}{j}$, say
\be \label{Fa}
\lim_{\g \to \infty} \frac{\chi^{1 \rm SF}(\g, \g/\alpha)}{ 2\, \g}= F(\alpha)\,.
\ee
The leading order (LO) contribution to the so-defined function $ F(\alpha)$ is $O(\alpha^2)$ and comes from the 
2PM-level term $  \chi_2(\g)/(h(\g,\nu) j^2)$ in the PM expansion of $\frac12 \chi(\g, j,\nu)$,
\be
\frac12 \chi(\g, j,\nu)= \frac{\chi_{1}^{\rm Schw}(\g)}{j} + \frac{ \chi_{2}^{\rm Schw}(\g)}{h(\g,\nu) j^2} + \ldots
\ee
when expanding $1/h(\g,\nu) = 1 - \nu (\g-1) + O(\nu^2)$.

The limiting behavior \eqref{Fa} would directly follow from the first conjecture made above, namely a tame HE behavior.
Indeed, we have proven above that the PM expansion coefficients $\chi_{n}(\g,\nu)$ and $q_{n}(\g,\nu)$
had a restricted dependence on the symmetric mass ratio $\nu$ described through the interplay of some
$\g$-dependent building blocks $\widehat \chi_{n}^{(p)}(\g)$ and $\widehat q_{n}^{(p)}(\g)$ with
some powers of the function $h(\g,\nu)$. More precisely, we obtained formulas of the following form
\be \label{chinhexp}
\chi_{n}(\g, \nu)=  \frac{\widehat \chi_{n}^{(n-1)}(\g)}{h^{n-1}(\g,\nu)}  +  \frac{\widehat \chi_{n}^{(n-3)}(\g)}{h^{n-3}(\g,\nu)} + \cdots
\ee
or
\bea \label{chinhexp2}
\chi_{n}(\g, \nu) &-& \chi_{n}^{\rm Schw}(\g) =\widehat \chi_{n}^{(n-1)}(\g) \left(\frac1{h^{n-1}(\g,\nu)}-1\right) \nonumber\\
  &+&  \widehat \chi_{n}^{(n-3)}(\g)\left(\frac1{h^{n-3}(\g,\nu)}-1\right) + \cdots
\eea
and
\bea\label{qnhexp}
q_{2}(\g, \nu)&=&  \widehat q_2^{(1)}(\g) \left(1- \frac{1}{h(\g,\nu)}  \right) , \nonumber\\
q_{3}(\g, \nu)&=&\widehat q_3^{(1)}(\g) \left(1- \frac{1}{h(\g,\nu)} \right) + \widehat q_3^{(2)}(\g) \left(1- \frac{1}{h^2(\g,\nu)} \right) ,\nonumber\\
q_{4}(\g, \nu)&=&\widehat q_4^{(1)}(\g) \left(1- \frac{1}{h(\g,\nu)} \right) + \widehat q_4^{(2)}(\g) \left(1- \frac{1}{h^2(\g,\nu)} \right) \nonumber\\
 &+&   \widehat q_4^{(3)}(\g) \left(1- \frac{1}{h^3(\g,\nu)} \right) .
\eea
The conjecture of a tame HE behavior would be related to assuming that the building blocks $\widehat q_n^{(p)}(\g)$
of the EOB potentials have a uniform HE behavior of the type
\be \label{HEhatqin}
\widehat q_n^{(p)}(\g)  \overset{\rm HE}{\sim} \g^2\,.
\ee
This behavior  holds for the building blocks $\widehat q_2^{(1)}(\g)$,  $ \widehat q_3^{(1)}(\g)$
entering the first two PM contributions, namely
\be\label{hatq21}
\widehat q_2^{(1)}(\g)= \frac32 (5 \g^2-1)\,,
\ee
\be\label{hatq31}
\widehat q_3^{(1)}(\g)= -\frac{2 \g^2-1}{\g^2-1}  \widehat q_2^{(1)}(\g)= -\frac32 \frac{(2 \g^2-1)(5 \g^2-1)}{\g^2-1}\,.
\ee
In addition,  our first conjecture, Eq. \eqref{bCTDbis}, for modifying the 3PM dynamics by softening its HE behavior,
would imply that the same behavior  holds for 
the other 3PM-level function $\widehat q_3^{(2)}(\g)$ (which is essentially
a different notation for the function denoted $C(\g)$ in Eq. \eqref{q3vsABC}).

When transcribed in terms of the related building blocks $\widehat \chi_{n}^{(p)}(\g)$,
one finds that the general conjectural HE behavior \eqref{HEhatqin} would imply the following uniform HE behavior
\be \label{HEhatchin}
\widehat \chi_{n}^{(p)}(\g) \overset{\rm HE}{\sim }\g^n.
\ee
In turn, when inserting  the HE behavior \eqref{HEhatchin} in the SF expansion of Eq. \eqref{chinhexp2} (with $1/h^p(\g,\nu) = 1 - p \nu (\g-1) + O(\nu^2)$), we find that 
the 1SF contribution to each coefficient $\chi_n(\g,\nu)$, defined as, 
\be
\chi_n(\g,\nu) =\chi_n^{\rm Schw}(\g)+  \nu \, \chi_n^{1 \rm SF}(\g) +O(\nu^2)\,,
\ee
 would then  behave as
\be \label{HEchinSF}
\chi_n^{1 \rm SF}(\g)  \sim \g^{n+1}\; {\rm as } \; \g \to \infty\,.
\ee
Finally, the latter HE behavior would be consistent with the existence of the limiting function $F(\alpha)$, \eqref{Fa},
if we assume (as holds within our presently assumed soft HE behavior) that the HE limit ($\g \to \infty$)
commutes with the PM expansion ({\it i.e.} the expansion in powers of $1/j$ defining the various coefficients 
$\chi_n(\g,\nu) $).  Furthermore the HE behavior \eqref{HEchinSF} is directly related to \eqref{HEhatqin},
which  predicts  that the 1SF expansion
 of the mass-shell potential $Q$ would be compatible, at each separate PM order, with the HE behavior found
 in Ref. \cite{Akcay:2012ea}, namely the existence of a finite limit for the ratio
 $\frac{  {\widehat Q}^{\, n \rm PM \,1SF} }{\g^3}$  when $\g \to \infty$.
 
 Summarizing: the conjectural scalings, Eqs. \eqref{HEhatqin}, \eqref{HEhatchin},  \eqref{HEchinSF},
 (based on the assumption of a tame HE behavior at each PM order) have been presented here as a simple
 way to transcribe within PM gravity the (independently derived) SF results, Eqs. \eqref{chi1SF}, \eqref{Fa}.
  However, the recent disproof \cite{Blumlein:2020znm,Cheung:2020gyp,Bini:2020wpo}
 of our (first conjectured) HE-soft 3PM dynamics, Eq. \eqref{bCTDbis}, shows that our search for 
 a unified understanding of the interplay between the SF expansion, the HE behavior and the PM expansion
must be done within a wider framework.    
  We have exemplified above, in Eq. \eqref{newQ3PM}, that another type of  conjecture 
  might reconcile the SF result of Ref. \cite{Akcay:2012ea} with the  logarithmically untame HE behavior of 
  the 3PM dynamics of Refs. \cite{Bern:2019nnu,Bern:2019crd}. We leave to future work a discussion
  of how the interplay between the various non commuting limits $\nu \to 0$, $\g \to \infty$, and $G \to 0$
  might work when using similar structures at higher PM orders.

%%%%%%%%%


\begin{thebibliography}{99}

\bibitem{Abbott:2016blz} 
  B.~P.~Abbott {\it et al.} [LIGO Scientific and Virgo Collaborations],
  ``Observation of Gravitational Waves from a Binary Black Hole Merger,''
  Phys.\ Rev.\ Lett.\  {\bf 116},  061102 (2016)
  %doi:10.1103/PhysRevLett.116.061102
  [arXiv:1602.03837 [gr-qc]].
  
\bibitem{Abbott:2016nmj} 
  B.~P.~Abbott {\it et al.} [LIGO Scientific and Virgo Collaborations],
  ``GW151226: Observation of Gravitational Waves from a 22-Solar-Mass Binary Black Hole Coalescence,''
  Phys.\ Rev.\ Lett.\  {\bf 116}, no. 24, 241103 (2016)
  %doi:10.1103/PhysRevLett.116.241103
  [arXiv:1606.04855 [gr-qc]].
  
\bibitem{Abbott:2017vtc}
  B.~P.~Abbott {\it et al.} [LIGO Scientific and VIRGO Collaborations],
  ``GW170104: Observation of a 50-Solar-Mass Binary Black Hole Coalescence at Redshift 0.2,''
  Phys.\ Rev.\ Lett.\  {\bf 118} (2017) no.22,  221101.
  %doi:10.1103/PhysRevLett.118.221101
  
  %\cite{Abbott:2017oio}
\bibitem{Abbott:2017oio} 
  B.~P.~Abbott {\it et al.} [LIGO Scientific and Virgo Collaborations],
  ``GW170814: A Three-Detector Observation of Gravitational Waves from a Binary Black Hole Coalescence,''
  Phys.\ Rev.\ Lett.\  {\bf 119}, no. 14, 141101 (2017)
  %doi:10.1103/PhysRevLett.119.141101
  [arXiv:1709.09660 [gr-qc]].
  
  %\cite{Laddha:2018rle}
\bibitem{Laddha:2018rle} 
  A.~Laddha and A.~Sen,
  ``Gravity Waves from Soft Theorem in General Dimensions,''
  JHEP {\bf 1809}, 105 (2018)
 % doi:10.1007/JHEP09(2018)105
  [arXiv:1801.07719 [hep-th]].
  
 %\cite{Corinaldesi:1956}
\bibitem{Corinaldesi:1956} 
  E.~Corinaldesi,
  ``The Two-body Problem in the Theory of the Quantized Gravitational Field,''
  Proceedings of the Physical Society, Section A, {\bf 69}, Issue 3, pp. 189-195 (1956)
  %doi:10.1088/0370-1298/69/3/301 
  
  %\cite{Barker:1966zz}
\bibitem{Barker:1966zz} 
  B.~M.~Barker, S.~N.~Gupta and R.~D.~Haracz,
  ``One-Graviton Exchange Interaction of Elementary Particles,''
  Phys.\ Rev.\  {\bf 149}, 1027 (1966).
 % doi:10.1103/PhysRev.149.1027
  
  %\cite{Barker:1970zr}
\bibitem{Barker:1970zr} 
  B.~M.~Barker and R.~F.~O'Connell,
  ``Derivation of the equations of motion of a gyroscope from the quantum theory of gravitation,''
  Phys.\ Rev.\ D {\bf 2}, 1428 (1970).
 % doi:10.1103/PhysRevD.2.1428
 
  %\cite{Corinaldesi:1971sz}
\bibitem{Corinaldesi:1971sz} 
  E.~Corinaldesi,
  ``Einstein-hoffmann-infeld equations and quantized gravitation,''
  Lett.\ Nuovo Cim.\  {\bf 2}, 909 (1971).
 % [Lett.\ Nuovo Cim.\  {\bf 2}, 909 (1971)].
  %doi:10.1007/BF02778151
  
  %\cite{Iwasaki:1971vb}
\bibitem{Iwasaki:1971vb} 
  Y.~Iwasaki,
  ``Quantum theory of gravitation vs. classical theory. - fourth-order potential,''
  Prog.\ Theor.\ Phys.\  {\bf 46}, 1587 (1971).
  %doi:10.1143/PTP.46.1587
  
  %\cite{Okamura:1973my}
\bibitem{Okamura:1973my} 
  H.~Okamura, T.~Ohta, T.~Kimura and K.~Hiida,
  ``Perturbation calculation of gravitational potentials,''
  Prog.\ Theor.\ Phys.\  {\bf 50}, 2066 (1973).
  %doi:10.1143/PTP.50.2066

\bibitem{Landau4}
	V. B. Berestetskii, E. M. Lifshitz, and L. P.  Pitaevskii,   
 {\it  Relativistic Quantum Theory ( Volume 4 part 1 of A Course of Theoretical Physics) }, (Pergamon Press, Oxford,1971).
 
 %\cite{Donoghue:1993eb}
\bibitem{Donoghue:1993eb} 
  J.~F.~Donoghue,
  ``Leading quantum correction to the Newtonian potential,''
  Phys.\ Rev.\ Lett.\  {\bf 72}, 2996 (1994)
  %doi:10.1103/PhysRevLett.72.2996
  [gr-qc/9310024].
  
  %\cite{Donoghue:1994dn}
\bibitem{Donoghue:1994dn} 
  J.~F.~Donoghue,
  ``General relativity as an effective field theory: The leading quantum corrections,''
  Phys.\ Rev.\ D {\bf 50}, 3874 (1994)
  %doi:10.1103/PhysRevD.50.3874
  [gr-qc/9405057].

%\cite{BjerrumBohr:2002kt}
\bibitem{BjerrumBohr:2002kt} 
  N.~E.~J.~Bjerrum-Bohr, J.~F.~Donoghue and B.~R.~Holstein,
  ``Quantum gravitational corrections to the nonrelativistic scattering potential of two masses,''
  Phys.\ Rev.\ D {\bf 67}, 084033 (2003)
  Erratum: [Phys.\ Rev.\ D {\bf 71}, 069903 (2005)]
  %doi:10.1103/PhysRevD.71.069903, 10.1103/PhysRevD.67.084033
  [hep-th/0211072].
  
%\cite{Bjerrum-Bohr:2013bxa}
\bibitem{Bjerrum-Bohr:2013bxa} 
  N.~E.~J.~Bjerrum-Bohr, J.~F.~Donoghue and P.~Vanhove,
  ``On-shell Techniques and Universal Results in Quantum Gravity,''
  JHEP {\bf 1402}, 111 (2014)
  %doi:10.1007/JHEP02(2014)111
  [arXiv:1309.0804 [hep-th]].

 %\cite{Neill:2013wsa}
\bibitem{Neill:2013wsa} 
  D.~Neill and I.~Z.~Rothstein,
  ``Classical Space-Times from the S Matrix,''
  Nucl.\ Phys.\ B {\bf 877}, 177 (2013)
  %doi:10.1016/j.nuclphysb.2013.09.007
  [arXiv:1304.7263 [hep-th]].

%\cite{Cachazo:2017jef}
\bibitem{Cachazo:2017jef}
F.~Cachazo and A.~Guevara,
``Leading Singularities and Classical Gravitational Scattering,''
JHEP \textbf{02}, 181 (2020)
%doi:10.1007/JHEP02(2020)181
[arXiv:1705.10262 [hep-th]].

%\cite{Guevara:2017csg}
\bibitem{Guevara:2017csg} 
  A.~Guevara,
  ``Holomorphic Classical Limit for Spin Effects in Gravitational and Electromagnetic Scattering,''
  JHEP {\bf 1904}, 033 (2019)
  doi:10.1007/JHEP04(2019)033
  [arXiv:1706.02314 [hep-th]].



%\cite{Damour:2017zjx}
\bibitem{Damour2018} 
  T.~Damour,
  ``High-energy gravitational scattering and the general relativistic two-body problem,''
  Phys.\ Rev.\ D {\bf 97}, no. 4, 044038 (2018)
 % doi:10.1103/PhysRevD.97.044038
  [arXiv:1710.10599 [gr-qc]].

%\cite{Bjerrum-Bohr:2018xdl}
\bibitem{Bjerrum-Bohr:2018xdl} 
  N.~E.~J.~Bjerrum-Bohr, P.~H.~Damgaard, G.~Festuccia, L.~Plant\'e and P.~Vanhove,
  ``General Relativity from Scattering Amplitudes,''
  Phys.\ Rev.\ Lett.\  {\bf 121}, no. 17, 171601 (2018)
  %doi:10.1103/PhysRevLett.121.171601
  [arXiv:1806.04920 [hep-th]].

%\cite{Cheung:2018wkq}
\bibitem{Cheung:2018wkq} 
  C.~Cheung, I.~Z.~Rothstein and M.~P.~Solon,
  ``From Scattering Amplitudes to Classical Potentials in the Post-Minkowskian Expansion,''
  Phys.\ Rev.\ Lett.\  {\bf 121}, no. 25, 251101 (2018)
  % doi:10.1103/PhysRevLett.121.251101
  [arXiv:1808.02489 [hep-th]].

 %\cite{KoemansCollado:2019ggb}
\bibitem{KoemansCollado:2019ggb} 
  A.~Koemans Collado, P.~Di Vecchia and R.~Russo,
  ``Revisiting the second post-Minkowskian eikonal and the dynamics of binary black holes,''
  Phys.\ Rev.\ D {\bf 100}, no. 6, 066028 (2019)
  %doi:10.1103/PhysRevD.100.066028
  [arXiv:1904.02667 [hep-th]].
  
   %\cite{Bern:2019nnu}
\bibitem{Bern:2019nnu} 
  Z.~Bern, C.~Cheung, R.~Roiban, C.~H.~Shen, M.~P.~Solon and M.~Zeng,
  ``Scattering Amplitudes and the Conservative Hamiltonian for Binary Systems at Third Post-Minkowskian Order,''
  Phys.\ Rev.\ Lett.\  {\bf 122}, no. 20, 201603 (2019)
  %doi:10.1103/PhysRevLett.122.201603
  [arXiv:1901.04424 [hep-th]].
  
  %\cite{Bern:2019crd}
\bibitem{Bern:2019crd} 
  Z.~Bern, C.~Cheung, R.~Roiban, C.~H.~Shen, M.~P.~Solon and M.~Zeng,
  ``Black Hole Binary Dynamics from the Double Copy and Effective Theory,''
  JHEP {\bf 1910}, 206 (2019)
 % doi:10.1007/JHEP10(2019)206
  [arXiv:1908.01493 [hep-th]].
  

%\cite{Amati:1990xe}
\bibitem{Amati:1990xe} 
  D.~Amati, M.~Ciafaloni and G.~Veneziano,
  ``Higher Order Gravitational Deflection and Soft Bremsstrahlung in Planckian Energy Superstring Collisions,''
  Nucl.\ Phys.\ B {\bf 347}, 550 (1990).
  %doi:10.1016/0550-3213(90)90375-N
  
 %\cite{DiVecchia:2019kta}
\bibitem{DiVecchia:2019kta}
P.~Di Vecchia, S.~G.~Naculich, R.~Russo, G.~Veneziano and C.~D.~White,
``A tale of two exponentiations in $ \mathcal{N} $ = 8 supergravity at subleading level,''
JHEP \textbf{03}, 173 (2020)
%doi:10.1007/JHEP03(2020)173
[arXiv:1911.11716 [hep-th]]. 
 
%\cite{Bern:2020gjj}
\bibitem{Bern2020} 
  Z.~Bern, H.~Ita, J.~Parra-Martinez and M.~S.~Ruf,
  ``Universality in the classical limit of massless gravitational scattering,''
  arXiv:2002.02459 [hep-th].  
  

%\cite{Guevara:2018wpp}
\bibitem{Guevara:2018wpp} 
  A.~Guevara, A.~Ochirov and J.~Vines,
  ``Scattering of Spinning Black Holes from Exponentiated Soft Factors,''
  JHEP {\bf 1909}, 056 (2019)
  %doi:10.1007/JHEP09(2019)056
  [arXiv:1812.06895 [hep-th]].
  
    %\cite{Chung:2018kqs}
\bibitem{Chung:2018kqs} 
  M.~Z.~Chung, Y.~T.~Huang, J.~W.~Kim and S.~Lee,
  ``The simplest massive S-matrix: from minimal coupling to Black Holes,''
  JHEP {\bf 1904}, 156 (2019)
  %doi:10.1007/JHEP04(2019)156
  [arXiv:1812.08752 [hep-th]].

%\cite{Guevara:2019fsj}
\bibitem{Guevara:2019fsj}
A.~Guevara, A.~Ochirov and J.~Vines,
``Black-hole scattering with general spin directions from minimal-coupling amplitudes,''
Phys. Rev. D \textbf{100}, no.10, 104024 (2019)
%doi:10.1103/PhysRevD.100.104024
[arXiv:1906.10071 [hep-th]].
 

%\cite{Arkani-Hamed:2019ymq}
\bibitem{Arkani-Hamed:2019ymq} 
  N.~Arkani-Hamed, Y.~t.~Huang and D.~O'Connell,
  ``Kerr black holes as elementary particles,''
  JHEP {\bf 2001}, 046 (2020)
  %doi:10.1007/JHEP01(2020)046
  [arXiv:1906.10100 [hep-th]].  
 
  
 %\cite{Siemonsen:2019dsu}
\bibitem{Siemonsen:2019dsu}
N.~Siemonsen and J.~Vines,
``Test black holes, scattering amplitudes and perturbations of Kerr spacetime,''
Phys. Rev. D \textbf{101}, no.6, 064066 (2020)
%doi:10.1103/PhysRevD.101.064066
[arXiv:1909.07361 [gr-qc]]. 

%\cite{Kosower:2018adc}
\bibitem{Kosower:2018adc} 
  D.~A.~Kosower, B.~Maybee and D.~O'Connell,
  ``Amplitudes, Observables, and Classical Scattering,''
  JHEP {\bf 1902}, 137 (2019)
  %doi:10.1007/JHEP02(2019)137
  [arXiv:1811.10950 [hep-th]].

%\cite{Maybee:2019jus}
\bibitem{Maybee:2019jus} 
  B.~Maybee, D.~O'Connell and J.~Vines,
  ``Observables and amplitudes for spinning particles and black holes,''
  JHEP {\bf 1912}, 156 (2019)
  %doi:10.1007/JHEP12(2019)156
  [arXiv:1906.09260 [hep-th]].  
   
  \bibitem{Bertotti1956}
   B. Bertotti, `` On gravitational motion'', Nuovo Cimento {\bf 4}, pp. 898-906 (1956) 
  %doi:10.1007/BF02746175
  
  %\cite{Portilla:1980uz}
\bibitem{Portilla:1980uz} 
  M.~Portilla,
  ``Scattering Of Two Gravitating Particles: Classical Approach,''
  J.\ Phys.\ A {\bf 13}, 3677 (1980).
  %doi:10.1088/0305-4470/13/12/017

  %\cite{Ledvinka:2008tk}
\bibitem{Ledvinka:2008tk} 
  T.~Ledvinka, G.~Schaefer and J.~Bicak,
  ``Relativistic Closed-Form Hamiltonian for Many-Body Gravitating Systems in the Post-Minkowskian Approximation,''
  Phys.\ Rev.\ Lett.\  {\bf 100}, 251101 (2008)
  %doi:10.1103/PhysRevLett.100.251101
  [arXiv:0807.0214 [gr-qc]].

%\cite{Westpfahl:1979gu}
\bibitem{Westpfahl:1979gu} 
  K.~Westpfahl and M.~Goller,
  ``Gravitational Scattering Of Two Relativistic Particles In Postlinear Approximation,''
  Lett.\ Nuovo Cim.\  {\bf 26}, 573 (1979).
  %doi:10.1007/BF02817047

%\cite{Bel:1981be}
\bibitem{Bel:1981be} 
  L.~Bel, T.~Damour, N.~Deruelle, J.~Ibanez and J.~Martin,
  ``Poincar\'e-invariant gravitational field and equations of motion of two pointlike objects: The postlinear approximation of general relativity,''
  Gen.\ Rel.\ Grav.\  {\bf 13}, 963 (1981).
 % doi:10.1007/BF00756073

 \bibitem{Westpfahl:1985}
K. Westpfahl, 
``High-Speed Scattering of Charged and Uncharged Particles in General Relativity,"
Fortschr. Physik {\bf 33}, 417 (1985).
%DOI: 10.1002/prop.2190330802

\bibitem{Westpfahl:1987}
K. Westpfahl, R. M\"ohles and H Simonis
``Energy-momentum conservation for gravitational two-body scattering in the post-linear approximation,"
Classical and Quantum Gravity,  {\bf 4}, L185 (1987).
%DOI: 10.1088/0264-9381/4/5/006

%\cite{Buonanno:1998gg} 
\bibitem{Buonanno:1998gg}
A.~Buonanno and T.~Damour,
``Effective one-body approach to general relativistic two-body dynamics,''
Phys.\ Rev.\ D \textbf{59}, 084006 (1999)
[arXiv:gr-qc/9811091].

%\cite{Buonanno:2000ef} 
\bibitem{Buonanno:2000ef}
A.~Buonanno and T.~Damour,
``Transition from inspiral to plunge in binary black hole coalescences,''
Phys.\ Rev.\ D {\bf 62}, 064015 (2000)
[arXiv:gr-qc/0001013].

%\cite{Damour:2000we}
\bibitem{Damour:2000we}
T.~Damour, P.~Jaranowski, and G.~Sch\"afer,
``On the determination of the last stable orbit for circular general relativistic binaries at the third post-Newtonian approximation,''
Phys.\ Rev.\ D {\bf 62}, 084011 (2000)
[arXiv:gr-qc/0005034].

%\cite{Damour:2016gwp}
\bibitem{Damour:2016gwp} 
  T.~Damour,
  ``Gravitational scattering, post-Minkowskian approximation and Effective One-Body theory,''
  Phys.\ Rev.\ D {\bf 94}, no. 10, 104015 (2016)
  %doi:10.1103/PhysRevD.94.104015
  [arXiv:1609.00354 [gr-qc]].

  
 
  %\cite{Vines:2017hyw}
\bibitem{Vines:2017hyw} 
  J.~Vines,
  ``Scattering of two spinning black holes in post-Minkowskian gravity, to all orders in spin, and effective-one-body mappings,''
  Class.\ Quant.\ Grav.\  {\bf 35}, no. 8, 084002 (2018)
  %doi:10.1088/1361-6382/aaa3a8
  [arXiv:1709.06016 [gr-qc]].
  
   
  %\cite{Bini:2017xzy}
\bibitem{Bini:2017xzy} 
  D.~Bini and T.~Damour,
  ``Gravitational spin-orbit coupling in binary systems, post-Minkowskian approximation and effective one-body theory,''
  Phys.\ Rev.\ D {\bf 96}, no. 10, 104038 (2017)
  %doi:10.1103/PhysRevD.96.104038
  [arXiv:1709.00590 [gr-qc]].

 
 %\cite{Bini:2018ywr}
\bibitem{Bini:2018ywr} 
  D.~Bini and T.~Damour,
  ``Gravitational spin-orbit coupling in binary systems at the second post-Minkowskian approximation,''
  Phys.\ Rev.\ D {\bf 98}, no. 4, 044036 (2018)
  doi:10.1103/PhysRevD.98.044036
  [arXiv:1805.10809 [gr-qc]].


%\cite{Vines:2018gqi}
\bibitem{Vines:2018gqi} 
  J.~Vines, J.~Steinhoff and A.~Buonanno,
  ``Spinning-black-hole scattering and the test-black-hole limit at second post-Minkowskian order,''
  Phys.\ Rev.\ D {\bf 99}, no. 6, 064054 (2019)
  %doi:10.1103/PhysRevD.99.064054
  [arXiv:1812.00956 [gr-qc]].


  %\cite{Antonelli:2019ytb}
\bibitem{Antonelli:2019ytb} 
  A.~Antonelli, A.~Buonanno, J.~Steinhoff, M.~van de Meent and J.~Vines,
  ``Energetics of two-body Hamiltonians in post-Minkowskian gravity,''
  Phys.\ Rev.\ D {\bf 99}, no. 10, 104004 (2019)
  %doi:10.1103/PhysRevD.99.104004
  [arXiv:1901.07102 [gr-qc]].
  
  %\cite{Bini:2019nra}
\bibitem{Bini:2019nra} 
  D.~Bini, T.~Damour and A.~Geralico,
  ``Novel approach to binary dynamics: application to the fifth post-Newtonian level,''
   Phys.\ Rev.\ Lett.\  {\bf 123},  231104  (2019)
  [arXiv:1909.02375 [gr-qc].]
 
%\cite{Akcay:2012ea}
\bibitem{Akcay:2012ea} 
  S.~Akcay, L.~Barack, T.~Damour and N.~Sago,
  ``Gravitational self-force and the effective-one-body formalism between the innermost stable circular orbit and the light ring,''
  Phys.\ Rev.\ D {\bf 86}, 104041 (2012)
  %doi:10.1103/PhysRevD.86.104041
  [arXiv:1209.0964 [gr-qc]].

%\cite{Damourtalks}
\bibitem{Damourtalks}
T. Damour, talks given at:   QCD Meets Gravity IV, 10-14 December 2018 (Nordita, Stockholm, Sweden; 
see https://agenda.albanova.se/conferenceDisplay.py?confId=6238 for slides);
 Multi-Loop 2019, 14-15 May 2019 (Paris, France; see https://multi-loop-2019.sciencesconf.org/program for slides); and 
at Physics and Astrophysics in the Era of Gravitational Wave Detection, 19-23 August 2019 (Niels Bohr Institute,
Copenhagen, Danemark; see https://indico.nbi.ku.dk/event/1114/contributions/ for slides).

%\cite{Kalin:2019rwq}
\bibitem{Kalin:2019rwq} 
  G.~K\"alin and R.~A.~Porto,
  ``From Boundary Data to Bound States,''
  JHEP {\bf 2001}, 072 (2020)
  %doi:10.1007/JHEP01(2020)072
  [arXiv:1910.03008 [hep-th]].
 
%\cite{Bjerrum-Bohr:2019kec}
\bibitem{Bjerrum-Bohr:2019kec} 
  N.~E.~J.~Bjerrum-Bohr, A.~Cristofoli and P.~H.~Damgaard,
  ``Post-Minkowskian Scattering Angle in Einstein Gravity,''
  arXiv:1910.09366 [hep-th].
 
\bibitem{Bohr1948}  
  N. Bohr, 
  ``The penetration of atomic particles through matter,''
  Kgl. Danske Videnskab. Selskab, Mat.-Fys. Medd., {\bf 18}, No 8, pp 1-144 (1948).

 
 %\cite{Damour:1988mr}
\bibitem{Damour:1988mr} 
  T.~Damour and G.~Sch\"afer,
  ``Higher Order Relativistic Periastron Advances and Binary Pulsars,''
  Nuovo Cim.\ B {\bf 101}, 127 (1988).
  %doi:10.1007/BF02828697

%\cite{Bini:2017wfr}
\bibitem{Bini:2017wfr} 
  D.~Bini and T.~Damour,
  ``Gravitational scattering of two black holes at the fourth post-Newtonian approximation,''
  Phys.\ Rev.\ D {\bf 96}, no. 6, 064021 (2017)
  %doi:10.1103/PhysRevD.96.064021
  [arXiv:1706.06877 [gr-qc]].  

 \bibitem{LandauQM}
  L. D. Landau and  E. M. Lifshitz,
  {\it  Quantum Mechanics, Volume 3 of A Course of Theoretical Physics}, (Pergamon Press, Oxford,1965).

 \bibitem{FordWheeler1959}
 K. W. Ford and J. A. Wheeler,
 ``Semiclassical description of scattering,''
  Annals of Physics {\bf 7}, 259 (1959)
  
   
  
\bibitem{KangBrown}
Ik-Ju Kang and Laurie M. Brown,
``Higher Born Approximations for the Coulomb Scattering of a Spinless Particle'',
Phys. Rev. {\bf 128}, 2828  (1962)
   
    %\cite{Weinberg:1965nx}
\bibitem{Weinberg:1965nx} 
  S.~Weinberg,
  ``Infrared photons and gravitons,''
  Phys.\ Rev.\  {\bf 140}, B516 (1965).
  %doi:10.1103/PhysRev.140.B516  

%\cite{DeWitt:1967uc}
\bibitem{DeWitt:1967uc} 
  B.~S.~DeWitt,
  ``Quantum Theory of Gravity. 3. Applications of the Covariant Theory,''
  Phys.\ Rev.\  {\bf 162}, 1239 (1967).
  %doi:10.1103/PhysRev.162.1239

%\cite{tHooft:1987vrq}
\bibitem{tHooft:1987vrq} 
  G.~'t Hooft,
  ``Graviton Dominance in Ultrahigh-Energy Scattering,''
  Phys.\ Lett.\ B {\bf 198}, 61 (1987).
  %doi:10.1016/0370-2693(87)90159-6

 %\cite{Kabat:1992tb}
\bibitem{Kabat:1992tb} 
  D.~N.~Kabat and M.~Ortiz,
  ``Eikonal quantum gravity and Planckian scattering,''
  Nucl.\ Phys.\ B {\bf 388}, 570 (1992)
  %doi:10.1016/0550-3213(92)90627-N
  [hep-th/9203082].
  
  %\cite{Saotome:2012vy}
\bibitem{Saotome:2012vy} 
  R.~Saotome and R.~Akhoury,
  ``Relationship Between Gravity and Gauge Scattering in the High Energy Limit,''
  JHEP {\bf 1301}, 123 (2013)
  %doi:10.1007/JHEP01(2013)123
  [arXiv:1210.8111 [hep-th]].

   %\cite{Akhoury:2013yua}
\bibitem{Akhoury:2013yua} 
  R.~Akhoury, R.~Saotome and G.~Sterman,
  ``High Energy Scattering in Perturbative Quantum Gravity at Next to Leading Power,''
  arXiv:1308.5204 [hep-th].
  
 %\cite{DiVecchia:2019myk}
\bibitem{DiVecchia:2019myk} 
  P.~Di Vecchia, A.~Luna, S.~G.~Naculich, R.~Russo, G.~Veneziano and C.~D.~White,
  ``A tale of two exponentiations in ${\cal N}=8$ supergravity,''
  Phys.\ Lett.\ B {\bf 798}, 134927 (2019)
  %doi:10.1016/j.physletb.2019.134927
  [arXiv:1908.05603 [hep-th]]. 
  
 %\cite{Krachkov:2015uva}
\bibitem{Krachkov:2015uva} 
  P.~A.~Krachkov, R.~N.~Lee and A.~I.~Milstein,
  ``Small-angle scattering and quasiclassical approximation beyond leading order,''
  Phys.\ Lett.\ B {\bf 751}, 284 (2015)
  %doi:10.1016/j.physletb.2015.10.049
  [arXiv:1507.04111 [physics.atom-ph]]. 
 
  
%\cite{Damour:2009sm}
\bibitem{Damour:2009sm} 
  T.~Damour,
  ``Gravitational Self Force in a Schwarzschild Background and the Effective One Body Formalism,''
  Phys.\ Rev.\ D {\bf 81}, 024017 (2010)
  %doi:10.1103/PhysRevD.81.024017
  [arXiv:0910.5533 [gr-qc]].
  
  
  %\cite{Barack:2019agd}
\bibitem{Barack:2019agd} 
  L.~Barack, M.~Colleoni, T.~Damour, S.~Isoyama and N.~Sago,
  ``Self-force effects on the marginally bound zoom-whirl orbit in Schwarzschild spacetime,''
  Phys. \ Rev. \ D {\bf 100}, 124015 (2019)
  [arXiv:1909.06103 [gr-qc].]

  
  %\cite{LeTiec:2011ab}
\bibitem{LeTiec:2011ab} 
  A.~Le Tiec, L.~Blanchet and B.~F.~Whiting,
  ``The First Law of Binary Black Hole Mechanics in General Relativity and Post-Newtonian Theory,''
  Phys.\ Rev.\ D {\bf 85}, 064039 (2012)
  %doi:10.1103/PhysRevD.85.064039
  [arXiv:1111.5378 [gr-qc]].
  
  %\cite{LeTiec:2011dp}
\bibitem{LeTiec:2011dp} 
  A.~Le Tiec, E.~Barausse and A.~Buonanno,
  ``Gravitational Self-Force Correction to the Binding Energy of Compact Binary Systems,''
  Phys.\ Rev.\ Lett.\  {\bf 108}, 131103 (2012)
  %doi:10.1103/PhysRevLett.108.131103
  [arXiv:1111.5609 [gr-qc]].

 %\cite{Barausse:2011dq}
\bibitem{Barausse:2011dq} 
  E.~Barausse, A.~Buonanno and A.~Le Tiec,
  ``The complete non-spinning effective-one-body metric at linear order in the mass ratio,''
  Phys.\ Rev.\ D {\bf 85}, 064010 (2012)
  %doi:10.1103/PhysRevD.85.064010
  [arXiv:1111.5610 [gr-qc]].
 
 %\cite{Blumlein:2019bqq}
\bibitem{Blumlein:2019bqq} 
  J.~Bl\"umlein, A.~Maier, P.~Marquard, G.~Sch\"afer and C.~Schneider,
  ``From Momentum Expansions to Post-Minkowskian Hamiltonians by Computer Algebra Algorithms,''
  Phys.\ Lett.\ B {\bf 801}, 135157 (2020)
  %doi:10.1016/j.physletb.2019.135157
  [arXiv:1911.04411 [gr-qc]]. 
  
  
%\cite{Ciafaloni:2014esa}
\bibitem{Ciafaloni:2014esa} 
  M.~Ciafaloni and D.~Colferai,
  ``Rescattering corrections and self-consistent metric in Planckian scattering,''
  JHEP {\bf 1410}, 085 (2014)
  %doi:10.1007/JHEP10(2014)085
  [arXiv:1406.6540 [hep-th]].
  
%\cite{Blumlein:2020znm}
\bibitem{Blumlein:2020znm} 
  J.~Bl\"umlein, A.~Maier, P.~Marquard and G.~Sch\"afer,
  ``Testing binary dynamics in gravity at the sixth post-Newtonian level,''
  arXiv:2003.07145 [gr-qc].

%\cite{Cheung:2020gyp}
\bibitem{Cheung:2020gyp} 
  C.~Cheung and M.~P.~Solon,
  ``Classical Gravitational Scattering at ${\cal O}(G^3)$ from Feynman Diagrams,''
  arXiv:2003.08351 [hep-th].
  
%\cite{Bini:2020wpo}
\bibitem{Bini:2020wpo}
D.~Bini, T.~Damour and A.~Geralico,
``Binary dynamics at the fifth and fifth-and-a-half post-Newtonian orders,''
[arXiv:2003.11891 [gr-qc]].  
  

%\cite{Blanchet:1987wq}
\bibitem{Blanchet:1987wq} 
  L.~Blanchet and T.~Damour,
  ``Tail Transported Temporal Correlations in the Dynamics of a Gravitating System,''
  Phys.\ Rev.\ D {\bf 37}, 1410 (1988).
 % doi:10.1103/PhysRevD.37.1410
 
  
  %\cite{Foffa:2011np}
\bibitem{Foffa:2011np} 
  S.~Foffa and R.~Sturani,
  ``Tail terms in gravitational radiation reaction via effective field theory,''
  Phys.\ Rev.\ D {\bf 87}, no. 4, 044056 (2013)
  %doi:10.1103/PhysRevD.87.044056
  [arXiv:1111.5488 [gr-qc]].
  
  %\cite{Damour:2014jta}
\bibitem{Damour:2014jta} 
  T.~Damour, P.~Jaranowski and G.~Sch\"afer,
  ``Nonlocal-in-time action for the fourth post-Newtonian conservative dynamics of two-body systems,''
  Phys.\ Rev.\ D {\bf 89}, no. 6, 064058 (2014)
  %doi:10.1103/PhysRevD.89.064058
  [arXiv:1401.4548 [gr-qc]].
  
  
 %\cite{Blanchet:2010zd}
\bibitem{Blanchet:2010zd} 
  L.~Blanchet, S.~L.~Detweiler, A.~Le Tiec and B.~F.~Whiting,
  ``High-Order Post-Newtonian Fit of the Gravitational Self-Force for Circular Orbits in the Schwarzschild Geometry,''
  Phys.\ Rev.\ D {\bf 81}, 084033 (2010)
 % doi:10.1103/PhysRevD.81.084033
  [arXiv:1002.0726 [gr-qc]].
  
  
  %\cite{Damour:2015isa}
\bibitem{Damour:2015isa} 
  T.~Damour, P.~Jaranowski and G.~Sch\"afer,
  ``Fourth post-Newtonian effective one-body dynamics,''
  Phys.\ Rev.\ D {\bf 91}, no. 8, 084024 (2015)
  %doi:10.1103/PhysRevD.91.084024
  [arXiv:1502.07245 [gr-qc]].
  
%\cite{Galley:2015kus}
\bibitem{Galley:2015kus} 
  C.~R.~Galley, A.~K.~Leibovich, R.~A.~Porto and A.~Ross,
  ``Tail effect in gravitational radiation reaction: Time nonlocality and renormalization group evolution,''
  Phys.\ Rev.\ D {\bf 93}, 124010 (2016)
  %doi:10.1103/PhysRevD.93.124010
  [arXiv:1511.07379 [gr-qc]].  
  
%\cite{Foffa:2019eeb}
\bibitem{Foffa:2019eeb} 
  S.~Foffa and R.~Sturani,
  ``Hereditary Terms at Next-To-Leading Order in Two-Body Gravitational Dynamics,''
  Phys. Rev. D \textbf{101}, no. 6, 064033 (2020)
%doi:10.1103/PhysRevD.101.064033
[arXiv:1907.02869 [gr-qc]].
 
  
  %\cite{Gruzinov:2014moa}
\bibitem{Gruzinov:2014moa} 
  A.~Gruzinov and G.~Veneziano,
  ``Gravitational Radiation from Massless Particle Collisions,''
  Class.\ Quant.\ Grav.\  {\bf 33}, no. 12, 125012 (2016)
  %doi:10.1088/0264-9381/33/12/125012
  [arXiv:1409.4555 [gr-qc]].
  
%\cite{Ciafaloni:2015xsr}
\bibitem{Ciafaloni:2015xsr}
M.~Ciafaloni, D.~Colferai, F.~Coradeschi and G.~Veneziano,
``Unified limiting form of graviton radiation at extreme energies,''
Phys. Rev. D \textbf{93}, no.4, 044052 (2016)
%doi:10.1103/PhysRevD.93.044052
[arXiv:1512.00281 [hep-th]].  

%\cite{Peters:1970mx}
\bibitem{Peters:1970mx} 
  P.~C.~Peters,
  ``Relativistic gravitational bremsstrahlung,''
  Phys.\ Rev.\ D {\bf 1}, 1559 (1970).
  %doi:10.1103/PhysRevD.1.1559  
  
  %\cite{DEath:1976bbo}
\bibitem{DEath:1976bbo} 
  P.~D.~D'Eath,
  ``High Speed Black Hole Encounters and Gravitational Radiation,''
  Phys.\ Rev.\ D {\bf 18}, 990 (1978).
  %doi:10.1103/PhysRevD.18.990
  
  %\cite{Kovacs:1977uw}
\bibitem{Kovacs:1977uw} 
  S.~J.~Kovacs and K.~S.~Thorne,
  ``The Generation of Gravitational Waves. 3. Derivation of Bremsstrahlung Formulas,''
  Astrophys.\ J.\  {\bf 217}, 252 (1977).
  %doi:10.1086/155576
  
  %\cite{Kovacs:1978eu}
\bibitem{Kovacs:1978eu} 
  S.~J.~Kovacs and K.~S.~Thorne,
  ``The Generation of Gravitational Waves. 4. Bremsstrahlung,''
  Astrophys.\ J.\  {\bf 224}, 62 (1978).
  %doi:10.1086/156350
  
%\cite{Aichelburg:1970dh}
\bibitem{Aichelburg:1970dh} 
  P.~C.~Aichelburg and R.~U.~Sexl,
  ``On the Gravitational field of a massless particle,''
  Gen.\ Rel.\ Grav.\  {\bf 2}, 303 (1971).
  %doi:10.1007/BF00758149 

 
  
\end{thebibliography}
\end{document}